\documentclass[aps,twocolumn,footinbib]{revtex4-1}
\usepackage{amsmath}
\usepackage{graphicx}
\usepackage{tabularx}
\usepackage{dcolumn}
\usepackage{bm}
\usepackage[T1]{fontenc}
\usepackage[french]{babel}
\usepackage{epstopdf} 
\usepackage{lipsum}
\usepackage{amssymb}
\usepackage{color}
\usepackage{xcolor}
\usepackage{amsmath}
\usepackage{subfigure}
\usepackage{natbib}
\usepackage[outercaption,wide]{sidecap}

\usepackage{graphicx}
\usepackage{color}
\usepackage{subfigure}

\newcommand{\beq}{\begin{equation}}
\newcommand{\eeq}{\end{equation}}

\newcommand{\ta}{\tilde{a}}
\newcommand{\tL}{\tilde{L}_{I}}
\newcommand{\brho}{\bar{\rho}}
\newcommand{\bzeta}{\bar{\zeta}}
\newcommand{\bPsi}{\bar{\Psi}}

\newcommand{\geff}{\gamma_{\rm eff}}
\newcommand{\tdeltast}{\tdelta^{*}}
\newcommand{\tdeltastst}{\tdelta^{**}}

\newcommand{\Li}{L_I}
\newcommand{\Lo}{L_{\mathrm{out}}}
\newcommand{\Rf}{R_{\mathrm{sheet}}}

\newcommand{\Ksub}{K_{\rm sub}}
\newcommand{\usub}{u_{\rm sub}}
\newcommand{\Keff}{K_{\rm eff}}

\newcommand{\sqq}{\sigma_{\theta\theta}}
\newcommand{\srr}{\sigma_{rr}}

\newcommand{\tdelta}{\tilde{\delta}}

\newcommand{\rmu}{{\rm u}}
\newcommand{\rmur}{{\rm u}_{\rm r}}

\newcommand{\ellbc}{\ell_{bc}}
\newcommand{\sigo}{\sigma_{0}}

\begin{document}

\title{Indentation %supported sheets without clamping}
%metrology 
of solid membranes on rigid substrates with Van-der-Waals attraction}
%graphene and other two dimensional materials.}

\author{Benny Davidovitch$^{1}$}
   % \email[Correspondence email address: ]{bdavidov@umass.edu}% Your name
    \affiliation{$^1$ Department of Physics, University of Massachusetts Amherst, Amherst, MA 01003}

\author{Francisco Guinea$^{2,3}$}
%\email{vincent.demery@espci.psl.eu}
\affiliation{$^2$ IMDEA Nanoscience, C/Faraday 9, 28049 Madrid, Spain}
\affiliation{$^3$ Donostia International Physics Center, Paseo Manuel de Lardiz\'abal 4, 20018 San Sebasti\'an, Spain}
%\affiliation{Gulliver, CNRS, ESPCI Paris PSL, 10 rue Vauquelin, 75005 Paris, France}
%\affiliation{Univ Lyon, ENS de Lyon, Univ Claude Bernard Lyon 1, CNRS, Laboratoire de Physique, F-69342 Lyon, France}   

%\maketitle
%\author{}
%\affiliation{}

%\date{\today}

\begin{abstract}
We revisit the indentation of %classical problem of indenting 
a thin solid sheet of size $\Rf$ suspended on a circular hole of radius $R \ll \Rf$ in a smooth rigid substrate, addressing the effects of boundary conditions at the hole's edge. Introducing a basic theoretical model for the Van-der-Waals (VdW) sheet-substrate attraction, we demonstrate the dramatic effect of replacing the clamping condition (Schwerin model) with a sliding condition, whereby the supported part of the sheet is allowed to slide towards the indenter and relax the induced hoop compression through angstrom-scale deflections from the thermodynamic equilibrium (determined by the VdW potential). We highlight the possibility that the indentation force $F$ may not exhibit the commonly anticipated cubic dependence on the indentation depth ($F\propto \delta^3$), in which the proportionality constant is governed by the sheet's stretching modulus and the hole's radius $R$, but rather a {\emph{pseduo-linear}} response, $F \propto \delta$, whereby the proportionality constant is governed by the bending modulus, the VdW attraction, and the sheet's size $\Rf \gg R$.           
\end{abstract}
\keywords{elasticity, geometry, thin sheets}
 \maketitle
 \section{Introduction}
 \subsection{Background}
The mechanics of a solid membrane is determined by a balance between its rigidity for in-plane (strain) and out-of-plane (bending) deformations.  %elastic rigidity and the out-of-plane bending rigidity.  
For Graphene and other nanometer-thick crystalline two-dimensional (2D) membranes ({\emph{e.g.}} transition metal dichacogenides and black phosphorus), the in-plane stretching modulus $Y$ is very large, whereas the bending modulus $B$ is small, such that the characteristic length $ \ell_{bend} \approx \sqrt{\frac{B}{Y}}$ is much smaller than the system size $\Rf$ \cite{LWKH08,BBK11,QS13,CSZS14,RCCG15,Wetal16}. %, and is much smaller than the lateral length, $\Rf$, whose typical values are few $ \mu m$'s or more .  
% $\kappa$ is the bending rigidity, $Y = 4 ( \lambda + \mu )^2 / ( \lambda + 2 \mu )$ is the two dimensional Young modulus, and $\lambda$ and $\mu$ are elastic Lam\'e coefficients. 
Given the huge characteristic values of the von-Karman ratio, $vK =  (\Rf/\ell_{bend})^2$, %where the lateral  $\ell_{sheet}$ is typically few $ \mu m$'s or more, 
it is commonly assumed that 
%at distances $\ell \gtrsim \ell_{elas}$ 
the bending rigidity does not affect the mechanics, and the response to exerted forces %of the solid membrane 
is determined solely by the in-plane stiffness. While such an anticipation is justified when the exerted loads are purely tensile (\emph{e.g.} isotropic stretching of the sheet), it is obviously wrong to totally ignore the bending rigidity in the presence of compressive loads, as can be easily demonstrated by subjecting sheets to uniaxial compression  \cite{Milner89,Bowden98,Pocivavsek08,Huang10}. Here, the low bending rigidity underlies an instability of the compressed planar state, and the consequent formation of a strain-free buckled shape (if the sheet is suspended) or a wrinkle pattern (if the sheet is supported on a substrate) reflects the relevance of the  
%In such deformations, the %curvature and correspondingly the 
%characteristic length scale %in such buckled or wrinkled states 
%is governed by the minimization of 
bending energy at scales much larger than $\ell_{bend}$. In this paper we study a conceptually similar, yet nontrivial 
effect of the low bending rigidity in %. We address  
indentation problems, where %-- depending on boundary conditions -- 
radial tension induces compression in the azimuthal (hoop) direction, thereby making the weak bending energy a crucial player in the mechanical response of the sheet.   
\\
%the exerted force gives rise to radial tension, but -- depending on boundary conditions -- compression may emerge in the azimuthal (hoop) direction, making the the weak bending energy a crucial player in the mechanical response of the sheet.       

%The relative importance of out of plane bending and in plane stiffnes depends on the length scale $ \ell_{elas} \approx \sqrt{\frac{\kappa}{Y}}$ where $\kappa$ is the bending rigidity, $Y = 4 ( \lambda + \mu )^2 / ( \lambda + 2 \mu )$ is the two dimensional Young modulus, and $\lambda$ and $\mu$ are elastic Lam\'e coefficients. For distances $\ell \gtrsim \ell_{elas}$ the behavior of the membrane is determined solely by the in plane stiffness. For graphene, and also for transition metal dichacogenides and black phosphorus, this length is a few nanometers or less, so that in micron size membranes the bending rigidity in negligible.
%\subsection{Indentation measurements.}
Indentation experiments on suspended samples became a primary tool for measuring the stretching modulii of 2D materials \cite{NBK07,FTZM07,LWKH08,JP14,Fetal15,Letal15}. In a typical set-up, the sheet is supported on a thick, rigid substrate ({\emph{e.g.}} SiO), which contains a hole of radius $R \sim 1 \mu m$. A localized force is exerted by an AFM tip at the center of the suspended part of the sheet, and the force $F$ is measured as a function of the deflection $\delta$. 
% 
%The  elastic properties of two dimensional materials are experimentally studied  through indentation experiments in suspended samples\cite{LWKH08}. The elastic energy of a suspended membrane ican be efficiently relaxed through the formation of wrinkles\cite{CM03,DSVAC11}, and wrinkles have been observed in suspended graphene and in other two dimensional materials\cite{Betal09,Vetal11,BAK11}. 
%
%The tendency of very thin sheets to relax compression through wrinkles is particularly relevant in indentation experiments, which became a primary tool for measuring the stretching modulus of graphene and other 2D solid membranes.   
In most experiments \cite{LWKH08,Letal15} the stretching modulus, $Y$, is extracted by fitting the force-displacement curve, $F ( \delta )$, to a prediction of a ``membrane elasticity'' model, whereby the suspended   
%This model  neglects the bending rigidity, and assumes a uniform "pre-stress'', $\sigma_0$, in the sheet. The 
sheet is assumed to be clamped to the substrate at the edge of the hole. %over which the membrane is suspended, $r = R$. 
This assumption implies that the indentation-induced stress field in the sheet is purely tensile, %the assumption of clamping %and a uniform pre-stress 
and consequently has a dramatic influence on the estimated value of the stretching modulus \cite{Vella17}. However, an unequivocal, independent support for the validity of the clamping assumption has been lacking. Furthermore, since layers of Graphene (like graphite) are known to slide easily on each other (due to a very low inter-layer shear modulus), one may suspect that the interaction of graphene with a substrate is even weaker, such that the ``no-sliding'' assumption may not be satisfied. 
%can be particularly controversial. 
 \subsection{Sliding, wrinkling, and response to applied forces.}
\begin{figure}
%\hspace{-3.5cm}
\includegraphics[width=0.5\textwidth]{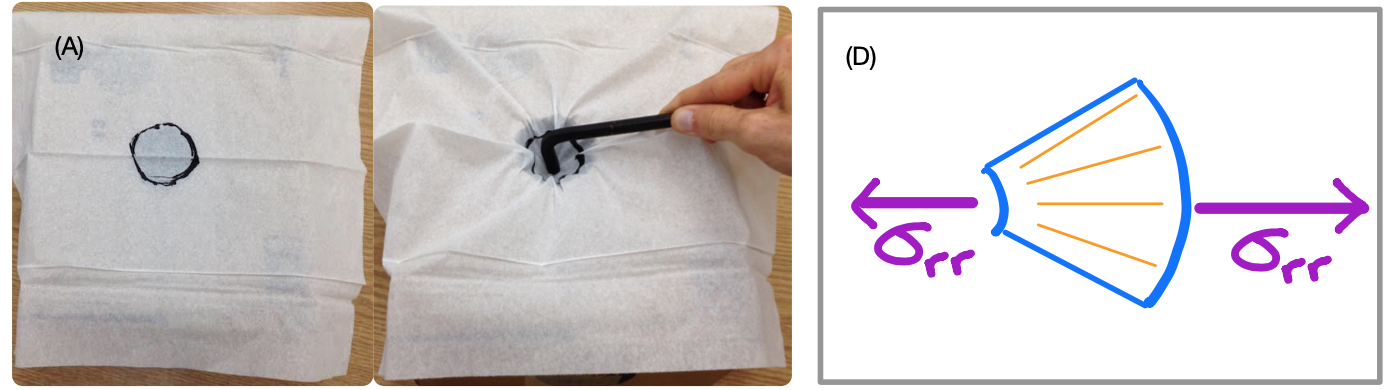}
\includegraphics[width=0.5\textwidth]{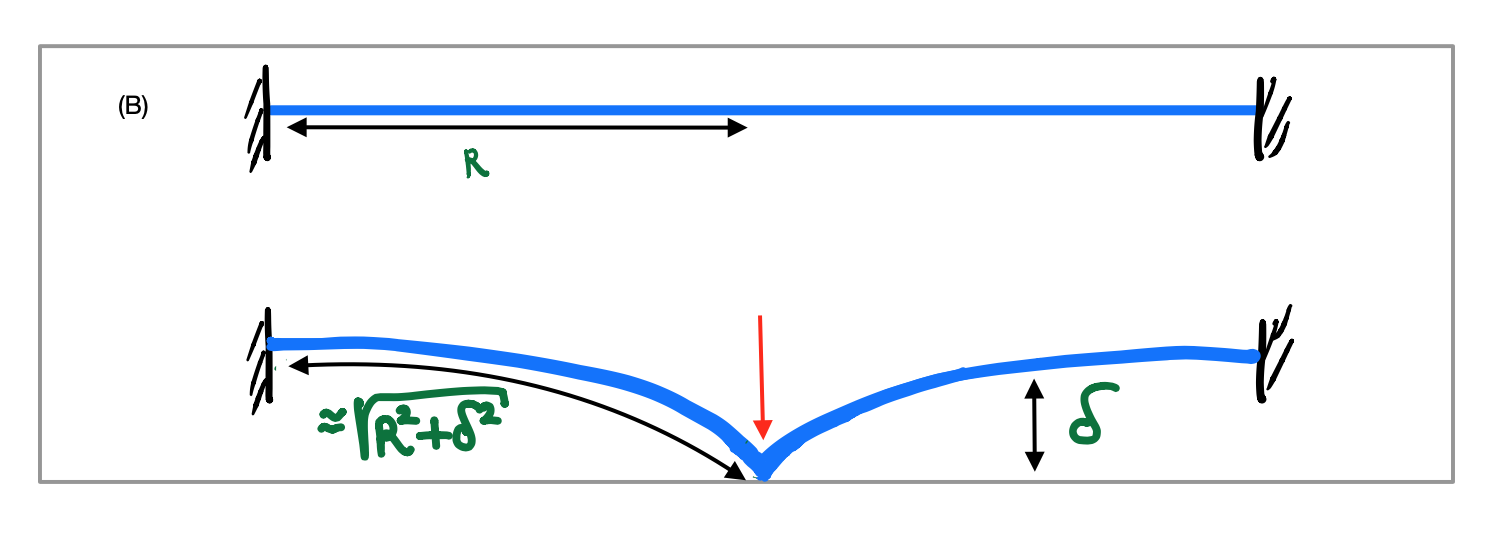}
\includegraphics[width=0.5\textwidth]{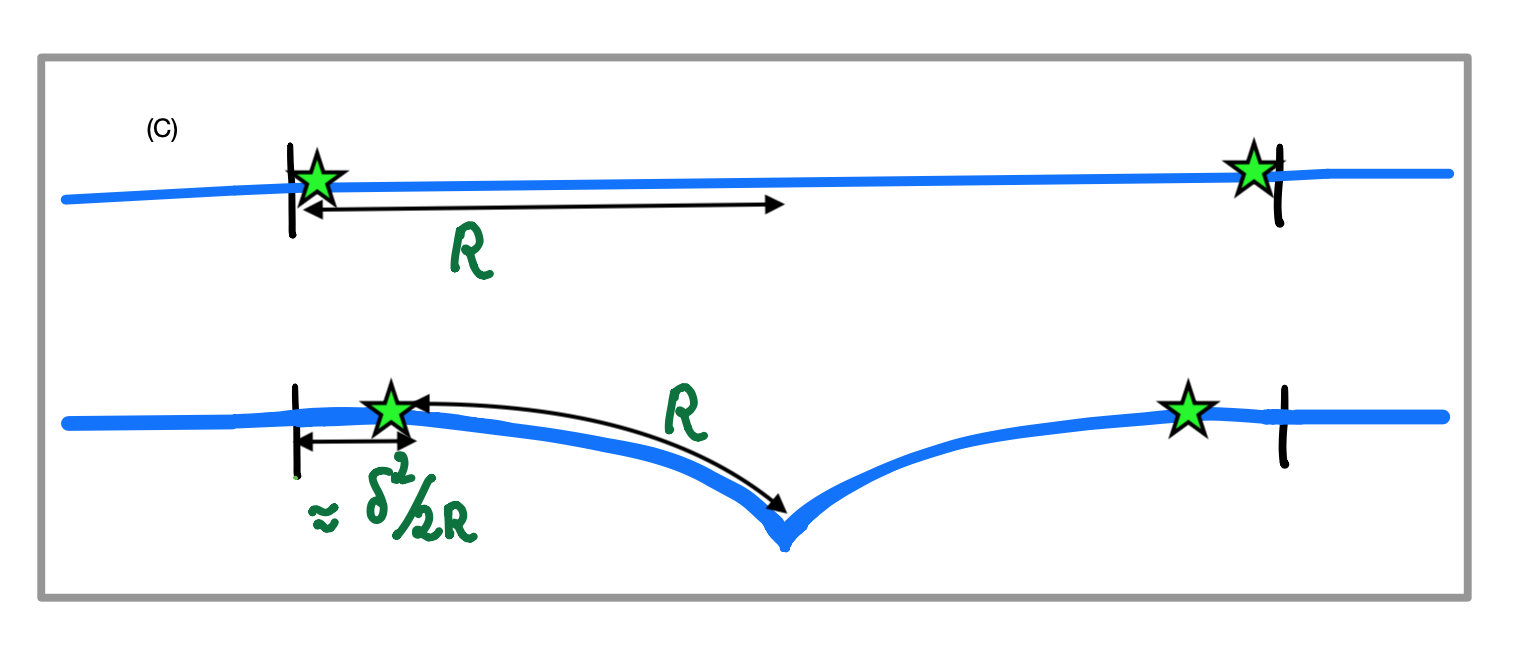}
%\vspace{-7cm}
\caption{(a) Pushing a tablecloth into a hole does not cause significant stretching of the fabric, but rather its sliding towards the indenter, and the formation of a pattern of ``radial buckles''. 
%The black circle marks the circle that was at the hole's edge prior to indentation. The mechanical response consists of sliding inwards and wrinkling. 
(b) Schematic sideview of indentation when a sheet is clamped to the hole's edge. The unavoidable stretching of radial lines 
%Radials are stretched to a length $\sim \sqrt{R^2 + \delta^2}$, yielding 
yields a tensile strain $\sim \sqrt{R^2 + \delta^2}/R - 1 \sim \delta^2/2R^2$. (c) If the sheet can slide inwards, the radial strain can be eliminated through 
%the elimination of radial strain implies 
a displacement $\rmur \!\sim\! -\delta^2/2R$ for $r>R$, and the ``bare" hoop compression thus acquired, $-\rmur/r$, is relieved by undulations into the normal direction. (d) A schematic top view of a small portion of  a radially-wrinkled sheet. Since wrinkles cause a collapse of the hoop stress, force balance in the radial direction implies that $\sigma_{rr} \propto 1/r$; Boundary conditions at the far edge give: $\sigma_{rr} \approx \sigma_0 \Rf /r$.}
\label{fig:demo}
\end{figure}

In order to understand the substantial effect of sliding on the indentation force, one must consider also the strength of the normal force that the substrate exerts on the supported part of the film. This interplay
%The interplay between the tangential force (which constrains sliding) and normal force (which constrains buckling of the sheet out of the substrate) 
can be demonstrated in a table-top example (Fig.~\ref{fig:demo}a): 
attempting to push a tablecloth into a hole in a frictionless table, one finds that the tablecloth responds by changing its morphology - sliding towards the indenter, and forming radially-oriented blisters %wrinkles
(by buckling out from the table's plane) that release the hoop compression induced by the inward sliding.   
Obviously, in such an experiment the fabric is not significantly stretched, indicating that 
%the indentation force, $F ( \delta )$, does not depend on its stretching modulus $Y$; Thus, 
the combined effect of %interplay between 
in-plane sliding and out-of-plane deflection 
%wrinkling 
may undermine  
%underlies a stretching-free response, revoking 
the use of indentation as a reliable probe for measuring the stretching modulus $Y$ of the sheet.  
%
%instead - the force is determined by a tension exerted at the far edge of the tablecloth (e.g. by its gravity), or by the energetic cost associated with forming wrinkles. In other words, the ability to slide and wrinkle underlies a geometry-dominated response to indentation, whose energetic cost is negligible in comparison to the stretching-dominated response.  
%
%
The theoretical model we introduce and analyze in this paper addresses the question that follows naturally from this simple observation: 
%precisely this problem: 
%but forsituations in which the attachment of the sheet 
{\emph{Under what conditions do sliding and deflection from the substrate
%wrinkling 
%eliminate 
%significantly alter 
curb
the effect of the stretching modulus $Y$ on the indentation force %mechanical response 
$F(\delta)$ ? }} \\

\begin{figure}
\includegraphics[width=0.47\textwidth]{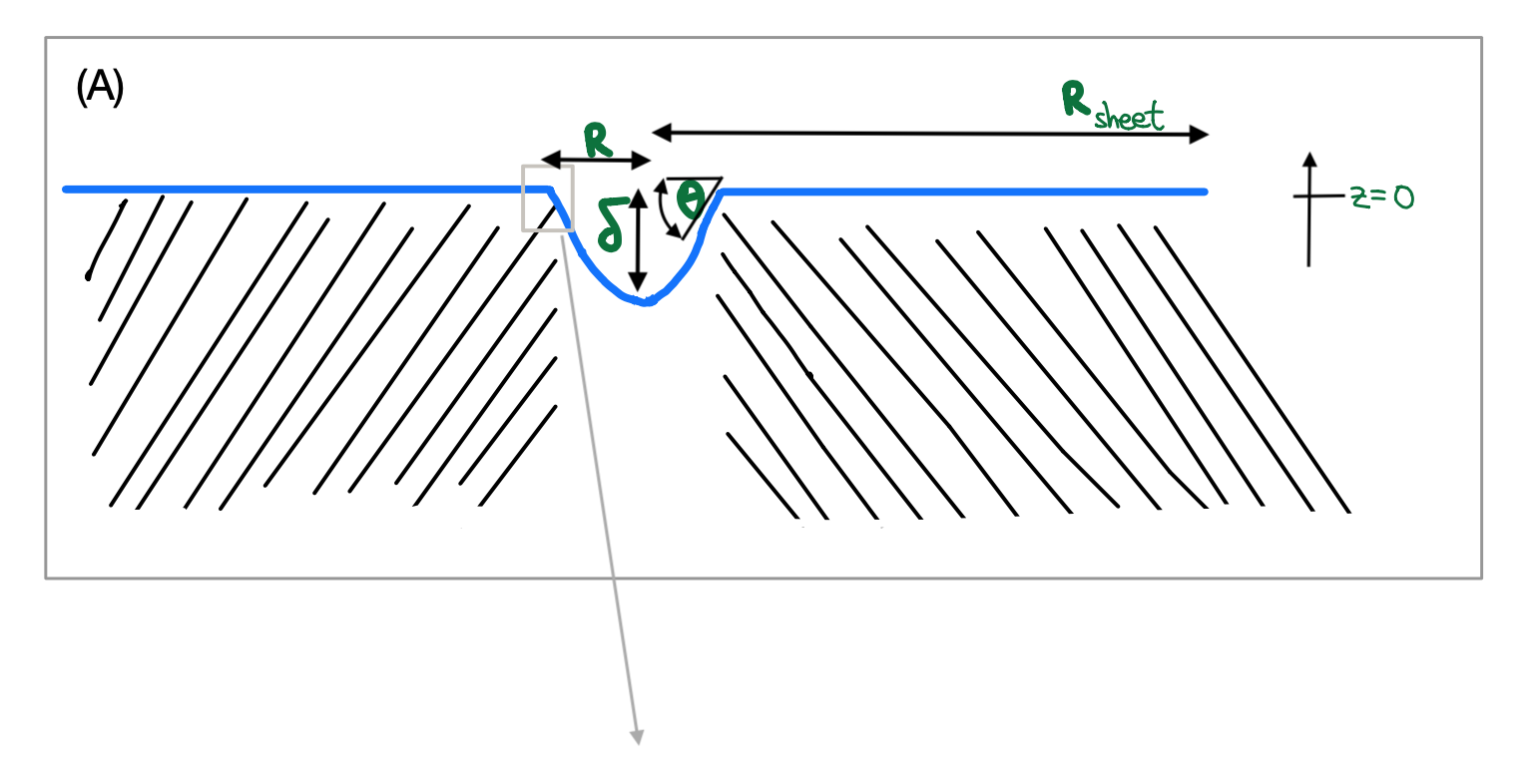}
\includegraphics[width=0.47\textwidth]{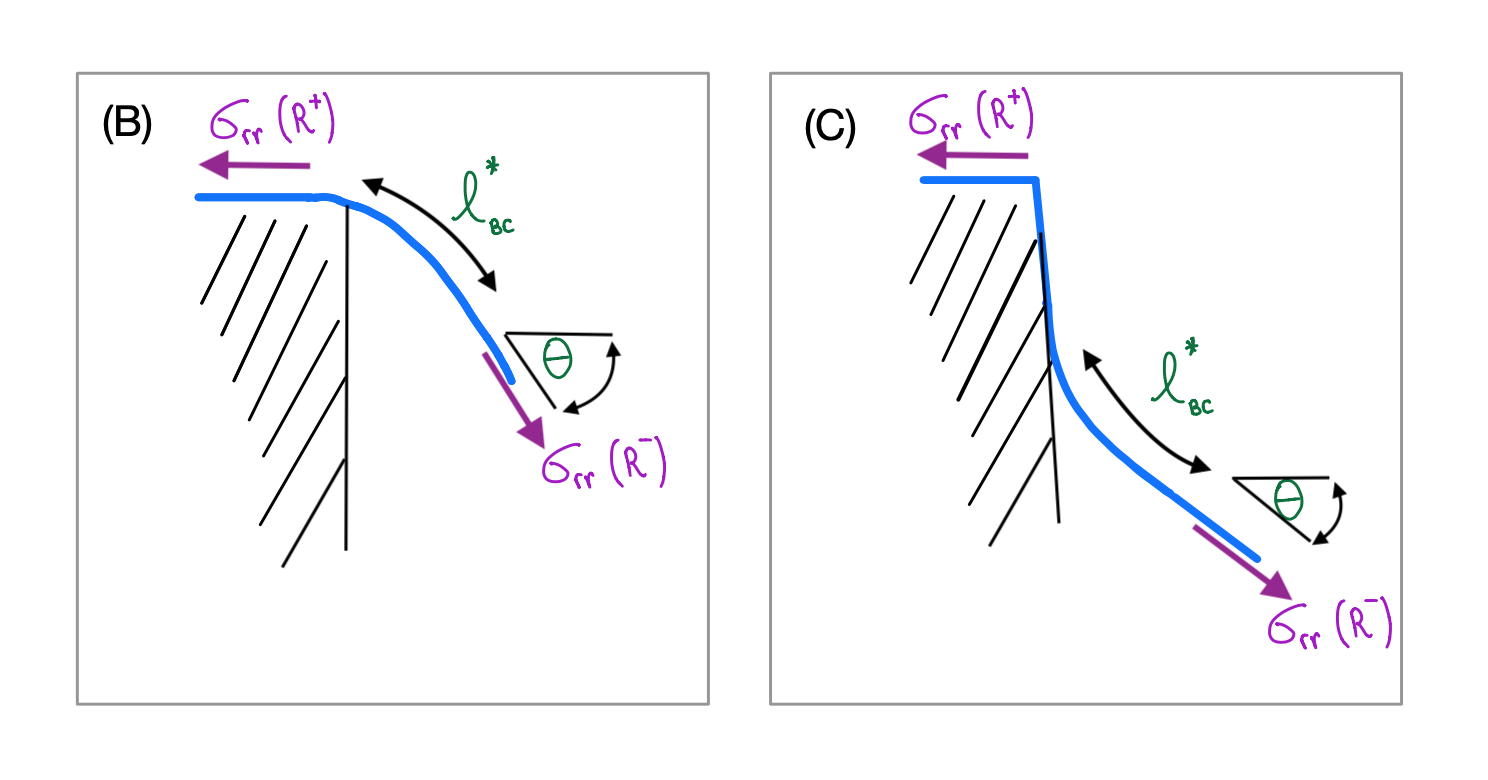}
\includegraphics[width=0.47\textwidth]{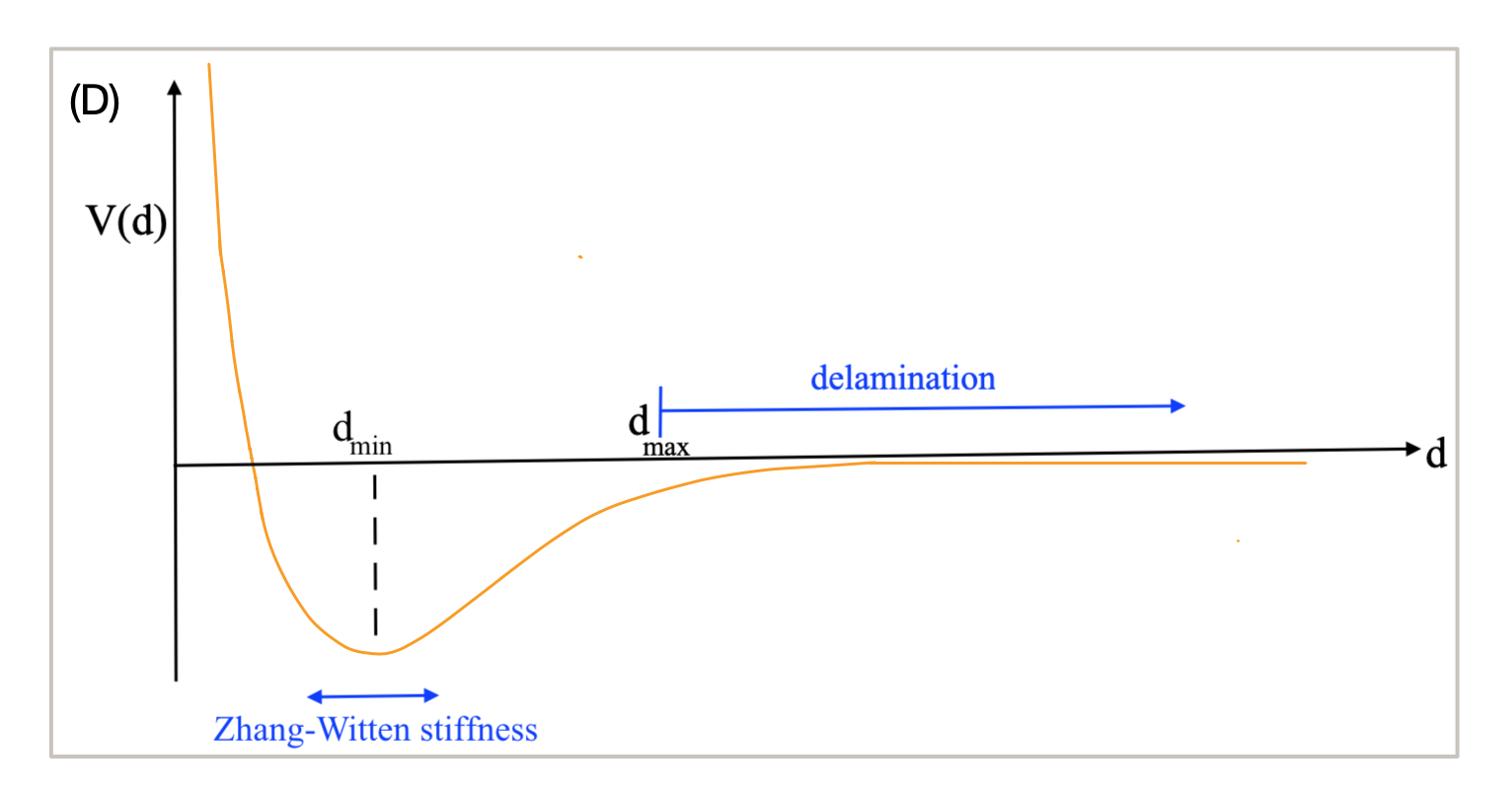}
%\vspace{-2cm}
\caption{Schematic of our model system and key physical mechanisms. (a) A sideview of a solid membrane (sheet), which is supported on a smooth, rigid substrate in $R<r<\Rf$, and suspended in $r<R$. A pointwise indenter pushes at the center, causing the sheet to deflect downwards. Even though a sheet with finite bending modulus $B$ cannot accommodate a discontinuity of the tangent to its plane, our model allows the sheet to make a finite angle $\theta$ with the horizontal at the hole's edge. The reasoning is illustrated in panels (b) and (c), depicting two possible scenarios at the vicinity of the hole's edges (see App.~\ref{app:krr-BC}). In both scenarios, the tangent ``jumps'' over a short distance $\ellbc^*$, Eq.~(\ref{eq:bendo-cap}), which constitutes a boundary layer of negligible energy cost. (d) We assume that the attractive force exerted by the substrate on the sheet is described by Zhang-Witten stiffness, Eq.~(\ref{eq:KsubWitten}), whereby the sheet-substrate distance remains in the VdW potential well. Delamination of the sheet from the substrate requires an energy barrier, and is not addressed within our model (see Sec.~\ref{sec:wrink-delam}).}
\label{fig:schem-slide}
\end{figure}

The mechanism for deflection from the substrate that we consider here, however, does not consist of blisters  
%delaminated zones (as in Fig.?) 
(which are penalized by 
surface energy and may be expected when the sheet-substrate attachment is sufficiently weak), %\cite{Vella2009,Valenkar2012}, 
but rather of 
small-amplitude wrinkles, such that the sheet-substrate distance $d$ remains close to its equilibrium value (see schematic Fig.~\ref{fig:schem-slide}d).  
A central conclusion of our study is that when sliding and wrinkling are effective, the indentation force $F$ scales as:  
%$F$ exhibits a linear dependence on $\delta$:
\begin{align}
F &\sim \geff \cdot \frac{\Rf}{R} \cdot \delta \ . 
\label{eq:pseudo-linear-1}
\end{align}
Underlying Eq.~(\ref{eq:pseudo-linear-1}), which we call a {\emph{pseudo-linear}} response (and is valid above a certain threshold),   
%such a response 
there is a highly non-linear geometric effect, comprising %reflecting 
a {\emph{global}} re-arrangement of the sheet in order to suppress the indentation-induced strain. The global nature of the pseudo-linear  response underlies its %is reflected through its 
dependence %not only on the radius of the suspended portion of the sheet ($R$), but also 
on the overall size of the sheet, $(\Rf)$,
in addition the hole's radius radius, $R$, %of the suspended portion of the sheet ($R$), but also  
%(including the large area supported by the substrate), 
and an effective tension, $\geff$, which is {\emph{independent}} on the stretching modulus $Y$, and may differ substantially from any pre-existing tension ($\sigma_0$) in the sheet. Specifically, $\geff$ may reflect the bending rigidity and the steepness of the substrate-membrane VdW potential.   
In contrast, %This equation describes a linear dependence of $F$ on the indentation depth $\delta$, which is strictly distinct from 
the standard linear response at infinitesimal indentation depth is %({\emph{i.e.}} 
$F \sim (\sigma_0/R) \delta$ (up to logarithmic corrections \cite{Vella17}), being fully determined by the pre-tension $\sigma_0$, and the size of the suspended portion. %part of the sheet. 
%and the hole's radius $R$.    
%where $\sigma_0$ is some ``pre-tension''), hence we refer to  
%Eq.~(\ref{eq:pseudo-linear-1}) as a {\emph{pseudo-linear}} response.    

Before delving into the details of our model, let us provide a heuristic argument for the mechanism by which sliding and wrinkling give rise to a pseudo-linear response~(\ref{eq:pseudo-linear-1}). This argument is inspired by the example of indenting an ultrathin polymer sheet that is floating on a liquid bath \cite{Vella15,Paulsen16,Vella18b,Ripp20}.

%What is the influence of sliding and wrinkling on the measured force $F$ at a given deflection $\delta$.  
 
%In this paper, we introduce a theoretical model that revisits this indentation set-up, taking into consideration the possibility that graphene can slide over and deflect slightly from the substrate.

%%%%%%%%%%%%%%%%%%%%%%%
%%%%%%%%%%%%%%%%%%%%%%%%%%%%%%%

%%%%%%%%%%%%%%%%%%%%%%%
%%%%%%%%%%%%%%%%%%%%%%%%%%%%%%%

\subsection{Heuristic argument: stretching versus asymptotically isometric response \label{subsec:Heuristic}} 
%Before dwelling into the details of our model, it is useful to 
Let us contrast the two limit cases in the above example of pushing a tablecloth through a hole: {\emph{(a)}} perfect clamping of the sheet at the edge of the hole ($r=R$); or {\emph{(b)}} free sliding and ``wrinkling'' of the sheet on the substrate. \\

{\emph{(a) clamping:}} Assuming that prior to indentation the sheet is subjected to a uniform tension $\sigma_0$, the elastic energy associated with the work $F\cdot \delta$ of the indenter can be estimated as: 
\begin{equation}
F \cdot \delta \sim U_{elas} \sim R^2 \left[\sigo + Y \left(\frac{\delta}{R}\right)^2 \right]  \left(\frac{\delta}{R}\right)^2  \ , 
\label{Eq:scaling1}
\end{equation}
where the stress in the sheet is estimated as the sum of the pre-tension $\sigo$ and the indentation-induced stress, $Y (\delta/R)^2$. Notice that the clamping assumption underlies our estimate of the radial strain, $\epsilon_{rr} \sim (\delta/R)^2$, as the indentation-induced extension of the radial distances (see Fig.~\ref{fig:demo}b). Note also that the bending energy is neglected, since we expect it to contribute only at some narrow, high-curvature zones, near the rim and around the indenter's tip. Equation~(\ref{Eq:scaling1}) shows that upon increasing $\delta$, the force transforms from a linear response, $F/\delta \sim \sigo$ (Column 2 of Table I), to a nonlinear, cubic response, $F / \delta^3  \sim Y/R^2$, which reflects the stretching modulus $Y$ \cite{Schwerin29} (Column 3 of Table I). 
Actual calculations \cite{Vella17} yield a quantitative description of the transition between the two regimes (gray curve in Fig.~\ref{fig:response-1}). 
\\ 
%and Fig.~\ref{fig:response-2}).   

{\emph{(b) sliding and wrinkling:}} 
Let us assume now that the sheet can slide freely on the substrate, such that material circles at radius $r$ undergo radial displacement $r \to r+ \rmur(r)$. An inward displacement ($\rmur<0$) enables the sheet to retain the length of radial lines, thus avoiding the indentation-induced tensile strain $(\delta/R)^2$ in the radial direction; a simple calculation shows that 
retention of the original length of radials, $\Rf$, requires a constant radial displacement % $\rmur(r)$  
outside the hole ($r>R$): 
%such that 
%$\sqrt{\delta^2 + (R + \rmur(R))^2} \approx R$, implying:  
\begin{equation} 
\rmur(r) \sim -\delta^2/R \  
\label{eq:radial-scale}
\end{equation}
(Fig.~\ref{fig:demo}c). 
Clearly, such an inward sliding causes a compression in the orthogonal planar (azimuthal) direction, since hoops of radius $r$ acquire a strain  $\rmur/r$. If the normal attractive force exerted by the substrate is very strong, such a compression cannot be relieved, and the indentation force $F(\delta)$ is qualitatively similar to the clamping case discussed in the above paragraph. However, if the sheet can deflect even slightly from the substrate, then the compressive strain $\rmu_r/r$ can be eliminated by forming azimuthal undulations  
%radial wrinkles, 
whose characteristic wavelength may be very small, being determined by the bending modulus and the strength of sheet-substrate attachment. This scenario is the essential mechanism by which the table cloth in Fig.~\ref{fig:demo}a responds to the indentation force.  

The elimination of tensile radial strain (by sliding) and compressive hoop strain (by deflection), suggests that the indentation force is not sensitive to the stretching modulus of the sheet. %, but rather by another resisting force. 
Understanding this type of response, which involves only minute, asymptotically vanishing level of strain, and is thus called ``asymptotically isometric'' \cite{Vella15,Vella18,Paulsen2019,Davidovitch19}, is the essence of our manuscript. At a heuristic level, one can make progress by considering a small tension $\sigo$ pulling radially on the sheet at its far edge, $r=\Rf$, where $\Rf \gg R$. The presence of boundary tension implies that the stress in the sheet is not totally eliminated by sliding and wrinkling, and the response becomes dominated by the dependence of the residual stress on the indentation depth $\delta$. Since wrinkles eliminate the azimuthal component of the stress tensor, force balance on infinitesimal annular zones implies that there is a residual radial stress in the sheet $\sigma_{rr}(r) \approx \sigo \Rf/r$ (Fig.~\ref{fig:demo}d). The consequent divergence %of the residual radial stress 
at $r\to 0$ is resolved by the presence of an unwrinkled core of radius $\Li$ in which the stress saturates to its ``bare'' value $\sim Y\cdot (\delta/\Li)^2$. Continuity of radial stress at the boundary, $r=\Li$, between the wrinkled zone and the unwrinkled core, gives: 
\begin{equation}
\Li \sim \frac{\sigo}{Y} \frac{R^2}{\delta^2}\Rf \ , 
\label{eq:Length-scale}
\end{equation}
(column 6 of Table II).

An interesting feature of such a sliding-wrinkling response is that only a negligible part of the indenter's work, $W_{inden} = F\cdot \delta$, is transmitted to the elastic energy of the sheet. In other words, the near-absence of strain, enabled by the combination of in-plane sliding 
%displacement (sliding) with 
and out-of-plane deflection, %(wrinkling), 
underlies a soft mode of an {\emph{asymptotically-isometric}} deformation, which eventually 
%and this soft mode 
controls the mechanical response to indentation.   
%and therefore we call such a response ``asymptotically-isometric". 
%In order to see this, 
This soft mode mechanics % of this soft mode 
may be realized by recalling that the only (finite) contribution to residual stress is the radial component $\sigma_{rr}$, whose integration yields $U_{elas} \sim \int_{\Li}^{\Rf} r \ dr \sigma_{rr}^2/Y \ \sim\  \sigma_0^2 \Rf^2/Y$, where we used Eq.~(\ref{eq:Length-scale}) and neglected logarithmic corrections and higher order terms associated with bending and other components of the stress. In contrast, the work done by the tensile load at the far edge against the indenter is $W_{ten} \sim 2\pi \Rf \sigo \rmu_r(\Rf) \sim \sigo \delta^2 \Rf/R $, where we used Eq~(\ref{eq:radial-scale}). 
Introducing dimensionless parameters for the indentation depth and the sheet's radius:
\begin{equation}
\tdelta = \frac{\delta}{R} \sqrt{\frac{Y}{\sigma_0}} \ \  ; \ \ {\cal R} = \frac{\Rf}{R} \ , 
\label{eq:DG-00}
\end{equation} 
we find that $W_{ten} > U_{elas}$ if:
%the tensile work $W_{ten}$ exceeds the elastic energy $U_{elas}$ if:
%Comparing the tensile work $W_{ten}$ with the elastic energy $U_{elas}$, we find that $W_{ten} \gg  U_{elas}$ as long as  $\delta \gg \delta^{**}$, where 
\begin{equation} 
\tdelta > \tdeltastst  \ \ \ {\rm where:} \ \ \tdeltastst \sim \sqrt{{\cal R}} %R \sqrt{\frac{\sigo}{Y}} \sqrt{\frac{\Rf}{R}}
\label{eq:deltastst}
\end{equation}
 %\  \to 0$. 
 For $\tdelta >  \tdeltastst$, the elastic energy stored in the sheet can be neglected, and the indentation force is $F \approx \partial W_{ten}/\partial \delta$, so that we readily obtain Eq.~(\ref{eq:pseudo-linear-1}), with $\geff = \sigma_0$.  \\

Crucially, Eq.~(\ref{eq:deltastst}) shows that %since $\tdeltastst$ is inversely proportional to the stretching modulus, 
the depth $\delta$ required to reach such a pseudo-linear response 
%and thus 
vanishes for a  ``nearly inextensible'' sheet ({\emph{i.e.}} $\sigma_0/Y \to 0$). In this regime, the combined effect of low energetic cost for bending and avoidance of indentation-induced strain, makes the solid sheet a ``bad capacitor'' of mechanical energy, and the work $W_{inden}=F\cdot \delta $ done by the indenter is transmitted almost entirely to the puller at the far edge \cite{Vella15,Vella18}: %($\sigma_0$).
\begin{eqnarray}
\tdelta < \tdeltastst:   \ \ \ F\cdot \delta  \ \ &\longrightarrow& \ \  W_{ten}  + U_{elas}    \nonumber \\
\tdelta > \tdeltastst:   \ \ \  F\cdot\delta  \ \ &\longrightarrow& \ \  W_{ten}  \ .   
\end{eqnarray}

\subsection{Overview \label{subsec:overview}}
We introduce a minimal model to study the interplay between stretching, sliding, and wrinkling, and the  
dependence of the indentation force %response to indentation 
on actual physical parameters -- external tension, bending and stretching modulii of the sheet, and the strength of sheet-substrate attachment.  %strength of adhesion between the sheet and substrate. 
Asymptotic analysis of this model enables us to elucidate qualitatively different types of  response to indentation by identifying distinct parameter regimes.   
\\

%In order to make quantitative predictions on the dependence of the response to indentation on actual physical parameters -- external tension, bending and stretching modulus of the sheet, and adhesion strength between the sheet and substrate -- we introduce a minimal model that exhibits the interplay between stretching, sliding, and wrinkling.   

%The above tablecloth example demonstrates that if the sheet is free to slide on the substrate, the indentation force may not reflect the stretching modulus, but rather the far-edge tension and/or the attachment strength between the sheet and substrate, which resists the formation of wrinkles. Inspired by this intuitive example, and recent studies on the poking of floating untrathin sheets and pressurized shells, we introduce a theoretical model for indentation in Hone's set-up, which account for sliding and wrinkling. 

\subsubsection{Model and analysis} 
Our model is depicted schematically in Fig.~\ref{fig:schem-slide}. 
We consider a disk-like sheet of radius $\Rf$, with bending rigidity $B$ and Young modulus $Y$, which is suspended on a flat rigid substrate with a hole of radius $R \ll \Rf$ around it center ($r\!=\!0$), and a point-like indenter, which induces an out-of-plane deflection of amplitude $\delta$ at $r\!=\!0$. 
We assume that the sheet is subjected to radial tension $\sigma_0$ at its far-edge $r = \Rf$, to which we will refer as ``pre-tension''. This may be an actual pre-tension ($\sigma_0=T_{\rm pre}$), applied prior to clamping the far-edge, or be exerted directly, such that the far edge, $r=\Rf$ is load-controlled rather than clamped. In our model, the normal force that resists deviations of the supported sheet from a planar state is characterized by a ``stiffness'' parameter $\Ksub$. Such a simplified response is known as Winkler foundation in the solid mechanics literature \cite{TimoshenkoBook}. The stiffness parameter $\Ksub$, together with the bending rigidity $B$ of the sheet, determine the deflections in the normal direction, which often take a periodic form that we call ``wrinkles'' -- the larger $\Ksub$ is, the smaller are the characteristic amplitude and wavelength of the emerging wrinkle pattern
\footnote{Familiar examples for $K$ are a liquid substrate (for a floating sheet), where deviations from planarity are resisted by the liquid gravity, hence $\Ksub = \rho_l g$  (where $\rho_l$ is the liquid density and $g$ is the gravity); 
%an homogenous compliant substrate, where  $K$ is determined by its Young's modulus, 
and a compliant substrate with a stiff near-surface layer of thickness $H$ and Young's modulus $E_s$, where $K \approx E_s/H$.}. For our primary interest here -- a highly rigid, undeformable substrate -- the stiffness $\Ksub$ was recognized by Zhang and Witten \cite{ZW07} as 
\begin{equation} 
{\rm Zhang\!-\!Witten \ stiffness:} \ \ \Ksub = V''(d_{\rm min}) \ , 
\label{eq:KsubWitten}
\end{equation} 
where $V(d)$ is the attractive substrate-sheet potential per unit area, and $d_{\rm min}$ is the thermodynamic equilibrium distance between the sheet and the substrate (Fig.~\ref{fig:schem-slide}d). The Zhang-Witten stiffness assumes that the substrate is infinitely rigid, and the energetic cost for forming wrinkles (in addition to bending energy) is associated with the slight deviation of the sheet-substrate distance from its favorable value in the absence of any external loads. The assumption underlying this picture is that the VdW interaction is sufficiently strong, such that the energy barrier ({\emph{i.e.}} the depth of VdW potential well) that is necessary for the sheet to delaminate from the substrate cannot be reached. Instead, the small-amplitude undulations keep the sheet {\emph{everywhere}} within the VdW potential well of the substrate, 
and the energetic penalty $\propto (d-d_{\rm min})^2$. 
In Sec.~\ref{sec:wrink-delam} we will elaborate on the important difference between the relaxation of compression through such small-amplitude undulations and the formation of delamination zones, for which the energy cost per area is $\sim V(d_{min})$, independent of the actual sheet-substrate distance.  
%which has been assumed in a recent work \cite{Dai2020}.  

%deflection of the sheet from the ground state distance from the substrate.    

%\subsubsection{Analysis} 
Our analysis is based on asymptotic analysis of the F\"oppl--von K\'arm\'an (FvK) equations, which describe the deformations of a thin solid sheet to exerted forces, assuming that the {\emph{local response}} is Hookean (namely, linear  stress-strain relationship), and that the deformed shape is characterized by small slopes. The FvK equations are {\emph{geometrically nonlinear}}, namely, the nonlinearity is {\emph{universal}} rather than material-dependent stemming from the coupling of out-of-plane deflections to in-plane strain. 

In order to understand the response of the sheet to exerted loads, it is imperative to distinguish between the response to compressive and tensile stresses. If the stress exerted on a small piece of the sheet is (uniaxially or biaxially) tensile, the piece will stretch along the tension direction(s); we call this {\emph{tensile strain}}. In contrast, if the piece is under a compressive stress, it may buckle to reduce the compression level, and this mechanism gives rise to wrinkle patterns. If the sheet is sufficiently thin, or more precisely {\emph{highly bendable}}, the residual compression depends on the bending modulus and the exerted loads through a dimensionless parameter, called {\emph{bendability}} \cite{Davidovitch11}. The method by which the residual compression level is found, along with geometric features of the wrinkled state, has been called a \cite{Davidovitch11} {\emph{far from threshold}} analysis; this is an expansion of FvK equations around the singular limit of {\emph{tension field theory}}, which pertains to a compression-free sheet with no bending resistance ({\emph{i.e.}} $B=0$) \cite{Wagner35,MansfieldBook,Stein61,Pipkin86,Steigmann90}. It is thus crucial to understand that despite the smallness of the amplitude, the mere existence of wrinkles has a strong effect on the stress field in the sheet, 
% in their effect on the stress field in the sheet is strong, 
and therefore cannot be considered as a perturbation to some compressed, pre-buckled state.      
\\

%which describes the mechanics of highly bendable sheets. In this approach, 
%one recognizes that despite the small amplitude of wrinkles, their mere existence has a strong effect the stress field in the sheet, 
% in their effect on the stress field in the sheet is strong, 
%and therefore cannot be considered as a perturbation to the compressed, pre-buckled state. In fact, for sufficiently small bending modulus, any compression is relaxed almost completely, such that the stress field approaches a compression-free form. This principle, which is also known as ``relaxed energy" or ``tension field theory", is paramount for the forthcoming analysis.   

%{\emph{near inextensibility}}: 
%Throughout this study, we will assume that 
%In order to keep the discussion as simple as possible, we will assume throughout most of our study 
%that the sheet is subjected to tension, $\sigma_0$, exerted at the far-edge $r=\Rf$ (or, alternatively, ``pre-tension" prior to clamping the far-edge). In our model, we will always assume that the characteristic {\emph{tensile strain}}, $\sigma_0/Y$ is very small. We will refer to such a condition as {\emph{``near inextensibility"}} of the solid sheet.  

%{\emph{high bendability}}: 
%In order to keep the discussion as simple as possible, it will be useful to further assume that the far-edge tension $\sigma_0$ is much larger than another characteristic scale of 

%%%%%%%%%%%%%%%%%%%%%%%%%%%%%%%%%%%%%%%   
\begin{table*}
\centering
\caption{\label{Tab-I}
Summary of central results under various types of conditions of clamping and sliding at the hole's edge, upon increasing indentation depth $\delta$ (left to right), assuming the substrate stiffness $\Ksub$ is sufficiently large ($\beta \gg \tdelta^2$), such that the supported part of the sheet cannot wrinkle. The first row summarizes the response in the ``no-sliding'' case ($\srr(R) = \sigma_0$); the second row describes the ``free sliding'' case ($\srr(\Rf) = \sigma_0$), where compression develops above a threshold value, $\tdelta_c\approx 3.3$. The third row summarizes the effect of wrinkle formation in the suspended portion of the sheet. % on the response. 
}
\vspace{0.2cm}
\begin{ruledtabular}
\begin{tabular}{||c|c|c|c|c|c||}
%{l*9 {c}}
%\begin{widetext}
%\begin{table}
%\begin{tabular}{||c|c|c|c|c|c||}
%\hline \hline
&linear constant &cubic constant &asymptotic slope &asymptotic &comments \\
&$\frac{F}{\delta}$ & $\frac{F}{\delta^3}$ & at core's edge&tensile core   &  \\ 
& & & normalized by $\frac{\delta}{R}$ & & \\
%& & & & $0 \le R \le L$& \\
\hline 
no sliding &$- \frac{2 \pi}{\log ( \tilde{\delta} )} \sigma_0$ &$ \frac{0.166 \times 2 \pi}{R^2} Y$ &$ 0.63 $ &$\Li = R$ &pure tension  \\
%$\sigma_{rr} (\! \Rf )\! =\! = \sigma_{rr} ( \!R ) \!= \!\sigma_0$   
& & & & &   (Sec.~\ref{subsec:II-clamping}) \\ \hline
%& & & &depend on $\nu$ & & \\ \hline
sliding %, $\sigma_{rr} ( \Rf ) = \sigma_0$ 
&$- \frac{2 \pi}{\log ( \tilde{\delta} )} \sigma_0$ &$\frac{0.101 \times 2 \pi}{R^2} Y$ &$ 0.83 $ &$\Li \approx 0.6 R$ &wrinkling instability \\ 
no wrinkling& &  & & &$\tilde{\delta} \gtrsim 3.3$ (Sec.~\ref{subsec:II-sliding})  \\
%&  \vline & | & & &(Sec.~\ref{sec:sliding}) \\ 
\hline 
sliding %, $\sigma_{rr} ( \Rf ) = \sigma_0$ 
& $- \frac{2 \pi}{\log ( \tilde{\delta} )} \sigma_0$  &$\frac{0.098 \times 2 \pi}{R^2} Y$ &$ 0.87$ &$\Li \approx 0.49 R$ & stable if $\beta \gg \tdelta^2$ \\ 
wrinkling  inside hole & & & & & (Sec.~\ref{subsec:II-wrinkling} )\\ 
%only inside hole& & & &  (Sec.~\ref{sec:wrinkles_suspended} ) \\ 
%\hline \hline
 \end{tabular}
 \end{ruledtabular}
%\caption{Summary of central results under various types of conditions of clamping and sliding at the hole's edge, upon increasing indentation depth $\delta$ (left to right), assuming the substrate stiffness $\Ksub$ is sufficiently large that $\beta \gg \tdelta^2$, and the supported part of the sheet cannot wrinkle. The first row summarizes the response in the ``no-sliding'' case ($\srr(R) = \sigma_0$); the second row describes the ``free sliding'' case ($\srr(\Rf) = \sigma_0$), where compression develops above a threshold value, $\tdelta_c\approx 3.3$. The third row describes the effect of wrinkles in the suspended portion of the sheet on the response.  } 
%Numerical values typically depend on the Poisson ratio $\nu$. The values in the table correspond to the case $\nu=1/3$.  }
 \end{table*}
% \end{widetext}
%%%%%%%%%%%%%%%%%%%%%%%%%%
%%%%%%%%%%%%%%%%%%%%%%%%%%%%%%
%%%%%%%%%%%%%%%%%%%%%%%%%%%%%%%
%%%%%%%%%%%%%%%%%%%%%%%%%%%%%%%
\begin{table*}
\caption{\label{Tab-II}
Summary of central results under various types of conditions of clamping and sliding at the hole's edge, upon increasing indentation depth $\delta$ (left to right), assuming $\beta <O(\tdelta^2)$, such that the substrate stiffness 
$\Ksub$ is sufficiently small to allow relief of compression by forming wrinkles on the supported portion of the sheet. 
%wrinkles to relieve %indentation-induced 
%compression on the substrate. 
%, namely $\sqrt{B K}/Y(\delta/R)^2 \gg 1$ for sufficiently large $\delta$ (while $\delta \ll R$). 
The upper row describes the response under the ``sliding" BC ($\srr(\Rf) = \sigma_0$) in the parameter regime $\beta \ll 1$, 
%a ``tensional wrinkled" state ($\sqrt{B K} \ll \gamma$), 
where the explicit values of the bending rigidity $B$ and substrate stiffness $\Ksub$ barely affect the stress field and  extent of the wrinkled zone. The middle row describes the response %under ``partial sliding", %on the substrate, 
when sliding is hindered by clamping the sheet at the far edge, such that $\rmur(\Rf) = (1-\nu)\sigma_0 \Rf$. The bottom row describes the response under sliding (or hindered sliding) conditions, but when $1 \ll \beta \ll \tdelta^2$, such that the radial stress is governed by the residual hoop compression, $\sqq \approx -2\sqrt{B\Ksub}$, rather than by the tensile load at the far edge. Here $\tdelta = \sqrt{Y/\geff} \cdot R$.}  
\vspace{0.2cm}
%\caption{\label{Tab-I}
%Summary of central results under various types of conditions of clamping and sliding at the hole's edge, upon increasing indentation depth $\delta$ (left to right), assuming the substrate stiffness $\Ksub$ is sufficiently large that $\beta \gg \tdelta^2$, and the supported part of the sheet cannot wrinkle. The first row summarizes the response in the ``no-sliding'' case ($\srr(R) = \sigma_0$); the second row describes the ``free sliding'' case ($\srr(\Rf) = \sigma_0$), where compression develops above a threshold value, $\tdelta_c\approx 3.3$. The third row describes the effect of wrinkles in the suspended portion of the sheet on the response. }
\begin{ruledtabular}
\begin{tabular}{||c|c|c|c|c|c|c||}
%\hline \hline
&linear &(sub) cubic & pseudo-linear & asymptotic slope &asymptotic &comments  \\ 
&constant & constant  & constant & at core's edge&tensile core   & \\ 
& $\frac{F}{\delta}$& $\frac{F}{\delta^3}$ & $\frac{F}{\delta}$ & normalized by $\frac{\delta}{R}$ & & \\
\hline \hline  
sliding &$- \frac{ 2 \pi}{\log ( \tilde{\delta} )} \sigma_0$ & $- \frac{0.22 \times 2 \pi}{\log ( \tilde{\delta} )} \frac{Y}{R^2}$  & $\sim 2 \pi \sigma_0 {\cal R}$  & $1$ 
%&$L\sim \frac{\sigma_0 \Rf^{4/3} R^{5/3} }{Y} \delta^{-2}$ 
&$\Li \sim R \cdot {\cal R} \tdelta^{-2}$ 
& wrinkling instability  \\
$\beta \ll 1$ & & & & & & at $\tilde{\delta} \approx 3.3$ (Secs.~\ref{subsec:both-wrinkling},\ref{subsec:geo-limit})  \\ \hline 
hindered sliding & $- \frac{2 \pi}{\log ( \tilde{\delta} )} \sigma_0$ & $\sim \frac{1}{\log {\cal R} } \frac{Y}{R^2}$ &none & $ 1 -O \left[ 1/\log{\cal R} \right] $ & $\Li \sim \frac{1}{\log {\cal R}} R$ & wrinkling instability  \\ 
$\beta \ll 1$
%$u_r ( \Rf ) \propto \frac{\gamma}{Y} \Rf$ 
& & & & & & at $\tilde{\delta} \approx 3.3$ (Sec.~\ref{sec:Expansion}) \\  \hline
sliding & as above & as above & as above & as above  & as above &$\geff = 2 \sqrt{B \Ksub}$ \\ 
%$\sqrt{B K}/\gamma \gg 1$ 
$1 \ll \beta \ll \tdelta^2$
& $\sigma_0 \to \geff$ & $\sigma_0\to \geff$ & $\sigma_0\to \geff$ & $\sigma_0\to \geff$ & $\sigma_0\to \geff$ &  
(Sec.~\ref{sec:residual_compression})  \\ %\\
%$\tilde{\delta}_{eff} =  \frac{\delta}{R} \sqrt{Y/\gamma_{eff}}$ \\  
%(ultra weak tension)  
%& $\tilde{\delta} \to \tilde{\delta}_{eff}$ &$\tilde{\delta} \to \tilde{\delta}_{eff}$ & & &  &(Sec.~\ref{sec:ultraweak})  \\
\hline
%sliding, $\sigma_{rr} ( R_{out} ) = \gamma$ &pseudo linear:  & &$\frac{\delta}{R} \left( 1 - \tilde{\delta}^2 \right)$ & $L \sim \frac{3.3 \gamma}{Y} \frac{R R_{out}}{\delta^2}$ & valid for &\ref{sec:wrinkles_attached} \\ wrinkles fully & $\frac{F}{\delta} \sim 2 \pi \left( \frac{R_{out}}{R} \right) Y$& & & & $\epsilon_k \ll 1, \tilde{\delta}^2 \gg \frac{R_{out}}{R}$ & \\ cover the substrate & & & & & & \\ (gemetrical limit) & & & & & & \\ \hline 
%\hline  \hline
 \end{tabular}
 \end{ruledtabular}
%
%``ultra weak" tension ($1 \ll \beta \ll \tdelta^2$), where the $B$ and $\Ksub$ induce a radial tension $\sim \geff =2\sqrt{B \Ksub}$, which governs the stress field and the extent of the wrinkled zone.}   
%Numerical values typically depend on the Poisson ratio $\nu$. The values in the table correspond to the case $\nu=1/3$.  }
 \end{table*}

\subsubsection{Classification of parameter regimes and central predictions}
%The mechanics of sheets is governed by FvK equations, which are {\emph{geometrically nonlinear}}; namely, the nonlinearity is universal, rather than material dependent, and stems from the coupling of out-of-plane deflections to the in-plane stress. As a result, 
Since the FvK equations are nonlinear, the stress cannot be considered a superposition of independent sources. Nevertheless, it is useful to identify three sources of stress that underlie the mechanical response. 
\begin{equation}
\sigma_0 \ \ ; \ \ Y (\frac{\delta}{R})^2 \ \ ; \ \ 2\sqrt{\Ksub B} 
\label{eq:stress-levels}
\end{equation}
The first source, $\sigma_0$, contributes a uniform isotropic tension to both radial and azimuthal (hoop) components of the stress tensor. The second source, $Y (\tfrac{\delta}{R})^2$, which is the only one that depends explicitly on the amplitude $\delta$, 
gives rise to radial tension ({\emph{i.e.}} stretching radial lines) and hoop compression (pulling latitudes inwards). The last term,
$2\sqrt{\Ksub B}$, characterizes the residual hoop compression in the presence of radial wrinkles, namely, it is the minimal possible value, to which the hoop compressive stress can be suppressed with the aid of wrinkles
%that relieve the hoop compression, $Y (\tfrac{\delta}{R})^2$, induced by indentation 
%\cite{Cerda03,Grason13,Hohlfeld15,Davidovitch19}. 
\cite{Cerda03,Davidovitch19}. 
\\
%; it stems from the resistance of the substrate to deflections, and consequently the existence of a minimal wrinkle wavelength. 

The characteristic scales of stress (\ref{eq:stress-levels}) form %a convenient choice of 
two dimensionless groups, in addition to $\tdelta$ and ${\cal R}$ (Eq.~\ref{eq:DG-00}), that we use to characterize the response to indentation at various parameter regimes: 
\begin{equation}
%\tdelta = \frac{\delta}{R} \sqrt{\frac{Y}{\sigma_0}} \ \ ; \ \ 
\epsilon = \frac{B/R^2}{Y(\delta/R)^2} \ \ ; \ \ \beta = \frac{2\sqrt{B\Ksub}}{\sigma_0}  \ . 
%\ \ ; \ \ {\cal R} = \frac{\Rf}{R}
\label{eq:DG-0}
\end{equation} 
The counterparts of the three dimensionless groups, $\tdelta, \epsilon, {\cal R}$, 
%are akin %(but not identical) to the three dimensionless parameters 
have been used to describe the indentation of a floating ultrathin polymer sheet \cite{Vella15,Vella18}, whereas the additional parameter $\beta$, which describes the ratio between the residual compression and isotropic pre-tension, has received less attention in those studies.  
%does not appear in previous studies of indentation problems.
%(but is nevertheless akin to a ``softness'' parameter that has been used to characterize wrinkling cascade problems \cite{Huang10}).  
The parameter $\tdelta$ -- a renormalized indentation depth -- is the ratio between the bare indentation-generated strain and the isotropic pre-tension in the sheet; 
the parameter $\epsilon$ is the inverse of the ``geometric bendability'' -- the ratio between a (minimal) bending-related strain and the bare indentation-induced strain; the parameter ${\cal R}$ is the ratio between the lateral sizes of the sheet and the hole. 

Throughout our study we will assume a highly bendable sheet, namely, $\epsilon \ll 1$, such that in-plane compression may be easily suppressed by wrinkling, and its size is large in comparison to the hole, namely ${\cal R} \gg 1$. Our primary interest is to understand the mechanics when $\tdelta$ is increased above a finite threshold value, $\tdelta_c \sim O(1)$, at which the indentation force is sufficiently strong to pull latitudes inwards and cause compression in part of the sheet. (Note that in the absence of pre-tension $\tdelta_c = 0$). In the rest of this introductory section we summarize the various types of response described by this model in terms of the parameters $\tdelta, \epsilon, \beta$, and ${\cal R}$. \\
%As $\tdelta$ is increased above this threshold value, we study the mechanics encapsulated by the response function, $F(\delta)$, at 
%various asymptotic regimes, defined through the other three parameters, $\epsilon, \beta$, and ${\cal R}$. Below, we elaborate on these parameter regimes. 
%\\
%Identifying the three characteristic stresses in (\ref{eq:stress-levels}), we can now distinguish between the types of mechanical response at various parameter regimes: \\

\begin{center} {\emph{Regime (i)}} \end{center}
\begin{equation}
%\sqrt{K\kappa} \ \gg  \ \sigma_0 \ , \ Y (\frac{\delta}{R})^2
\beta \gg \tdelta^2
\label{eq:regime-i}
\end{equation}
%At this parameter regime, the attachment force between the substrate and the sheet is so strong that wrinkling the supported portion of the sheet requires very high energetic cost. Hence the supported part of the sheet is subjected to hoop compression, which is determined by $\sigma_0$ and $Y (\tfrac{\delta}{R})^2$ alone. 
%In order to study this parameter regime, it is useful to introduce the dimensionless amplitude $\tdelta$: 
%\begin{equation}
% \tdelta = \frac{\delta}{R} \sqrt{\frac{Y}{\sigma_0}}  \ , 
% \label{eq:tdelta-1}
%\end{equation} 
%which reflects the ratio between the two {\emph{relevant}} characteristic stresses. For $\tdelta \ll 1$ the effect of 
%pre-tension, $\sigma_0$, dominates, such that the sheet is everywhere under nearly uniform and anisotropic tension. 
In this parameter regime, the sheet-substrate attachment is so strong %very strongly attached to the substrate attachment, such 
that the supported portion of the sheet cannot relieve compression through wrinkling, even though the sheet is highly bendable.
%of the sheet).   

If the sheet is not clamped to the hole's edge and can freely slide on the substrate, we find that for $\tdelta>\tdelta_c \approx 3.3$,  azimuthal (hoop) compression develops around $r=R$. For $\tdelta  \gg \tdelta_c$, the indentation-induced load dominates, and the stress becomes highly nonuniform and anisotropic, whereby the hoop-compressed zones extend upon increasing $\delta$. In the suspended part, the hoop compression can be effectively suppressed through the formation of radial wrinkles, but the supported part remains compressed.  
%In order to understand the response to hoop compression, one must distinguish between the suspended and supported parts of the sheet, and their respective "bendabilities": 
%\begin{equation}
%\epsilon^{-1}_{sus} = \frac{Y (\tfrac{\delta}{R})^2 R^2}{\kappa} \sim (\frac{\delta}{t})^2 \gg 1 \  \ \ ; \ \ 
%\epsilon^{-1}_{sup} = \frac{Y (\tfrac{\delta}{R})^2 R^2}{\sqrt{K\kappa}}  = \frac{Y\delta^2}{\sqrt{K\kappa}} \ll 1 \ . 
%\label{eq:bendabilities-1}
%\end{equation}
%The suspended part is {\emph{highly bendable}}, where most of the bare compression is relieved by radial wrinkles, and the residual hoop compression is determined by a bendability parameter, $\epsilon^{-1}_{sus}$, which compares a characteristic tensile stress and bending force in this part of the sheet. 
%In contrast, the bendability of the supported part of the sheet is affected by the impeding effect of attachment to substrate, as reflected in the relevant bendability parameter, $\epsilon^{-1}_{sup}$. In this part, despite its thinness, the sheet is characterized by low bendability, implying that compression cannot be relieved by wrinkles.    

The study of this parameter regime is the subject of the first parts of Sec.~\ref{sec:clamping_sliding} (A-C), and the results are summarized in Table I. Here, the central prediction of our study is a suppression of the force $F(\delta)$ due to sliding and wrinkling (second and third rows in Table I). Nonetheless, since the supported part of the sheet cannot wrinkle, the qualitative behavior -- transition from $F \sim \delta$ at $\tdelta \ll 1$ to $F\sim \delta^3$ at $\tdelta \gg 1$ -- is similar to the indentation of a sheet clamped at the hole's edge (first row in Table I). Note that the prefactor of the $\delta^3$ term changes significantly as function of the boundary conditions. Hence, even in this relatively simple regime, the extraction of the value of the Young modulus from an indentation experiment requires a careful consideration of the boundary conditions.    
\\

\begin{center} {\emph{Regime (ii)}} \end{center}
\begin{equation}
%\sqrt{K\kappa} \ \ll  \ \sigma_0 \ , \ Y (\frac{\delta}{R})^2 \ . 
\beta \ll 1
\label{eq:regime-ii}
\end{equation}     
In this parameter regime, described %which we address 
in Sec.~\ref{sec:clamping_sliding}.E-F, %~\ref{sec:clamping_sliding}.F,  
the sheet-substrate attachment is sufficiently low such that it is energetically favorable to suppress hoop compression through radial wrinkles in both suspended and supported parts of the sheet. As a consequence,  
%The basic physics is again governed by the dimensionless amplitude $\tdelta$ (Eq.~\ref{eq:tdelta-1}). but 
%Here, the effect of wrinkles makes 
the response to indentation %-- described in the last part of Sec.~\ref{sec:clamping_sliding} (E-F) -- 
is qualitatively different from regime {\emph{(i)}}, and  
%Specifically, the effect of the bending rigidity is approximately uniform throughout the whole sheet, and the residual hoop compression can be described through a {\emph{single}} bendability parameter: 
%\begin{equation}
%\epsilon^{-1} = \frac{Y (\tfrac{\delta}{R})^2 R^2}{\kappa} \sim (\frac{\delta}{t})^2 \gg 1  \ . 
%\label{eq:bendabilities-2}
%\end{equation} 
%Our primary study of regime {\emph{(ii)}} is described in Sec.~\ref{sec:Expansion}, and 
is summarized in the first two rows of 
Table II. A central prediction %of this study %which echoes the response of a ultrathin polymer sheets floating on liquid bath \cite{Vella15}, 
is the emergence of a {\emph{pseudo-linear}} response, Eq.~(\ref{eq:pseudo-linear-1}), at sufficiently large indentation depth, $\tdelta \geq  O({\cal \sqrt{R}}) \gg 1$, which was motivated by our 
%This response, whose origin was described in the 
heuristic discussion in Sec.~\ref{subsec:Heuristic}. 
%\\
%, reflects the {\emph{asymptotically isometric}} response of a sheet that is {\emph{nearly inextensible}} ({\emph{i.e.}} $\sigma_0/Y \ll 1$), yet {\emph{highly bendable}} ({\emph{i.e.}} $\epsilon \ll 1$).  
%
%
In Sec.~\ref{sec:Expansion} we discuss a situation where the sliding of the sheet on the substrate is hindered by clamping at the far edge, $r=\Rf$. %, which may be relevant for experimental set-ups. 
We find that far-edge clamping implies bi-axial tension %that in an annular zone 
at the vicinity of $\Rf$, % is under biaxial tension, 
even for large indentation depth ($\tdelta \gg {\cal R})$, % = \Rf/R$), 
and thus eliminates the pseudo-linear response (second row of Table~II). 
Nevertheless, the ability to relax hoop compression through wrinkling gives rise to dramatic suppression of the cubic response, $F\sim \delta^3$, when the clamping is at the sheet's edge ($r=\Rf$) in comparison to clamping %a situation where the sheet is clamped 
at the hole's edge ($r=R$). For $\tdelta \gg 1$, we find that the asymptotic ratio $F/\delta^3$ is proportional to $1/\log({\cal R})$.     
\\

\begin{center} {\emph{Regime (iii)}} \end{center}
\begin{equation}
%\sqrt{K\kappa} \ \gg  \ \sigma_0 \  \ 
1 \ll \beta \ll \tdelta^2
\label{eq:regime-iii}
\end{equation}  
In this parameter regime, the pre-tension $\sigma_0$ is irrelevant, and the substrate response is governed by a competition between the characteristic stress, $2\sqrt{B\Ksub}$, associated with the residual hoop compression in the wrinkled zone, and the bare indentation-induced stress, $Y(\tfrac{\delta}{R})^2$. We discuss this regime in Sec.~\ref{sec:residual_compression}, and show that the residual hoop compression gives rise to a comparable, {\emph{bending-induced}} radial tension \cite{hure12,Davidovitch19}. 
%As a consequence, the mechanical response in regime {\emph{(iii)}} resembles the response in regime {\emph{(ii)}} upon replacing: $\sigma_0 \to \sqrt{K\kappa}$. 
This leads us to introduce an ``effective tension'' \cite{Tobasco20}: 
\begin{equation} 
\geff
 \equiv \max\{\sigma_0,2\sqrt{B\Ksub}\} \ . 
\label{eq:gamma-eff}
\end{equation} 
Redefining the dimensionless amplitude: 
\begin{equation}
 \tdelta = \frac{\delta}{R} \sqrt{\frac{Y}{\gamma_{\rm eff}}}  \ , 
 \label{eq:tdelta-2}
\end{equation} 
we can characterize the mechanical response in regime {\emph{(iii)}} through a simple generalization of the predictions for regime {\emph{(ii)}, upon substituting (in all expressions that involve $\tdelta$): $\gamma_{\rm eff} = \sqrt{B\Ksub}$, rather than $\gamma_{\rm eff}=\sigma_0$. This is the content of the last row in Table II.     
\\

\section{Clamping versus sliding and wrinkling \label{sec:clamping_sliding}}
We start by considering a perfectly axisymmetric deformation in response to indentation, namely, no wrinkles are allowed on the suspended or supported parts of the sheet. In Subsec.~\ref{subsec:II-clamping} and \ref{subsec:II-sliding} we address two types of boundary conditions (BCs). The first type is clamping at the hole's edge ($r=R$) with a ``pre-tension''
$\sigma_0$. 
%which is assumed to be isotropic and uniform at the exterior of the hole ($r>R$).  This type of BCs 
%which has been assumed by most workers in this field (\cite{Hone08},\blue{OTHERS,McEuen}), and whose theoretical description through FvK equations was reported recently \cite{Vella17}.   
% reflects conditions of ``no-sliding" of the sheet on the substrate, and consequently an inability of the sheet to accommodate a far-field tension $\sigma_0$  exerted at the far-edge ($r= R_{out} \gg R$) through gradients of the stress tensor at $r>R$.  
The second type of BCs allows for sliding of the sheet on the substrate, while a given tensile load, $\srr(\Rf) = \sigma_0$, is exerted at the far edge of the sheet.  
%, such that the radial tension at the hole's edge, $T = \sigma_{rr}(R)$, may be different from $\sigma_0$ exerted at the far-edge $r= R_{out}$. 
We show that the freedom to slide on the substrate significantly suppresses the indentation force. 
%Interestingly, we find that the "linear" response (i.e. for infinitesimal $\tilde{\delta}$ where pre-tension predominates, is actually sub-linear, with a logarithmic correction.    
Importantly, we find that if sliding is allowed, the sheet becomes azimuthally compressed in the vicinity of the hole's edge, if the dimensionless indentation depth, $\tdelta$, exceeds a critical value $\delta_c \approx 3.3$. 
% (which depends on the Poisson ratio $\nu$). 
This indicates an instability to the formation of radial wrinkles, which we address in Subsec.~\ref{subsec:II-wrinkling}, assuming a sufficiently strong attachment to the substrate (regime {\emph{(i)}}, Eq.~\ref{eq:regime-i}), such that wrinkles can form only at the suspended part of the sheet. We use this case to introduce the basic principles of the {\emph{far from threshold}} approach \cite{Davidovitch11}, through which we characterize the emerging wrinkle pattern, and show how the formation of wrinkles underlies further, albeit modest suppression of the indentation force. In Subsec.~\ref{subsec:threshold-2} we relax the condition of infinitely strong sheet-substrate attachment, and find a second threshold, such that for $\tdelta>\tdeltast(\beta)$, hoop compression is sufficiently strong to give rise to radial wrinkles also on the supported part of the sheet. In Subsecs.~\ref{subsec:both-wrinkling} and \ref{subsec:geo-limit} we address the parameter regime {\emph{(ii)}} (Eq.~\ref{eq:regime-ii}), where wrinkles expand throughout the supported part of the sheet and further suppress the indentation force, culminating with a transition to the pseudo-linear response, Eq.~(\ref{eq:pseudo-linear-1}). 
%sections, we employ this approach to study parameter regimes {\emph{(ii)}} (Eq.~\ref{eq:regime-ii}) and  {\emph{(iii)}} (Eq.~\ref{eq:regime-iii}), where wrinkles can form also in the supported part of the sheet, suppressing further the indentation force.      

\subsection{Clamping at the hole's edge}
\label{subsec:II-clamping}
Following Ref.~\cite{Vella17}, we address a circular sheet clamped with pre-tension $\sigma_0$ at the hole's edge, $r=R$. 
%Physically, this means that the sheet is perfectly stuck to the substrate with no sliding at all. Thus, 
%Since radial displacement is not allowed, stress is uniform and isotropic outside the hole, such that the far - field radial tension $\gamma$ induces an isotropic uniform stress field in the film outside the hole that is not affect by the indentation process. 
The axial symmetry of the set-up calls for the use of polar coordinates. We denote the out-of-plane displacement by $z(r)$, and by $\psi(r)$ the radial derivative of the Airy stress function, % \cite{Landau}, 
where the stress components are:
\begin{equation}
\sigma_{rr} = \frac{\psi}{r} \, , \, \sigma_{\theta\theta} = \psi'  \ .  
\label{eq:define-Airy} 
\end{equation}
The hoop component of the strain tensor $\varepsilon_{\theta\theta}$ and the consequent radial displacement field $\rmu_r$, satisfy: 
\begin{gather} 
\varepsilon_{\theta\theta} = \frac{\rmu_r}{r} = \frac{1}{Y} \left( \sigma_{\theta \theta} - \nu \sigma_{rr} \right) = \frac{1}{Y} \left( \psi' - \nu \frac{\psi}{r} \right) \nonumber \\
 \Rightarrow \rmu_r = \frac{1}{Y} \left( r \psi' - \nu \psi \right)
 \label{eq:disp-strain-stress-t}
 \end{gather}
whereas the radial strain is:
\begin{gather} 
\varepsilon_{rr} = \frac{\partial \rmu_r}{\partial r}  + \frac{1}{2} (\frac{\partial z}{\partial r})^2 =
 \frac{1}{Y} \left( \sigma_{rr} - \nu \sigma_{\theta \theta}  \right) \nonumber \\ 
 = \frac{1}{Y} \left(\frac{\psi}{r} -\nu  \psi' \right) \ . 
 \label{eq:disp-strain-stress-r}
\end{gather}
\\
The $2^{nd}$ FvK equation, expressing in-plane force balance (as well as compatibility of the stress and strain tensors with the displacement field), is:
\begin{align}
r \frac{d}{dr} \left[ \frac{1}{r} \frac{d}{dr} \left( r \psi \right) \right] &= - \frac{1}{2} Y \left( \frac{dz}{dr} \right)^2
\label{eq:FvK-dim-2}
\end{align}
and the $1^{st}$ FvK equation, which expresses force balance in the normal direction ($\approx \hat{z}$) is: 
\begin{align}
\frac{1}{r} \frac{d}{dr} \left( \psi \frac{dz}{dr} \right) &= \frac{F}{2 \pi r} \delta ( r ) \ .  
\label{eq:FvK-dim-1}
\end{align}
In the last equation we neglected a bending force, $B\partial^4z/\partial r^4$, due to the radial curvature of the sheet. As we explain in App.~\ref{app:krr-BC}, this term is significant only at the vicinity of the hole's edge, and its omission -- together with a suitable choice of BCs at the hole's edge -- is justified in all parameter regimes addressed in our paper (see schematic Fig.~\ref{fig:schem-slide}b-c).
%where we assume no external normal forces, except the indentation force $F$. 

Let us turn now to describe the BCs at the vicinity of the indenter ($r \to 0$), and the hole's edge:  
\begin{align}
r \!\to\! 0  &:  %\, \,  \, \, 
(i) \, %\, 
z  \!=\!- \delta  \, \, \, %\, \, \,  \, \, \, 
(ii) \,  %\, 
\rmu_r \!=\! \lim_{r \rightarrow 0} \frac{1}{Y} \left( r \psi' \!-\! \nu \psi \right) \!=\! 0 \nonumber \\
%\end{align}
%and at the clamped edge, $r = R$,
%\begin{align}
r \!=\! R &:  %\, \, \, \, 
(iii) \, %\, 
z \!=\! 0 \, \, \, %\, \, \, \, \, \, 
(iv) \, \rmu_r  \!=\! \frac{1}{Y}  \left(\! r \psi' \!-\! \nu \psi \!\right) \!=\! ( 1\! -\! \nu ) \frac{\sigma_0}{Y} R \ . 
\label{eq:BC-clamping}
\end{align}
Note that clamping at $r = R$ in the presence of a pre-tension, $\sigma_0$, means that the radial displacement $\rmu_r (R )$  (rather than the load) is set to a fixed non-zero value, determined by the pre-indentation condition, as reflected in BC {\emph{(iv)}}, Eq.~(\ref{eq:BC-clamping}).
\\

Throughout this paper, we denote by $\psi , r , z$ dimensional values of the potential, radial (in-plane) length, and deflection (out-of-plane) length, respectively, and $\Psi , \rho$, and $\zeta$, for their dimensionless counterparts:
\begin{align}
\rho &= \frac{r}{R} , \, \, \, \Psi = \frac{\psi}{\sigma_0 R} , \, \, \, \zeta = \sqrt{\frac{Y}{\sigma_0}} \frac{z}{R} \ . 
\label{eq:dimensionless-var}
\end{align}
Additionally, we define a dimensionless version of the force $F$: 
\begin{align}
{\cal F} & %= \frac{1}{2 \pi R} \frac{F}{\sigma_0 \sqrt{\frac{\sigma_0}{Y}}} 
=  \frac{1}{2 \pi R} \sqrt{\frac{Y}{\sigma_0^3}}  \ F  \ . 
% , \, \, \, \tilde{\delta} =  \frac{1}{R} \sqrt{\frac{Y}{\sigma_0}} \ \delta \ . 
 \label{eq:dimensionless-param}
\end{align}
The dimensionless form of the FvK equations (\ref{eq:FvK-dim-2},\ref{eq:FvK-dim-1}) is: 
\begin{align}
\rho \frac{d}{d\rho} \left[ \frac{1}{\rho} \frac{d}{d\rho} \left( \rho \Psi \right) \right] &= - \frac{1}{2}  \left( \frac{d \zeta}{d \rho} \right)^2 \ , 
\label{eq:FvK-nondim-2}
\end{align}
\begin{align}
\frac{1}{\rho} \frac{d}{d \rho} \left( \Psi \frac{d \zeta}{d \rho} \right) &= \frac{\cal F}{2 \pi \rho} \delta ( \rho ) \ . 
\label{eq:FvK-nondim-1}
\end{align}
The dimensionless version of the BCs (\ref{eq:BC-clamping}) is: 
\begin{align}
\rho \rightarrow 0 \ &:  \, \, \, \,  (i ) \, \zeta = -\tdelta \, \, \, \, \, \, \, \, \,(ii ) \lim_{\rho\to 0} \left( \rho \Psi' - \nu \Psi \right) = 0 \nonumber \\
\rho =1 \ &:  \, \, \, \,  (iii) \, \zeta = 0 \, \, \, \, \, \, \, \, \, (iv) \, (\Psi' - \nu \Psi ) = 1 - \nu
\label{eq:BC-nondim-clamping}
\end{align}

Although Eqs.~(\ref{eq:FvK-nondim-2},\ref{eq:FvK-nondim-1}) are nonlinear, there exists a transformation 
\cite{Bhatia68}, which allows an analytic solution (up to integrals that can be evaluated numerically).
%The solution involves variable transformation:  \begin{equation}  \Phi \equiv \rho \Psi \, \, , \, \, \eta \equiv \rho^2\ , \label{eq:transformation-1} \end{equation} such that: $\Psi = \tfrac{\Phi}{\sqrt{\eta}} \ , \ \tfrac{d \Psi}{d \rho} = \left( 2 \frac{d \Phi}{d \eta} - \frac{\Phi}{\eta} \right)$, and the BCs~(\ref{eq:BC-nondim-clamping}) become:  \begin{align} \eta =0 \ &:  \, \, \, \,  (i ) \, \zeta = -\tdelta \, \, \, \, \, \, \, \, \,(ii ) \Phi = 0 \nonumber \\ \eta =1 \ &:  \, \, \, \,  (iii) \, \zeta = 0 \, \, \, \, \, \, \, \, \, (iv) \, 2\Phi'  = (1-\nu) + (1+\nu) \Phi  \ . \label{eq:BC-nondim-trans-clamping} \end{align}  
The analytic solution \cite{Vella17}, which we repeat in App.~\ref{app:axisymmetric}, enables us to express the force, $F$, the shape, $z(r)$, and the stress components, $\sigma_{rr}(r), \sigma_{\theta\theta}(r)$, for any value of $\tdelta$. These are shown, respectively, in the gray curves in Figs.~\ref{fig:response-1}-\ref{fig:shape}.  Let us discuss briefly some key features of of these results. 
\\

At sufficiently large values of $\tdelta$, the force $F(\delta) \sim \delta^3$, 
%as was anticipated already by the qualitative discussion in Subsec.~\ref{subsec:Heuristic}, 
reflecting a transition from pretension-dominated stress (for $\tdelta \ll 1$) % in Fig.~\ref{fig:stress-edge}) 
to indentation-dominated stress: $\sim Y (\delta/R)^2$ (for $\tdelta \gg 1$). %in Fig.~\ref{fig:stress-edge}). 
The radial and hoop components of the stress, shown, respectively, by the dashed and solid curves in Fig.~\ref{fig:stress-profile-small} ($\delta = 3$) and Figs.~\ref{fig:hoop-stress-profile},\ref{fig:radial-stress-profile} ($\tdelta =10$),  %, for $\tdelta = 10^3$ \blue{???}) 
indicate that an indented sheet clamped at the hole's edge is under pure tension at any indentation depth, in agreement with the discussion in 
Subsec.~\ref{subsec:Heuristic}. 

An interesting feature of the indentation force  %, shown in Fig.~\ref{fig:response-2}, 
is the absence of a true linear response (Fig.~\ref{fig:response-1}b). Instead, for $\tdelta \ll 1$ the response is sub-linear, with $F \sim -1/\log(\tdelta)$. Such a sub-linear response appears also for the sliding BCs, albeit with a different numerical pre-factor (blue curve in Fig.~\ref{fig:response-1}). This peculiar feature emanates from the assumption of a point-wise indentation, and is intimately related to the  (integrable) divergence of the stress components at $r \to 0$ ($\srr,\sqq \sim r^{-1/3}$ \cite{Vella17}), which is observed in Fig.~\ref{fig:stress-profile-small}. For an indenter with a finite tip's radius $R_{tip}$, a linear response is recovered, with a numerical pre-factor that scales as $-[\log(R_{tip}/R)]^{-1}$ \cite{Vella17}. 

Finally, it is noteworthy that the deformed shape (Fig.~\ref{fig:shape}) defers substantially from an ideal cone; this is signified by the slope in the vicinity of the hole's edge $(\tfrac{dz}{dr})_{r=R}$, which is only 63\% of the slope of an ideal cone (first row of Table I). We will show later that the slope at the vicinity of the edge depends strongly on the boundary conditions and other physical parameters, and may thus serve as an indirect experimental probe of the actual boundary conditions associated with a given set-up.     
 
%%%%%%%%%%%%%%%%%%%%%%%%%%%%%%%%%
\begin{figure*}
\includegraphics[width=1.0\textwidth]{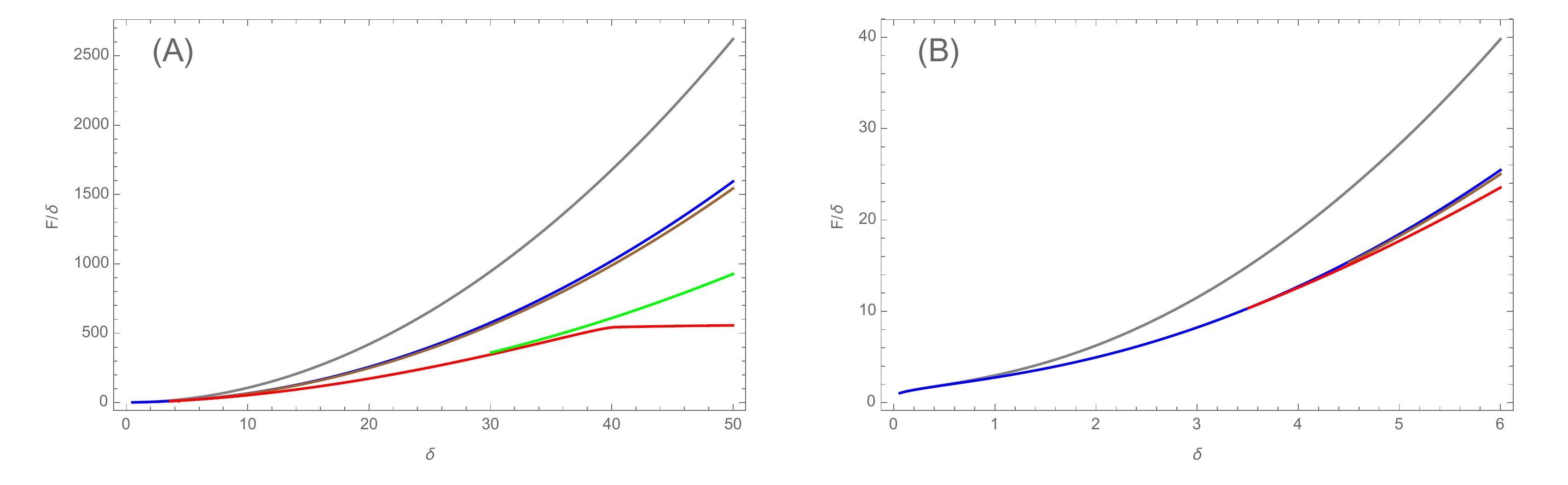}
\caption{``Spring constant" $F/\delta$, {\emph{versus}} indentation amplitude $\delta$, where the force $F$ is normalized by $R^2\sigma_0$, and $\delta$ is normalized by $R\sqrt{\sigma_0/Y}$. Different colors represent different boundary conditions and physical parameters. Gray: clamping at the hole's edge. Blue: sliding of the sheet on the substrate, assuming the deformed shape is perfectly axisymmetric. Brown: sliding of the sheet on the substrate, where wrinkles are allowed to relax hoop compression only in the suspended part of the sheet (Regime {\emph{(i)}}, Eq.~\ref{eq:regime-i}). Green: sliding is hindered by clamping at the far edge, $r=\Rf \ (\approx 90 R$), and wrinkles are allowed to relax hoop compression in both suspended and supported parts of the sheet (Regime {\emph{(ii)}}, Eq.~\ref{eq:regime-ii}). Red: sliding throughout the whole sheet (same value of $\Rf \approx 90 R$), where wrinkles are allowed to relax compression in both suspended and supported parts of the sheet (Regime {\emph{(ii)}}, Eq.~\ref{eq:regime-ii}). For the problems that assume clamping (at the hole' edge (gray curve) or the sheet's edge (green curve)), $\sigma_0$ is the ``pretension" in the sheet, whereas for the sliding problems (blue, brown, and red curves), $\sigma_0$ is the tensile load exerted at the far edge. Regardless of the various BCs, the ``linear'' response at $\delta \ll R\sqrt{\sigma_0/Y}$, shown in panel B, is actually sub-linear, whereby $F/\delta \sim 1/|\log\delta| \to 0$ \cite{Vella17}. When wrinkles are not allowed on substrate (gray, blue, and brown curves) the asymptotic response at $\delta \gg R\sqrt{\sigma_0/Y}$ is cubic ($F/\delta\sim \delta^2$). When wrinkles can form on the substrate, the response become eventually ``pseudo-linear'' ($F/\delta \sim const$) after wrinkles can reach the sheet's edge (red curve), and sub-cubic ($F/\delta \sim \delta^2/|\log\delta|$) if the sheet's edge is clamped (green curve).        
}\label{fig:response-1}
\end{figure*}
%%%%%%%%%%%%%%%%%%%%%%%%%%%%%%%%%%
%\begin{figure}
%\includegraphics[width=0.9\textwidth]{lindentation_wrinkles.pdf}
%\caption{At small values of the (dimensionless) indentation amplitude, $\tdelta$, the response is sub-linear $F \propto -1/\log(\tdelta)$, where the pre-factor is $\approx 0.86$ for the clamped BC's at the edge (gray curve), and $\approx 0.62$ for the sliding BC's ((blue curve). The sub-linear response stems from the assumption of point-wise indenter; for a finite-size indenter, one recovers a linear response with a pre-factor that exhibits logarithmic dependence on the size of the indenter's tip \cite{Vella17}.}
%\label{fig:response-2}
%\end{figure}
%%%%%%%%%%%%%%%%%%%%%%%%%%%%%%%%%
\begin{figure*}
\includegraphics[width=1.0\textwidth]{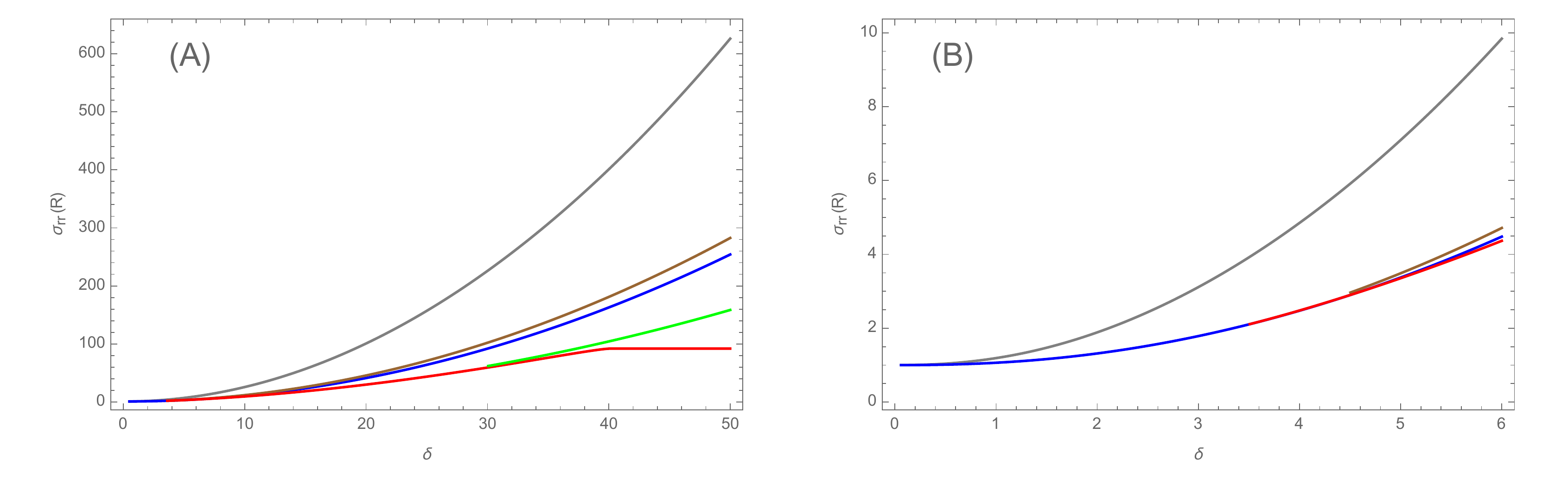}
\caption{The radial stress at the hole's edge, $\srr(R)$, normalized by $\sigma_0$, {\emph{versus}} indentation amplitude $\delta$, normalized by $R\sqrt{\sigma_0/Y}$. Colors of curves represent different boundary conditions and physical parameters, as in Fig.~\ref{fig:response-1}.}
\label{fig:stress-edge}
\end{figure*}
%%%%%%%%%%%%%%%%%%%%%%%%%%%%%%%%%%

%Notice the introduction of the tension, $\gamma$. Importantly, this is defined as $\gamma = \sigma_{rr} ( r = R , \delta = 0 )$. Namely, it is the pre-tension, meaning prior to indentation. 

\subsection{Sliding at the hole's edge and the buckling threshold in the suspended zone}
%\\ (no wrinkling)}
\label{subsec:II-sliding}
Now we address the axisymmetric (unwrinkled) state of the indented sheet in a set-up, where sliding of the sheet is allowed at the hole's edge (and on the substrate). Clearly, the only difference between this case and the above analysis of clamping at $r=R$, is encapsulated by the BC {\emph{(iv)}} in Eq.~(\ref{eq:BC-nondim-clamping}). For simplicity, we assume a fixed tensile load at the far edge, $\sigma_{rr}(\Rf) = \sigma_0$. We note that, as long as $\Rf \gg R$, replacing this BC with clamping at the far edge (with pre-tension $\sigma_0$), gives rise to practically indistinguishable results.

In order to derive the appropriate BC at the hole's edge, we must consider the stress field in the supported part of the sheet. In this annular zone, $R<r<\Rf$, the sheet is subjected to radial tension $\sigma_0$ at $r=\Rf$ and an unknown radial tension $\srr(R)$ at $r=R$. This problem is readily recognized as the Lam\'e problem, and its classical solution yields the hoop and radial stress components \cite{TimoshenkoBook}: 
%\begin{eqnarray}
\begin{equation}
\label{eq:Lame-stress} 
%\!\!\!\!\!\!\!\!\! 
\small{R\!<\!r\!<\!\Rf:} \left\{ 
\begin{array}{c}
  \srr (r) = \sigma_0 + \left(\srr(R) - \sigma_0\right)\frac{R^2}{r^2}     \\
   \sqq (r) = \sigma_0 - \left(\srr(R) - \sigma_0\right)\frac{R^2}{r^2}   
%  &\srr& (r) = \sigma_0 + \left(\srr(R) - \sigma_0\right)\frac{R^2}{r^2} \label{eq:Lame-stress}   
%\\
%\  \ \ \ \ ; \ \ \ \ \ \   
%&\sqq& (r) = \sigma_0 - \left(\srr(R) - \sigma_0\right)\frac{R^2}{r^2} %\nonumber 
\end{array}
\right.
\end{equation}
%\end{eqnarray}
(where we simplified the general solution for ${\cal R} = \Rf/R \gg 1$), 
allowing one to express the radial displacement, $\rmu_r(r)$ as a function of $\Rf,r,\sigma_0$, and $\sigma_{rr}(r)$: 
\begin{equation}\
%{\rm axisymmetric state} \  
\!\!\!R<r<\Rf: \ \rmu_r(r) = \frac{r}{Y} [2 \sigma_0 - (1+\nu)\sigma_{rr}(r)]  \ . 
\label{eq:dis-stress-Lame} 
\end{equation}    
Obviously, integrity of the sheet requires continuity of the radial displacement at the hole's edge, namely: 
\begin{equation}\rmu_r(r \to R^+) = \rmu_r(r \to R^-) \ ,  \label{eq:matchingur}
\end{equation}
and the analogous relationship for the radial component of the stress reads:   
\begin{equation}\srr(r \to R^+) = \srr(r \to R^-) \ . \label{eq:matchingsrr}
\end{equation}
In App.~\ref{app:krr-BC} we will elaborate further on the continuity of radial displacement and stress at the hole's edge and the validity of the corresponding equations (\ref{eq:matchingur},\ref{eq:matchingsrr}). 
%We note by passing that although the hoop stress may exhibit discontinuity (although it typically does not), since this does not violate any force balance. Indeed, the 
%
%In writing Eq.~(\ref{eq:matchingsrr}) we assume that the substrate-sheet contact zone at the vicinity of the hole's edge does not give rise to tangential forces on the sheet. We %will assume this scenario throughout our analysis and 
%will elaborate on the validity of this assumption in Sec.~\ref{sec:discussion}.  

%Integrity of the sheet and radial force balance imply continuity of the radial displacement $\rmu_r(r)$ and the radial stress $\sigma_{rr}(r)$ at the hole's edge, $r=R$. 
Equations~(\ref{eq:dis-stress-Lame},\ref{eq:matchingur},\ref{eq:matchingsrr}), together with Eqs.~(\ref{eq:define-Airy},\ref{eq:disp-strain-stress-t}) yield: $\psi/R + \psi' = 2\sigma_0$. Turning to dimensionless representation we obtain the BCs:   
\begin{align}
\rho \rightarrow 0 \ &:  \, \, \, \,  (i ) \, \zeta = -\tdelta \, \, \, \, \, \, \, \, \,(ii ) \lim_{\rho\to 0} \left( \rho \Psi' - \nu \Psi \right) = 0 \nonumber \\
\rho =1 \ &:  \, \, \, \,  (iii) \, \zeta = 0 \, \, \, \, \, \, \, \, \, (iv) \, \Psi + \Psi' = 2
\label{eq:BC-nondim-sliding-axi}
\end{align} 
The solution of the FvK equations (\ref{eq:FvK-nondim-2},\ref{eq:FvK-nondim-1}), with the BCs~(\ref{eq:BC-nondim-sliding-axi}) can be obtained in a similar way to the solution in the preceding subsection (see App.~\ref{sub-app:sliding}), 
allowing us to express the force, $F$, the shape, $z(r)$, and the stress components, $\sigma_{rr}(r), \sigma_{\theta\theta}(r)$, for any value of $\tdelta$. These are shown, respectively, in the blue curves in Figs.~\ref{fig:response-1}-\ref{fig:shape}. 

One may notice that the qualitative behavior of the axisymmetric state with sliding BC's is very similar to the edge-clamped set-up. Considering the stress and force as functions of the dimensionless parameter $\tdelta$ (Figs.~\ref{fig:response-1},\ref{fig:stress-edge}, respectively), their magnitudes scale similarly in both set-ups, for $\tdelta \ll 1$ as well as for $\tdelta \gg 1$, whereas the numerical pre-factors become smaller once sliding is allowed. Intuitively, sliding allows the sheet to moderately relax the stretching in the suspended part at the expense of more stretching at the supported part. An interesting observation is the pronounced %relatively strong 
effect of the change in BCs on the shape (Fig.~\ref{fig:shape}). Specifically, we found that for $\tdelta \gg 1$, the contact angle of the membrane at the edge approaches the asymptotic value: $(\tfrac{dz}{dr})_{r=R} \rightarrow 0.82 \frac{\delta}{R}$ -- an increase of over 25\% of the slope under clamped BC's. 
%We will see in later sections that if wrinkles are allowed to form throughout the sheet, the shape approaches asymptotically an ideal cone: $(\tfrac{dz}{dr})_{r=R} \rightarrow \frac{\delta}{R}$  as $\tilde{\delta} \rightarrow \infty$.

A dramatic feature of the stress profile is the emergence of an annular zone on both sides of the edge, where the hoop stress is compressive (blue curve in Fig.~\ref{fig:hoop-stress-profile}), for $\tdelta \gtrsim 3.3$. 
While a compressive stress may not be relieved in the supported part due to a strong attachment to the substrate (regime {\emph{(i)}}, Eq.~\ref{eq:regime-i}), the existence of compression in the suspended part of a thin sheet clearly gives rise to a wrinkling instability. Understanding the wrinkle pattern under such physical conditions is the subject of the next subsection.  
 
%%%%%%%%%%%%%%%%%%%%%%%%%%%%%%%%%%%%%%%%%%%%%%%%

%%%%%%%%%%%%%%%%%%%%%%%%%%%%%

\begin{figure}
\includegraphics[width=0.5\textwidth]{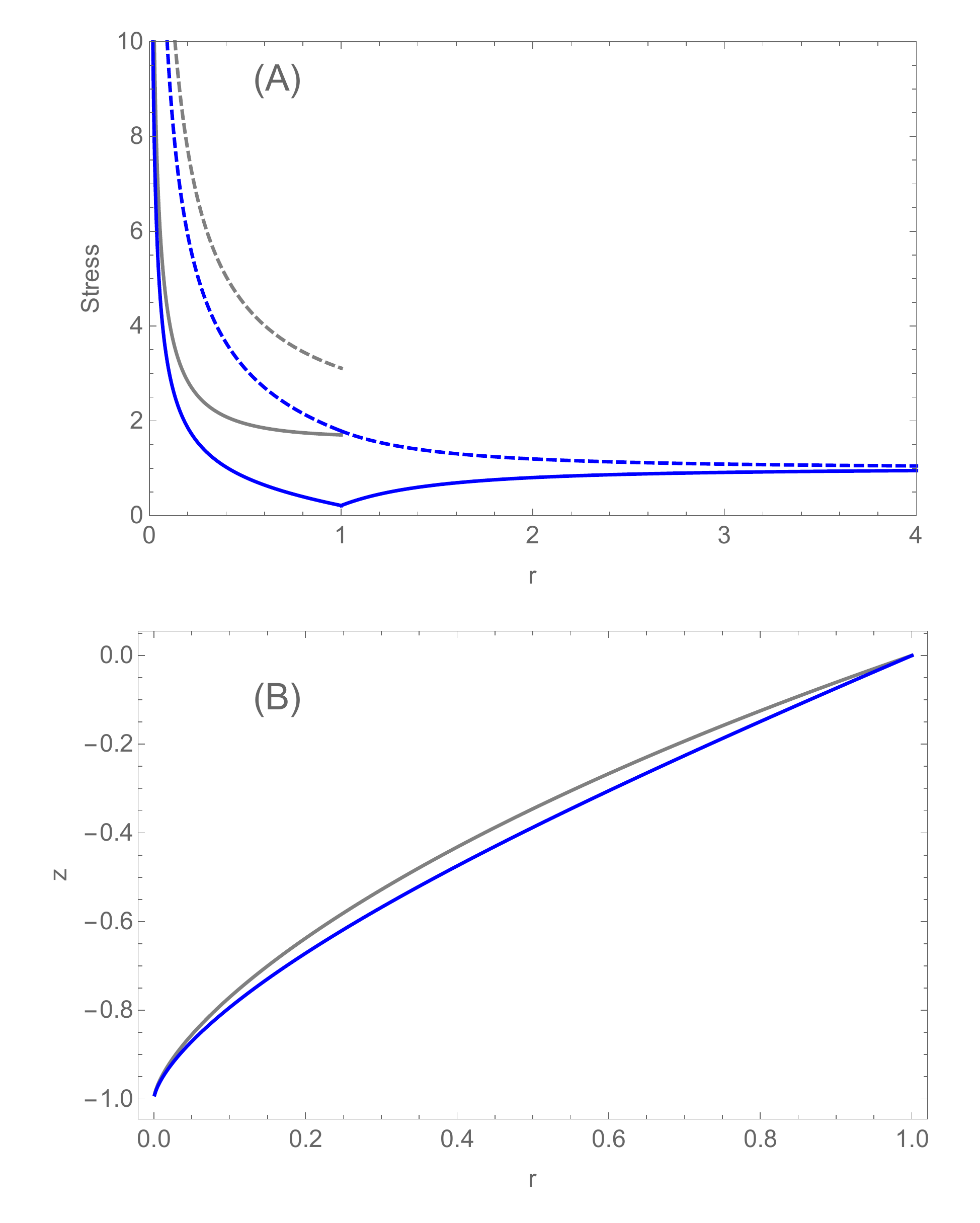}
\caption{(a) The hoop stress (solid) and radial stress (dashed) for a sheet that is clamped (gray) or slide (blue) at the hole's edge. Here, the value of the dimensionless indentation depth $\tdelta = 3$, for which both cases the sheet is under pure tension, and the axisymmetric response is stable. Both stress components are normalized by a constant $\sigma_0$ (see text). (b) The profile of the suspended sheet. Here, radial distances are normalized by the hole's radius $R$, whereas vertical distances are normalized by the indentation depth $\delta$. Note the deviations in both cases from a perfect conical shape.}  
\label{fig:stress-profile-small}
\end{figure}
%%%%%%%%%%%%%%%%%%%%%%%%%%%%%%%%%

%\subsection{How wrinkles affect the mechanical response: A primer} 
%\subsection{Sliding at the hole's edge \\ (wrinkling in suspended sheet)}
\subsection{Wrinkling in the suspended zone}
\label{subsec:II-wrinkling}
We come to study the simplest case in which wrinkles affect the mechanical response, where the hoop compression induced by sliding at the hole's edge gives rise to wrinkles in the suspended portion of the sheet ($r<R$), but not at the supported part ($r>R$). 
%(Recall that assuming $\beta \gg \tdelta^2$ implies that the VdW potential is sufficiently strong to prevent even a slight deviation of the sheet-substrate separation from the energetically-favorable distance $d=d_0$). 
In order to study the effect of wrinkles on the stress and thereby on the indentation force, we employ {\emph{tension field theory (TFT)}} \cite{Wagner35,MansfieldBook,Stein61,Pipkin86,Steigmann90}. In this approach, one assumes that wrinkles suppress almost entirely compressive stress, such that one of the two principal components of the  
%the (diagonal form of the) 
stress tensor in the wrinkled zone is positive, %consists of a single positive component, which
corresponding to tensile stress {\emph{along}} wrinkles, whereas the other principal component vanishes, signifying the direction along which wrinkles undulate.  
%all other components vanish. 
The stress field in the whole sheet is then obtained by matching the displacement field and the compression-free stress in the wrinkled zone to the adjacent, purely tensile zones, where both principal stress components are non-negative.
%Realizing the substantial effect that sliding and wrinkling may have on the response to indentation, we employ the far-from-threshold approach to obtain a quantitative prediction for the simplest scenario, where the sheet-substrate attachement is sufficiently strong to prevent any wrinkling on the substrate (but not sliding), such that wrinkles can relax compression only inside the hole, at $r<R$. 

Applying the TFT methodology to our indentation problem, the axial symmetry of the set-up suggests that for $\tdelta > 3.3$, confinement of latitudes occurs in an annular zone, $\Li<r<\Lo$, where $\Li <R$, and $\Lo >R$. In this subsection we assume that in the supported part, $R<r<\Lo$, the large effective stiffness $\Ksub$, Eq.~(\ref{eq:KsubWitten}), %(due to attachment to the substrate) 
prohibits the formation of wrinkles, such that the sheet must accommodate the indentation-induced hoop compression; however, in the suspended part, $\Li<r<R$, the formation of radial wrinkles underlies {\emph{collapse}} of hoop compression. Hence, the sheet is naturally divided into three parts: {\emph{(i)}} $R<r<\Rf$ -- where the supported sheet undergoes a planar axisymmetric deformation; 
%and slides on the substrate; 
{\emph{(ii)}} $\Li<r<R$ -- where the suspended sheet is wrinkled; {\emph{(iii)}} $r<\Li$ -- where the suspended sheet is unwrinkled and the stress is purely tensile. In the sequel, we will describe the stress and deformation in each zone %, boundary conditions, 
and the matching among them.    
\\

\noindent 
\underline{Zone {\emph{(i)}} $R<r<\Rf$:} 
Similarly to Subsec.~\ref{subsec:II-sliding}, the state of the sheet in this part is determined by solving the planar Lam\'e problem, subject to radial tensile load at the far edge, $\srr(\Rf) = \sigma_0$, and a radial tension $\srr(R)$, which must be determined by matching the three zones. In this zone, the stress components are given by Eq.~(\ref{eq:Lame-stress}) through the unknown $\srr(R)$, and the ratio between the radial displacement and stress is given by Eq.~(\ref{eq:dis-stress-Lame}), which we repeat here for completeness: %whose evaluation at $r=R$ yields the equation: 
\begin{equation}\
\rmu_r(r) = \frac{r}{Y} [2 \sigma_0 - (1+\nu)\srr(r)]    \ . 
\label{eq:dis-stress-Lame-1} 
\end{equation} 

\noindent 
\underline{Zone {\emph{(ii)}} $\Li<r<R$:} Here, the formation of wrinkles underlies a collapse of the hoop compression, such that we need to solve the radial force balance equation with $\sqq=0$. (More precisely, a TFT solution is the leading order in a ``high bendability'' expansion, $\epsilon \to 0$, of the FvK equations \cite{Davidovitch11}, % {Davidovitch12}, 
rather than a standard expansion around the compressed, axisymmetric state \cite{TimoshenkoBook}). Technically,  in the wrinkled zone radial force balance is obtained by satisfying Eq.~(\ref{eq:define-Airy}) with $\psi(r) = constant$, whereas the relationship $\varepsilon_{\theta\theta} = \rmur/r$ (\ref{eq:disp-strain-stress-t}) between the hoop strain and radial displacement is ``ignored''  since it merely determines a comparable contribution to the hoop strain ($\varepsilon_{\theta\theta} = -\nu \varepsilon_{rr}$), which is missing from the RHS of Eq.~(\ref{eq:disp-strain-stress-t}) due to the excess length in the wrinkly undulations \cite{Davidovitch11}. Equation (\ref{eq:FvK-dim-2}) also relies on (\ref{eq:disp-strain-stress-t}) and its validity is thus limited to an axisymmetric, unwrinkled state, hence it is likewise ignored. Requiring continuity of radial displacement and stress, Eqs.~(\ref{eq:matchingur},\ref{eq:matchingsrr}), 
%(which condition follows directly from force balance in the radial direction at the vicinity of the hole's edge), 
as well as continuity of the deflection $z(R) = 0$, and employing Eqs.~(\ref{eq:define-Airy},\ref{eq:disp-strain-stress-r},\ref{eq:FvK-dim-1}), we obtain the stress components, deflection $z(r)$ and radial displacement in the wrinkled zone: 
 \begin{eqnarray}
 \srr(r) &=& \srr(R)\frac{R}{r} \, , \, %\, \, \, 
 \sqq(r) = 0 \ \Rightarrow \ \psi(r) = R \cdot \srr(R) 
 \nonumber \\ 
 z(r) &=& \arctan\theta \cdot ( r - R ) \approx \theta \cdot ( r - R )  \ ,  \nonumber  \\
 \rmur(r) &=& \!- \!\frac{1}{2} \theta^2\cdot (r\!-\!\Li) \!+\! \frac{1}{Y}{\srr(R)} R \log \left(\! \frac{r}{\Li} \!\right) 
\nonumber \\
&+&  \rmur(\Li) 
 \label{eq:wrinkled-I}
 \end{eqnarray}
where $\theta \ll 1$ is the angle between the suspended sheet and the planar substrate at $r=R$, and $\rmur(\Li)$ is the radial displacement at the %inner 
edge of the wrinkled zone, $r=\Li$. We re-emphasize that although both radial displacement and Airy potential in the wrinkled zone are given by axisymmetric functions, the presence of symmetry-breaking wrinkles is reflected in the violation of the relationship (\ref{eq:disp-strain-stress-t}) between , $\rmur(r)$ and $\psi(r)$. 
\vspace{0.2cm} 

\noindent 
\underline{Zone {\emph{(iii)}} $0<r<\Li$:}  
In the purely tensile core the state is again axisymmetric, and the FvK equations there, expressed through the dimensionless functions 
$\zeta(\rho)$ and $\Psi(\rho)$~(Eq.~\ref{eq:dimensionless-var}) are correspondingly given by Eqs.~(\ref{eq:FvK-nondim-2},\ref{eq:FvK-nondim-1}), with the strain-displacement relationship for both parts of the strain tensor (\ref{eq:disp-strain-stress-t},\ref{eq:disp-strain-stress-r}). 
%exactly as we had in the previous subsections for an axially-symmetric deformation with point force at $r=0$. 
Exploiting once again the continuity of the radial stress $\srr(r) = \psi(r)/r$ and the deflection $z(r)$, 
%and defining a dimensionless version of the unknown $L$, as $\tL = L/R$:     
%at the outer edge of the tensile core, $r=L$, 
we obtain the BCs: % at $\rho=0$ and $\rho=\tL$:   
\begin{eqnarray}
%r \to 0 \ :  \ \ \ \ \ \ 
\zeta(0)  = - \tdelta  \ \ \ &;& \  \ \  %\rmu_r(0) = \lim_{r \rightarrow 0} \frac{1}{Y} 
\lim_{\rho \to 0}\left( \rho \Psi' - \nu \Psi\right) = 0 \nonumber \\
%\end{align}
%and at the clamped edge, $r = R$,
%\begin{align}
%r = L \ :  \ \ \ \ \ \ 
\zeta(\tL) =\ta (\tL-1) \ \ \ &;& \ \   \Psi(\tL) = \Psi(1) = \frac{\srr(R)}{\sigma_0}  
% \rmur(L)  = \frac{1}{Y}  \left[ r \psi'(L) - \nu \psi(L) \right] \ . 
\label{eq:BC-sliding-wrinkling-I}
\end{eqnarray}
where $\tL$ and $\ta$ are dimensionless versions of the core radius and the slope at the hole's edge:  
\begin{equation}
\tL = \Li/R \ \ ; \ \ \ta = \sqrt{Y/\sigma_0} \theta \ . 
\label{eq:new-nondim-param}
\end{equation}
Similarly to the previous subsections, we find that the nonlinear Eqs.~(\ref{eq:FvK-nondim-2},\ref{eq:FvK-nondim-1}) with the BCs~(\ref{eq:BC-sliding-wrinkling-I}) can be solved analytically. Namely, for a given value of the control parameter $\tdelta$ and given values of the three unknowns, $\Psi(1), \theta$, and $\tL$, there is a single analytic solution that fully characterizes the function $\Psi(\rho)$, the related components of the stress (\ref{eq:define-Airy}), and the deflection $\zeta(\rho)$, in the interval $0<\rho<\tL$. Since the state in this core zone is axisymmetric (unwrinkled), Eq.~(\ref{eq:disp-strain-stress-t}) implies that the radial displacement at $r=L$ satisfies:
% related to the given by Eq.~(\ref{eq:disp-strain-stress-t}), and we may thus evaluate at $\rho \to \tL$ through:  
\begin{gather}
\rmu_r(\Li) = \frac{1}{Y} \left( L \psi'(\Li) - \nu \psi(\Li) \right)  \nonumber \\
= \frac{\sigma_0}{Y} R \left( \tL \Psi'(\tL) - \nu \Psi(\tL) \right) \ . 
 \label{eq:ur-core}
\end{gather}       
%\vspace{0.2cm} 

\noindent 
\underline{Matching conditions:} 
In addressing the zones {\emph{(i-iii)}}, we only used the continuity of radial stress (hence $\Psi(\rho)$), and the deflection $\zeta(\rho)$. In order to determine the three unknown variables, $ \Psi(1),%\srr(R), 
\theta$, and $\tL$, we must invoke three other matching conditions. Two of them are continuity of the slope, $\zeta'(\rho)$, and hoop stress, $\sqq(r) = \psi'(r) = \sigma_0 \Psi'(\rho)$, at the borderline between the tensile core and the wrinkled zone, %which yields the 
yielding two equations: 
\begin{equation}
\zeta'(\tL) = \ta \ \  \ \ ; \  \ \ \ \Psi'(\tL) = 0 \ . 
\label{eq:twomore}
\end{equation}   
(As was noted in the similar problem of indenting a floating sheet \cite{Vella15}, these two equations do not follow from local force balance at $r=\Li$ {\emph{per se}}, but rather from minimization of the total energy of a wrinkled state, which is realized when the hoop stress is continuous throughout the sheet). The last matching condition is the continuity of radial displacement at the hole's edge, $\rho=1$, which is obtained through Eqs.~(\ref{eq:dis-stress-Lame-1},\ref{eq:wrinkled-I},\ref{eq:ur-core}, \ref{eq:twomore}), yielding: 
%recasting the dimensionless form: 
\begin{equation}
\Psi(1) \cdot (1-\log\tL)  - \frac{1}{2}\ta^2\cdot (1-\tL)  = 2  \ .
\label{eq:thirdmatching}
\end{equation} 
\\
With the three equations~(\ref{eq:twomore},\ref{eq:thirdmatching}), and the four BCs~(\ref{eq:BC-sliding-wrinkling-I}), the FvK equations (\ref{eq:FvK-nondim-2},\ref{eq:FvK-nondim-1}), which are two coupled $2^{nd}$ order ODEs, yield a single solution for 
$\Psi(\rho), \zeta(\rho)$ in the interval $0<\rho<\tL$, as well as the three unknowns, $\tL, \ta, \Psi(1)$. The details of the analytic solution are given in App.~\ref{sub-app:wrinkling}. Together with Eqs.~(\ref{eq:Lame-stress},\ref{eq:dis-stress-Lame-1},\ref{eq:wrinkled-I},\ref{eq:new-nondim-param}), this solution fully characterizes the displacement and stress fields %in the sheet 
for any value of the dimensionless control parameter $\tdelta >\tdelta_c$.

%However, there is an important difference with the two cases studied in Sec.\ref{sec:clamping_sliding}-- 
%here, the extent of the tensile core is an unknown ($L$), rather than a where $L$ is not a control parameter, but rather must be found as part of the solution. As we will see below, following SI in \cite{VHMRD15}, it is convenient to redefine the dimensionless variables and parameters in the tensile core, such that the only unknown associated with the tensile core is $L/R$. Therefore, we will express first the matching/boundary conditions in terms of the original (dimensional) variables, and only later we introduce the dimensionless version of the equations.      
 
 % where $\arctan ( a )$ is the contact angle of the sheet at the hole's edge. 
%The above solution also yields the radial displacement: 
 %\begin{align}
 %u_r ( r ) &= - \frac{1}{2} a^2 r^2 + \frac{T}{\gamma} R \log \left( \frac{r}{C_0} \right)
 %\end{align}
% where $C_0$ is some constant.
%\begin{equation}
%\frac{R}{Y} [2 \sigma_0 - (1+\nu)\srr(R)]  = - \frac{1}{2} \arctan(\theta)^2 (R-L) + \frac{\srr(R)}{Y} R \log \left( \frac{R}{L} \right) + \rmur(L) \ . 
%\label{eq:BC-11a}
%\end{equation}

%The solution of the FvK equations (\ref{eq:FvK-nondim-2},\ref{eq:FvK-nondim-1}), with the BCs~(\ref{eq:BC-nondim-sliding-axi}) can be obtained in a similar way to the previous solution (see App.~\ref{sub-app:sliding}), allowing us to express the force, $F$, the shape, $z(r)$, and the stress components, $\sigma_{rr}(r), \sigma_{\theta\theta}(r)$, for any value of $\tdelta$. 
The brown curves in Figs.~\ref{fig:response-1}-\ref{fig:stress-edge} show the force, $F$, and the radial stress at the hole's edge, $\srr(R)$, upon increasing $\delta$ above $\tdelta_c \approx 3.3$, and the brown curved in Figs.~\ref{fig:hoop-stress-profile}-\ref{fig:shape} show the hoop and radial stresses, $\sigma_{\theta\theta}(r),\sigma_{rr}(r)$, and the  
shape, $z(r)$, at the suspended part, %and the stress components, $\sigma_{rr}(r), \sigma_{\theta\theta}(r)$, 
for $\tdelta=10$. %various value of $\tdelta$. 
One may notice that for any $\tdelta > 3.3$, wrinkling in the suspended part of the sheet reduces slightly further the force (in comparison to the unstable, axisymmetric deformation with sliding, represented by the blue curve).  
%for $\tdelta \gg 1$, the force scales as $F\sim %YR^2 \tdelta^3$, similarly to the response we found in the preceding sections, but the numerical pre-factor is smaller by approximately 45\% in comparison to the clamped, uncompressed sheet (third column in Table 1). 
Note also that the formation of wrinkles acts to slightly increase the angle $\theta$ at the hole's edge %is slightly larger 
in comparison to an unwrinkled deformation and   
%Finally, 
%that the formation of wrinkles acts 
to extend the azimuthally-confined zone (fourth and fifth columns in Table 1, respectively). 
%for $\tdelta \gg 1$ we find that the ratio $L/R \to 0.49$, whereas in the absence of wrinkling $L/R \to 0.6$. 
Intuitively, since wrinkles suppress the energetic cost of hoop strain, it is favorable to extend this zone. 
%Finally, the angle $\theta$ at the hole's edge is larger in comparison to an unwrinkled deformation (fourth column in Table 1).  

One should note the discontinuity exhibited in Fig.~\ref{fig:hoop-stress-profile} by the hoop stress at the hole's edge. Such a discontinuity does not violate any force balance, and is therefore physically allowed. More precisely, while this discontinuity emerges naturally in TFT, which describes the infinite bendability limit (of a hypothetical sheet with no bending rigidity, {\emph{i.e.}} $\epsilon \!=\! 0$),  we do expect the formation of a  ``boundary layer'' at the vicinity of the hole's edge, whose length vanishes as $\epsilon \to 0$, over which the ``jump'' in the hoop stress occurs (similarly, but not identically, to the boundary layer that regularizes a ``jump'' in the radial stress, see App.~\ref{app:krr-BC}). Nevertheless, the consequent effect on the elastic energy is negligible, {\emph{i.e.}} ``sub-dominant'' in the terminology of the far-from-threshold approach \cite{Davidovitch11}.    

Taken together, these results demonstrate the wrinkles-assisted %reflect the wrinkles-enabled 
suppression of the energetic cost of the deformation induced by indentation, and consequently a reduction of the force constant $F ( \delta ) / \delta^3$ in comparison to %is lower than 
the analogous nonlinear force constant for the unwrinkled state. These observations reinforce our qualitative discussion in Sec.\ref{subsec:Heuristic}, indicating that the formation of wrinkles implies a non-perturbative modification to the stress field, and thereby to the indentation force.  

\begin{figure*}
\includegraphics[width=1.0\textwidth]{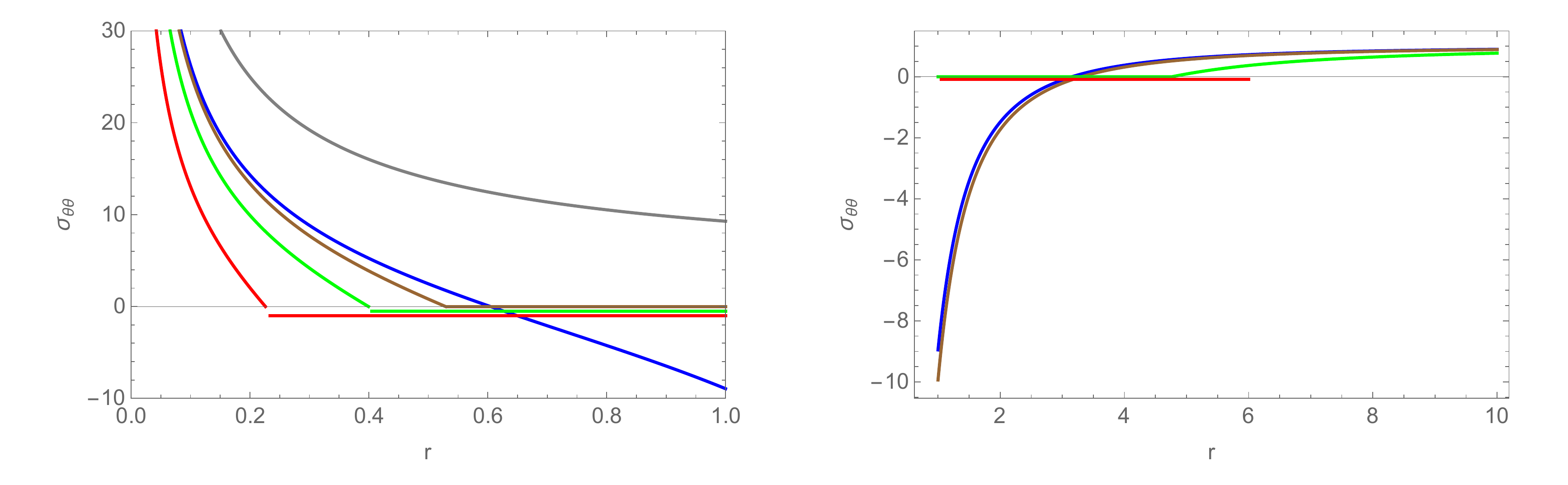}
\caption{The hoop stress %for the various types of boundary conditions 
for a dimensionless indentation depth $\tdelta = 10$ (left-suspended part, right - supported part). Distances are normalized by the hole's radius $R$ and stress is normalized by $\sigma_0$ (see text). The colors correspond to the various types of BCs, noted already in the caption of Fig.~\ref{fig:response-1}: gray (clamping at the hole's edge); blue (axisymmetric (unstable) response under sliding at the hole's edge); brown (wrinkling at the suspended part of the sheet only); green (wrinkling in both suspended and supported parts of the sheet, for a sheet with ${\cal R} \approx 90$ ($\tdeltastst ({\cal R}=90)<10$, Eq.~(\ref{eq:deltastst}));  
red (wrinkling in both suspended and supported parts of the sheet, for a sheet with ${\cal R} = 6$ ($\tdeltastst ({\cal R}=6)>10$)). Note that only the brown curve is discontinuous at the hole's edge.}  
\label{fig:hoop-stress-profile}
\end{figure*}
%%%%%%%%%%%%%%%%%%%%%%%%%%%%%%%%%
%\begin{figure}
%\includegraphics[width=0.9\textwidth]{lindentation_wrinkles.pdf}
%\caption{A sheet under uniaxial compression -- delamination versus wrinkling}  
%\label{fig:schem-uniaxial}
%\end{figure}

\begin{figure*}
\includegraphics[width=1.0\textwidth]{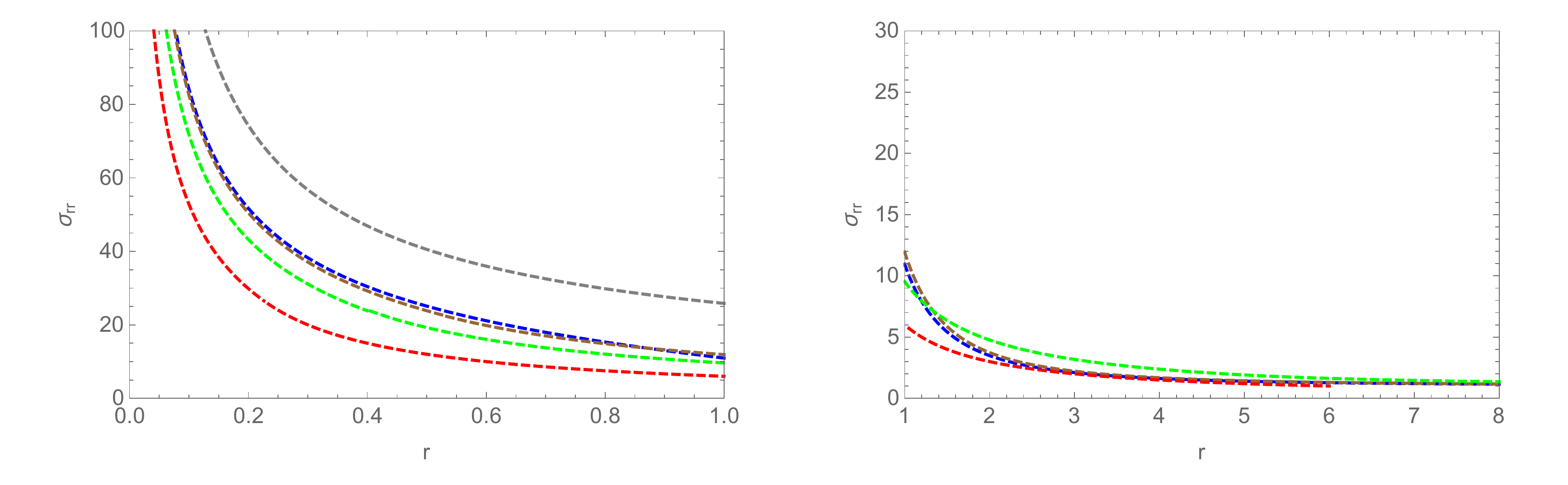}
\caption{Same as Fig.~\ref{fig:hoop-stress-profile}, but for the radial stress. Note that all curves are continuous at the hole's edge.}
\label{fig:radial-stress-profile}
\end{figure*}
%%%%%%%%%%%%%%%%%%%%%%%%%%%%%%%%%

\subsection{Buckling threshold in the supported zone \label{subsec:threshold-2}}      
In the previous subsection we let wrinkles suppress hoop compression only in the suspended part of the sheet,
%through the formation of wrinkles, 
whereas the supported part of the sheet remains unwrinkled. %and must accommodate the hoop compression induced by indentation. 
In order to identify the parameter regime at which such a scenario may be realized, we note that the supported sheet is subjected to hoop compression at the vicinity of the hole's edge that keeps increasing in magnitude and spatial extent upon increasing the indentation depth. Physically, such a state is mechanically stable if the hoop compression is below the threshold value, $\approx 2\sqrt{B\Ksub} = \beta \sigma_0$ (Eq.~\ref{eq:DG-0}), 
at which the supported sheet buckles. 
%For a sheet of bending modulus $B$ supported on a substrate with stiffness $\Ksub$, this threshold is known to be $\approx 2\sqrt{B\Ksub} = \beta \sigma_0$ (Eq.~\ref{eq:DG-0}).
%where we used the definition of the dimensionless parameter,  
This criterion is well known for uniaxial deformations \cite{Milner89,Bowden98,Pocivavsek08,Huang10} and was shown to be relevant also for more complicated, non-uniaxial confinement problems \cite{Davidovitch19,Bella17}). Considering our solution in Subsec.~\ref{subsec:II-wrinkling}, we note that the hoop compression at the edge ($\sqq(R) = \psi'(r \to R^+) $) is approximately $ 0.11 \cdot Y(\delta/R)^2 = 0.11 \sigma_0 \tdelta^2$ (where we assumed $\tdelta \gg 1$ for simplicity). Hence, we obtain that the indentation depth, $\tdeltast(\beta)$, at which the {\emph{supported}} part of the sheet becomes wrinkled is given by:   
%if the sheet-substrate attachment is strong ($\beta \gg 1$), the supported part of the sheet becomes wrinkled when the (dimensionless) indentation depth, 
%$\tdelta$, exceeds a threshold value: $\tdeltast$, where: 
\begin{equation}
%{\rm For} \ \beta \gg 1 : \ \ \ \ 
\tdeltast (\beta) \approx \left\{ \begin{array}{cc}
 \sqrt{0.11}\cdot  \beta^{1/2}    &   \beta \gg 1 \\
    \tdelta_c \approx 3.3 &  \beta \ll 1
\end{array} \right.
\label{eq:threshold-stst}
%\sqrt{\frac{\sqrt{B\Ksub}}{Y}} \ \ \ ; \ \ \  \delta^{**}  \sim R \sqrt{\sigma_0}{Y} \sqrt{\frac{\sqrt{B\Ksub}}{\sigma}
\end{equation}
%For smaller values of $\beta$, we expect that the threshold $\tdeltast (\beta)$ becomes smaller, approaching the trivial constant value  $\tdeltast (\beta) \to  \tdelta_c \approx 3.3$ as $\beta \to 0$. 
%Physically, 
Note that for $\beta <1$, the resistance to buckling in the supported part is sufficiently low, such that both supported and suspended parts of the sheets become wrinkled almost simultaneously, as soon as indentation-induced hoop compression emerges at $\tdelta %the threshold value 
\gtrsim \tdelta_c \approx 3.3$. %at which hoop compression appears in the vicinity of the hole's edge. 
\\

Equation~(\ref{eq:threshold-stst}) shows that the analysis in Subsec.~\ref{subsec:II-wrinkling} describes the parameter regime $\beta \gg 1 \ \& \  \tdelta_c < \tdelta \ll \tdeltast (\beta)$, namely, where the sheet-substrate attachment is sufficiently strong to prevent wrinkling in the supported part, for sufficiently small indentation depth. This is precisely regime {\emph{(i)}} we described in Subsec.~\ref{subsec:overview}. 

In the rest of this section, we will turn our attention to %focus on 
regime {\emph{(ii)}}, $\beta \ll 1 \ \& \  \tdelta > \tdelta_c$, at which both suspended and supported parts of the sheet become wrinkled at $\tdelta \gtrsim \tdelta_c$, and the sheet-substrate attachment %to is weak and 
does not affect the residual stress field. % and the mechanical response to indentation. 
In Sec.~\ref{sec:residual_compression} we will address regime {\emph{(iii)}}, $\beta \gg 1 \ \& \  \tdelta \gg \tdeltast(\beta)$, at which the residual compression in the wrinkled, supported part of the indented sheet must be taken into consideration.    

\subsection{Wrinkling in both suspended and supported zones \label{subsec:both-wrinkling}} 
Considering the parameter regime {\emph{(ii)}}, $\beta\ll 1$ and $\tdelta > \tdelta_c \approx 3.3$, 
%we continue to assume that the sheet is not clamped at the hole's edge and can freely slide on the substrate, and -- in distinction from the preceding subsection (where $\beta \gg \tdelta^2$) -- can form 
%radial wrinkles at negligible energetic cost, thereby effectively ``waste'' any excess length in latitudes that are pulled inwards due to the indenter. 
%(This last condition is guaranteed by our consideration of the parameter regime $\beta \ll 1$). 
%Following 
we follow our analysis in Subsec.~{\ref{subsec:II-wrinkling}}, noting  
%we consider $\tdelta > \tdelta_c \approx 3.3$, for which the indented sheet is under compression in an annular zone, $\Li<r<L_O$, around the hole's edge ({\emph{i.e.}} $\Li<R$ and $L_O>R$). 
%The fact 
%that since $\beta \ll 1$ allows us to 
that since $\beta \ll 1$, the direct effect of the sheet-substrate attachment on the stress field in the sheet is negligible, and therefore 
%hence we can continue to use 
the standard TFT approach of Subsec.~{\ref{subsec:II-wrinkling}} can be employed also here. Namely -- in the wrinkled zone, $\Li<r<L_O$, the stress field is given by a tensile radial stress, $\srr(r)>0$, and negligible hoop and shear stresses, $\sqq(r), \sigma_{r\theta}(r) \approx 0$. %(More precisely, for a given indentation depth, $\tdelta$, the radial stress $\srr(r)$ is approaching a finite limit as the parameters $\epsilon$ and $\beta$ in Eq.~(\ref{eq:DG-0}) vanish, whereas the residual compression, $\sqq(r)$ and shear stress both vanish in this asymptotic limit \cite{Hohlfeld15}). 

Similarly to Subsec.~{\ref{subsec:II-wrinkling}}, we proceed by considering the displacement and stress fields in the three parts of the sheet: {\emph{(i)}} $R<r<\Rf$ -- where the sheet is nearly planar, but (unlike Subsec.~{\ref{subsec:II-wrinkling}}) it develops radial wrinkles in $R<r<L_O$ and is axisymmetrically deformed only at $L_O<r<\Rf$, where both radial and hoop stress components are tensile; {\emph{(ii)}} $\Li<r<R$ -- where the suspended sheet is wrinkled; {\emph{(iii)}} $r<\Li$ -- where the suspended sheet is unwrinkled and both hoop and radial stresses are tensile. %Considering first 
For the last two parts, we notice that the displacement and stress are given by expressions identical to their counterparts in Subsec.~{\ref{subsec:II-wrinkling}}, namely, Eqs.~(\ref{eq:wrinkled-I}) and the BCs~(\ref{eq:BC-sliding-wrinkling-I}) for the nonlinear FvK equations (\ref{eq:FvK-nondim-2},\ref{eq:FvK-nondim-1}) in the unwrinkled core, albeit with a different triplet of constants $\Psi(1),\ta, \tL$, that must be determined by matching the radial displacement and stress at the hole's edge %$\rmur(R)$
with the wrinkled portion of the sheet at $r>R$. Thus, among the three equations that specify the constants $\Psi(1),\ta, \tL$, the two equations that reflect these continuity conditions %continuity of the displacement and radial stress at $r=\Li$ 
are identical to their counterparts in Eq.~(\ref{eq:twomore}).  

In order to find the remaining equation that relates the constants $\Psi(1),\ta, \tL$, we turn to discuss the exterior zone, 
%the nearly planar part of the sheet at 
$r>R$. Once again, we find a direct mapping %the problem is identical 
to the Lam\'e problem of an annulus under co-axial, co-planar tensile loads, $\srr(R) = \Psi(1)\cdot\sigma_0$ and $\sigma_{rr}(\Rf) = \sigma_0$. For ${\cal R} \gg 1$ and $\Psi(1) >2$ (for which the Lam\'e solution, Eq.~(\ref{eq:Lame-stress}) is unstable), the TFT solution is given by \cite{Davidovitch11}: 
%(that should be contrasted with the axisymmetric solution to the Lam\'e problem, Eq.~\ref{eq:Lame-stress}):
\begin{gather}
%\!\!\!\!\!
{R\!<\!r\!<\!L_O} \left\{
\begin{array}{c}
 \srr (r) \!=\! \srr(R)\frac{R}{r}       \\
    \sqq (r) \!=\! 0   
\end{array} \right.
\label{eq:Lame-TFT-2-a}  \\
%&:& \  \srr (r) \!=\! \srr(R)\frac{R}{r}  \label{eq:Lame-TFT-2-a}  \\
%&\mbox{}& \  \sqq (r) \!=\! 0   %\ \ \ ;  \ \ \ 
%\nonumber \\
%\end{eqnarray} 
%\begin{eqnarray} 
%\!\!\!\!\! 
{L_O\!<\!r\!<\!\Rf} 
\left\{
\begin{array}{c}
 \srr (r) \!=\! \sigma_0 + \left(\srr(L_O) - \sigma_0\right)\frac{L_O^2}{r^2}      \\
  \sqq (r) \!= \!\sigma_0 - \left(\srr(L_O) - \sigma_0\right)\frac{L_O^2}{r^2}      
\end{array}
\right.
\label{eq:Lame-TFT-2-b}
%
%\!&:&   \ \srr (r) \!=\! \sigma_0 + \left(\srr(L_O) - \sigma_0\right)\frac{L_O^2}{r^2} \label{eq:Lame-TFT-2-b} \\
%&\mbox{}& \ \sqq (r) \!= \!\sigma_0 - \left(\srr(L_O) - \sigma_0\right)\frac{L_O^2}{r^2}  \nonumber 
\end{gather}
\begin{equation} 
{\rm where:}\  L_O = \frac{\srr(R)}{2\sigma_0}R=\frac{\Psi(1)}{2}R  \ , 
%\nonumber
\label{eq:Lame-TFT-2-c}
\end{equation} 
%\!\!\! L_O<r<\Rf &:& \ \  \ \srr (r) = \sigma_0 + \left(\srr(L_O) - \sigma_0\right)\frac{L_O^2}{r^2} \  ;  \    \sqq (r) = \sigma_0 - %\left(\srr(L_O) - \sigma_0\right)\frac{L_O^2}{r^2} \ ,
%\label{eq:Lame-TFT-2}
%\end{eqnarray} 
and %relationship between 
the radial displacement at the wrinkled zone, $R<r<L_O$, is given by: %and the tensile loads, $\srr(R), \sigma_0$, 
%akin to Eqs.~(\ref{eq:dis-stress-Lame},\ref{eq:dis-stress-Lame-1}), 
\begin{equation}
\rmu_r(r) = r\frac{\srr(r)}{Y} [- \nu - \log(\frac{L_O}{r})] \ . 
\label{eq:dis-stress-Lame-2} 
\end{equation} 
Comparing Eq.~(\ref{eq:dis-stress-Lame-2}) with its counterpart, Eq.~(\ref{eq:dis-stress-Lame}) in Subsec.~\ref{subsec:II-wrinkling}, reveals a dramatic effect associated with the expansion of wrinkles on the supported part upon increasing indentation depth, $\tdelta$. While Eq.~(\ref{eq:dis-stress-Lame}) shows that $\rmur(R)$ is proportional to the radial stress at the hole's edge, $\srr(R)$, Eqs.~(\ref{eq:Lame-TFT-2-c},\ref{eq:dis-stress-Lame-2}) show that in the presence of wrinkles %the radial displacement is larger and 
the ratio $\rmur(R)/\srr(R)\propto\log(\srr(R)/\sigma_0)$.
%between the tensile loads exerted on the supported part of the sheet. 
As we will show now, this effect has a strong impact on indentation mechanics, associated with 
the continuity equation for radial displacement at the hole's edge: 
%requires one to equate Eqs.~(\ref{eq:dis-stress-Lame-2}) and (\ref{eq:ur-core}) at $r=R$, yielding: 
\begin{equation}
\Psi(1)\cdot \log\frac{\Psi(1)}{2\tL} - \frac{1}{2}\ta^2\cdot(1-\tL) = 0 \ ,  
\label{eq:thirdmatching-1}
\end{equation}   
which follows from Eqs.~(\ref{eq:dis-stress-Lame-2}) and (\ref{eq:ur-core})).  

Let us inspect Eq.~(\ref{eq:thirdmatching-1}), contrasting it with its counterpart, Eq.~(\ref{eq:thirdmatching}) in Subsec.~\ref{subsec:II-wrinkling}.   
Considering the asymptotic limit $\tdelta \to \infty$, one may easily notice that a solution of the form $\Psi(1) \sim \tdelta^2 \ , \ \ta \sim \tdelta \ ,  \ \tL \sim O(1)$ is consistent with Eq.~(\ref{eq:thirdmatching}), but not with Eq.~(\ref{eq:thirdmatching-1}). Instead, a consistent asymptotic solution of Eq.~(\ref{eq:thirdmatching-1}) has the form: $\Psi(1) \sim \tdelta^2/\log(\tdelta) \ , \ \ta \sim \tdelta \ , \ \tL \sim 1/\log(\tdelta)$. Obtaining the numerical values of the pre-factors in these asymptotic relations requires the use of Eqs.~(\ref{eq:wrinkled-I},\ref{eq:BC-sliding-wrinkling-I},\ref{eq:FvK-nondim-2},\ref{eq:FvK-nondim-1}), and the detailed calculation is described in App.~\ref{sub-app:wrinkling}. 

The results are shown in the green curves in Figs.~\ref{fig:response-1},\ref{fig:stress-edge} and Figs.~\ref{fig:hoop-stress-profile}-\ref{fig:shape}. We note that the presence of wrinkles on the substrate underlies a sub-cubic asymptotic response, namely $F/\delta^3 \sim 1/\log(\tdelta) \to 0$ as $\tdelta \to \infty$, reflecting a logarithmic suppression of the radial stress at the hole's edge with respect to the bare indentation-induced stress: $\srr(R) \!\sim\! \tfrac{1}{\log\tdelta}\!\cdot\! Y\!\cdot\!(\delta/R)^2$. The invasion of wrinkles into the supported zone of the sheet affects strongly also the displacement field, where the slope at the hole's edge now approaches asymptotically the ``natural'' cone angle: $a \to \tfrac{\delta}{R}\cdot[1- O(1/\log\tdelta)]$, and the size of the unwrinkled core vanishes, $\Li = R\cdot \tL \sim R/\log\tdelta$, as is described in the first row of Table II.    

%show the force, $F$, the shape, $z(r)$, and the stress components, $\sigma_{rr}(r), \sigma_{\theta\theta}(r)$, for any value of $\tdelta$. One may notice that -- for any $\tdelta > 3.3$, wrinkling in the suspended part of the sheet reduces further the force; for $\tdelta \gg 1$, the force scales as $F\sim YR^2 \tdelta^3$, similarly to the response we found in the preceding sections, but the numerical pre-factor is smaller by approximately 45\% in comparison to the clamped, uncompressed sheet (third column in Table 1). One may also note that the angle $\theta$ at the hole's edge is larger in comparison to an unwrinkled deformation (fourth column in Table 1).  
%Finally, the formation of wrinkles acts to extend the azimuthally-confined zone (fifth column in Table 1). 
%for $\tdelta \gg 1$ we find that the ratio $L/R \to 0.49$, whereas in the absence of wrinkling $L/R \to 0.6$. 
%Intuitively, since the formation of wrinkles suppresses the energetic cost of hoop strain, it is energetically favorable to extend this zone.     
%%%%%%%%%%%%%%%%%%%%%%%%%%%%%%%%%
\begin{figure}
\includegraphics[width=0.5\textwidth]{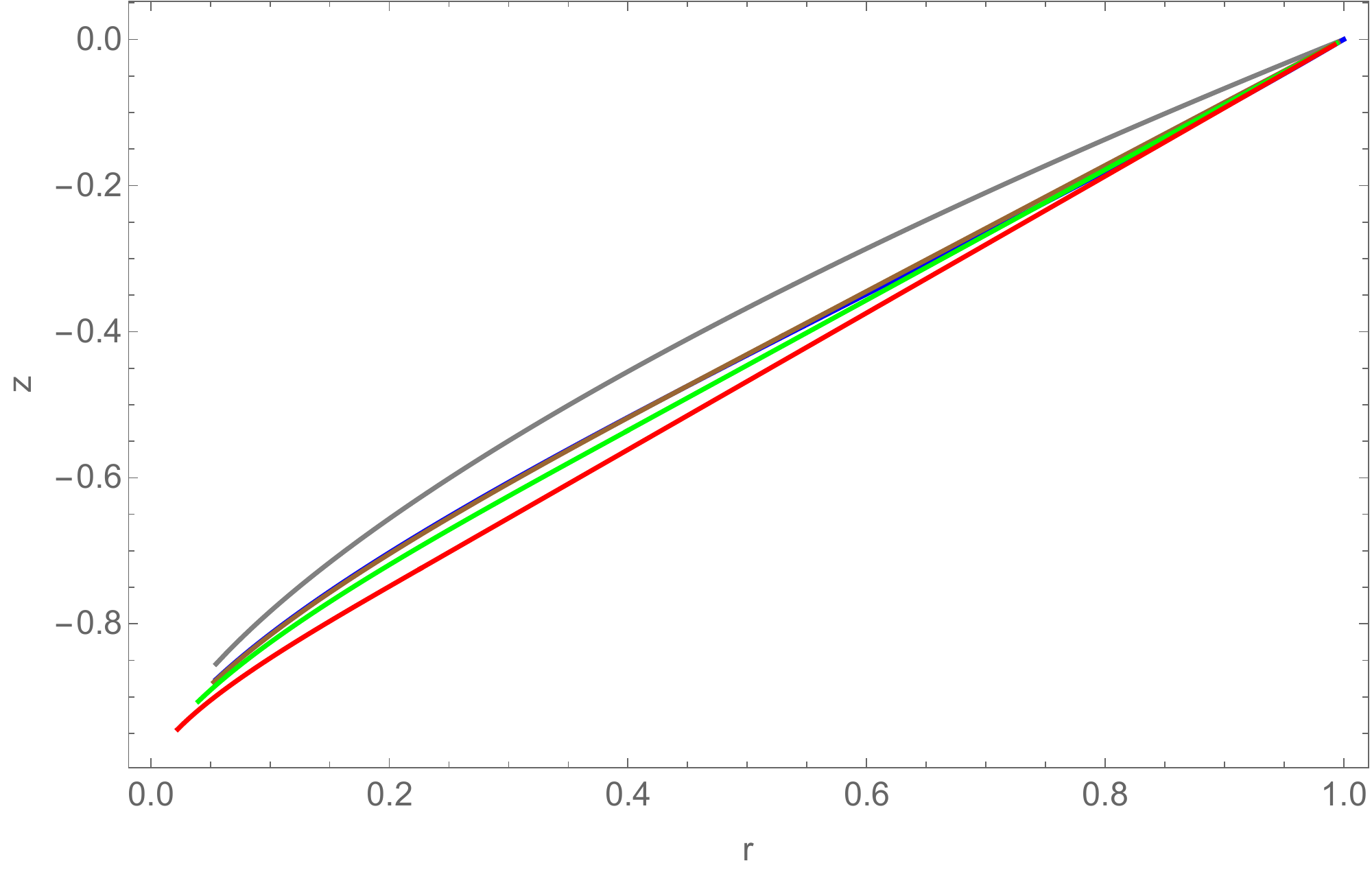}
\caption{A side view of the deformation in the suspended part for $\tdelta =10$. Radial distances are normalized by the hole's radius $R$, and vertical distances are normalized by the indentation depth $\delta$. The colors represent the same types of BCs as in previous figures.} 
%ifferent boundary conditions and physical parameters are represented by the same colors as in Fig.~\ref{fig:response-1}. The thick portions of the (brown, green, red) curves indicate zones where radial wrinkles collapse hoop compression.}  
\label{fig:shape}
\end{figure}
%%%%%%%%%%%%%%%%%%%%%%%%%%%%%%%%%

\subsection{The geometric limit: pseudo-linear response \label{subsec:geo-limit}}}
In the preceding section we saw that if the sheet-substrate attachment is sufficiently weak ($\beta \ll 1$),  radial wrinkles expand in the supported part of the sheet, occupying an annular zone whose external radius, $L_O \sim R \cdot \tdelta^2 /\log\tdelta$. If $\tdelta$ is sufficiently large, wrinkles approach the edge of the sheet, causing yet another dramatic change in the distribution of stress in the sheet and its response to the indentation force. (A similar phenomenon has been found for the indentation of a floating sheet \cite{Vella18b,Ripp20}). For a given value of the parameter ${\cal R}$, our numerical results in the preceding section allow us to estimate the value $\tdeltastst$ at which wrinkles reach the far edge: 
\begin{gather}
{\cal R} \approx 0.12 (\tdeltastst)^2 \log\tdeltastst  \ \  \Rightarrow \nonumber \\
\tdeltastst({\cal R}) \approx 2.07 \sqrt{{\cal R}\log{\cal R}} \cdot [1 + 
O(\log{\cal R})]  \ . 
\label{eq:delta-sttd}
\end{gather} 
For $\tdelta > \tdeltastst$, the supported part of the sheet is fully wrinkled, and the stress field for any $R<r<\Rf$ is described by %the first row of 
Eq.~(\ref{eq:Lame-TFT-2-a}). Together with the BC $\srr(\Rf)=\sigma_0$, we find that: % yields: 
\begin{equation}
\srr(r) = \sigma_0 \frac{\Rf}{r} \ \ ; \ \ \sqq(r) = 0 \ , 
\label{eq:TFT-22}
\end{equation}  
as we described already in the introductory section~\ref{subsec:Heuristic}. In this regime, the value of the unknown $\Psi(1)$ is directly given by Eq.~(\ref{eq:TFT-22}): 
\begin{equation}
\Psi(1) = {\cal R} \ , 
\label{eq:thirdmatching-1a}  
\end{equation}  
and the deformed state is fully described by solving the FvK equations~(\ref{eq:FvK-nondim-2},\ref{eq:FvK-nondim-1}) with the BCs~(\ref{eq:BC-sliding-wrinkling-I}), along with replacing Eq.~(\ref{eq:thirdmatching-1}) by (\ref{eq:thirdmatching-1a}), and the two additional equations in (\ref{eq:twomore}). The solution of these equations is described in App.~\ref{sub-app:wrinkling}. We note that  this solution merely determines the numerical pre-factors in the scaling laws we already found in Subsec.~\ref{subsec:Heuristic}, specifically the pseudo-linear response, $F(\delta) \propto \delta$, Eq.~(\ref{eq:pseudo-linear-1}, with $\geff = \sigma_0$).    

The results of this calculation, for a dimensionless indentation depth $\tdelta = 10$ and ${\cal R} = 6$ (such that $\tdelta > \tdeltastst({\cal R})$, are shown through the red curves in Figs.~\ref{fig:response-1},\ref{fig:stress-edge} and Figs.~\ref{fig:hoop-stress-profile}-\ref{fig:shape}. 
%Figs.~\ref{fig:response-1}-\ref{fig:stress-profile}.    
%The above results (for $\frac{R_{out}}{R} = 10^3$) prove the prediction of our qualitative discussion in Sec.\ref{sec:FT_geometric}. 
As we noted already in Subsec.~\ref{subsec:Heuristic}, the pseudo-linear response reflects an asymptotically-isometric mechanics, whereby the indentation force %effectively 
``decouples'' from the stretching modulus of the sheet, transmitting work to the puller at the far edge of the sheet, $r=\Rf$. 
%being proportional (with a proportionality constant of $2 \pi$) to $\gamma \frac{R_{out}}{R}$. 
Echoing an observation made already for indenting floating polymer sheets \cite{Vella18b,Ripp20}, our results show that after wrinkles reach the far edge the shrinkage of the tensile core zone with $\tdelta$ becomes much more pronounced ($\tL \sim \tdelta^{-2}$ {\emph{vs.}} 
%``speeded up'' from a logarithmical dependence 
$\tL \sim 1/\log\tdelta$ for $\tdelta <\tdeltastst({\cal R})$). % to $\tL \sim \tdelta^{-2}$). 
In the asymptotically isometric regime, $\tdelta \gg \tdeltastst({\cal R})$ the suspended portion %art of the sheet 
approaches the shape of a perfect cone, with      
%shrinks to zero (as $\tilde{\delta}^{-2}$), and that the shape of the film throughout most of the sheet in the hole (including the contact angle) is the "trivial" cone, with 
a slope $\frac{\delta}{R}$, superimposed with radial wrinkles. 
\\
%Results for the size of the tensile region inside the suspended area, and the contact angle at the edge of the contact area are shown in Fig.[\ref{fig:wrinkles_substrate_5}].

%%%%%%%%%%%%%%%%%%%%%%%%%%%%%%%%%%%%%%%%
\section{%Partial wrinkling: 
The role of sheet-substrate attachment \label{sec:residual_compression}} 
%%%%%%%%%%%%%%%%%%%%%%%%%%%%%%%%%%%%%%%%%
In the previous section we avoided the need to address {\emph{explicitly}} the effect of sheet-substrate attachment 
%on the response 
by considering the two opposite limits %identified through the parameter $\beta$ (\ref{eq:DG-0}). These limits
of strong and weak attachment, namely, the parameter regimes {\emph{(i)}} ($\beta \!\gg\! \tdelta^2\!\gg\! 1$) and {\emph{(ii)}} ($\beta \!\ll\! 1$), respectively.   
%which were presented in the introduction (Subsec.~\ref{subsec:overview}) as regime {\emph{(i)}} ($\beta \gg \tdelta^2\gg 1$) and regime {\emph{(ii)}} ($\beta \ll 1$), respectively. 
Notwithstanding the striking difference between these regimes, %markedly from each other 
%in mechanical response to indentation 
(compare brown {\emph{vs.}} green and red curves in Figs.~\ref{fig:response-1}-\ref{fig:stress-edge} and Figs.~\ref{fig:hoop-stress-profile}-\ref{fig:shape}), 
%\ref{fig:stress-profile}), 
in each of them the mechanical response is not affected by the actual values of the bending modulus $B$ and stiffness $\Ksub$, but only by the tensile load $\sigma_0$ exerted at the far edge, the stretching modulus $Y$, and the indentation depth $\delta$ (as well as ${\cal R}$).  
%geometric features, namely, the radii $R,\Rf$ of the hole and the sheet, respectively). 
In contrast, in the intermediate regime {\emph{(iii)}}, $1 \!\ll \!\beta\! \ll \!\tdelta^2$, 
%where the sheet-substrate attachment is ``moderate'', and 
the stress and indentation force depend explicitly on $B$ and $\Ksub$.  

In order to elucidate %better understand 
this distinction
%we describe in the first part of this section the nature of the wrinkle pattern and the energetic hierarchy that undelies it. %, which underlies the mechanics. 
%In the subsequent part we show how understanding the energetic hierarchy of wrinkle patterns allows us to describe the response in regime {\emph{(iii)}} by identifying a bending-induced ``effective tension``, $\geff = 2\sqrt{\Ksub B}$, that is induced by bending and substrate deformation.     
%
%\subsection{Wrinkling morphology and energetic hierarchy}
%%%%%%%%%%%%%%%%%%%%%%%%%%%%
%In order to understand the absence of the bending modulus $B$ and substrate stiffness $\Ksub$ from the formulas obtained in Sec.~\ref{sec:clamping_sliding},
%we its negligibility in comparison to $U_{\rm dom}$, %the high bendability limit, 
%one may envision 
let us consider a narrow annulus %confined latitude of 
of radius $r$ as an elastic ring of bending modulus $B$ that is forced to contract due to radial displacement $\rmur(r)<0$ (which is given for each parameter regime by the corresponding expressions in Subsecs.~\ref{subsec:II-wrinkling},\ref{subsec:both-wrinkling},\ref{subsec:geo-limit}). 
%in the azimuthally-confined zone ({\emph{i.e.}} where $\rmur(r) <0$) 
%
%whose radius is forced to contract inward by  displacement $\rmur(r)<0$ (which is given for each parameter regime by the corresponding expressions in Subsecs.~\ref{subsec:II-wrinkling},\ref{subsec:both-wrinkling},\ref{subsec:geo-limit}). 
If the contracted ring is forced to retain a circular shape, it must acquire %implies 
a ``bare'' hoop strain, $\varepsilon_{\theta\theta}= \tfrac{\rmur(r)}{r}\!<\!0$, and thereby a compressive stress, 
\begin{gather}
\sqq^{(bare)}(r) \approx Y \frac{\rmur(r)}{r} \sim -Y(\delta/R)^2 \ , 
\label{eq:bare-stress} 
\end{gather}
and correspondingly an energetic penalty $\sim Y  (\rmur(r)/r)^2$. If out-of-plane deflections are allowed,  
%the ring can buckle out-of-plane, 
the ring may respond as an {\emph{elastica}} -- developing wrinkles of wavelength $\lambda$ and amplitude $A$, such that $(\pi A/\lambda)^2 \approx -\tfrac{\rmur(r)}{r}$.
 %$A \approx  \sqrt{\rmur(r)/\pi^2 r} \cdot \lambda $, 
 %thereby keeping its 
Such a deformation retains the %hoop 
arclength nearly intact, suppressing the hoop stress to a residual value \cite{Paulsen16,Davidovitch19}.  
\begin{gather}
\sqq^{(res)}(r) \approx - 2B/\lambda^2 \ , 
\label{eq:res-stress} 
\end{gather}
whose magnitude will be shown to be much smaller than $\sqq^{(bare)}(r)$.  
The %wrinkle 
wavelength $\lambda$ and consequently the residual hoop stress, is determined by a ``local $\lambda$ law'' \cite{Cerda03,Paulsen16}: 
\begin{gather}
\lambda \approx 2\pi (B/\Keff)^{1/4}   \ , 
\label{eq:local-lam-law-1} 
\end{gather}
where %$B$ is the bending moduls, and 
$\Keff$ is an 
%balance between the energy/area of bending,  
%hoop strain. The residual hoop compression in the wrinkled state of the ring is $\approx 2B$. bending, 
%$B\cdot (A/\lambda^2)^2$, and 
``effective stiffness'', which may be associated with the resistance of the supporting substrate ($\Keff \sim \Ksub$) or with the presence of radial tension that resists a large wrinkle amplitude ($\Keff \sim  \srr(r)/r^2$).      
% associated with the supporting substrate,  
%that affects wrinkling in the supported part of the sheet, 
%$\Ksub\cdot A^2$, or a tension-induced resistance, $\sim \srr(R)(A/R)^2 \sim YA^2/\delta^2$ that affects wrinkling in the suspended part \cite{Cerda03,Paulsen16}. (In the above, we assumed $A \ll \lambda \ll r$ \cite{Davidovitch11}). This balance yields the ``local $\lambda$ law'' \cite{Paulsen16}: %Balancing these energetic costs one obtains:   
%substrate attachment, $\Ksub\cdot A^2$. (In the above, we assumed $A \ll \lambda \ll r$ \cite{Davidovitch11}). Balancing these energetic costs one obtains that: 
Implementing this rule we find different values of $\lambda$  
%a distinction between the wrinkle wavelength 
(and consequently the residual stress and energy) in the suspended and supported parts of the sheet
\begin{eqnarray}
r\!>\!R\! &:&  \ \ \  \lambda \sim \left({B}/{\Ksub}\right)^{1/4}   \nonumber \\
r<R\! &:&  \ \ \ \lambda \sim (\frac{BR^4}{Y \delta^2})^{1/4}  \sim R\sqrt{t/\delta} \  
\label{eq:lambda-1}
\end{eqnarray}           
%which is an example of a ``local $\lambda$ law'' \cite{Cerda03,Paulsen16}.
\footnote{The second line of Eq.~(\ref{eq:lambda-1}) is valid for $\tdelta <\tdeltastst(\beta)$, such that $\srr(R) \sim Y(\delta/R)^2$. For $\tdelta > \tdeltastst(\beta)$, $\lambda$ at the suspended part is estimated by substituting in Eq.~(\ref{eq:local-lam-law-1}) $\Keff \sim \sigma_0 {\cal R}/R^2$. 
}.
For the suspended part, $r\!\!<\!\!R$, Eqs.~(\ref{eq:bare-stress},\ref{eq:res-stress},\ref{eq:lambda-1}) show that the residual, wrinkle-induced hoop compression $\sqq^{(res)}                                                                                      (r)$ %in the suspended part, $r<R$, 
is much smaller than its counterpart $\sqq^{(bare)}(r) \sim Y(\delta/R)^2$, and therefore the formation of wrinkles is energetically favorable in $r<R$, regardless of the value of $\beta$. Turning now to the supported part, and addressing first the parameter regimes {\emph{(i)}} ($\beta \!\gg\! \tdelta^2 \! \gg\! 1$), and {\emph{(ii)}} ($\tdelta \!>\! \tdelta_c  \ \& \ \beta\!\ll\! 1$), %we find that 
an analogous comparison of $\sqq^{(res)}(r)$ and $\sqq^{(bare)}(r) \sim Y(\delta/R)^2$ yields precisely the same conclusion we reached already in Subsec.~\ref{subsec:threshold-2}, namely, the stress in the supported part is given by the axisymmetric Lam\'e solution in the former regime and by the TFT solution in the latter. However,           
when inspecting regime {\emph{(iii)}}, $\tdelta^2 \!\gg\! \beta \!\gg \!1$, we find that the supported portion, $r\!\!>\!\!R$, consists of a zone close to the hole's edge, where $|\sqq^{(res)}(r)|\!\!\ll\!\! |\sqq^{(bare)}(r)|$, and another zone, away from the hole's edge, where $|\sqq^{(res)}(r)|\!\!\gg\!\! |\sqq^{(bare)}(r)|$. This observation reflects the complexity of the mechanical response in this parameter regime, where %one must take into account {\emph{explicitly}} 
the value of the residual hoop compression, $\sqq^{(res)}(r)$, must be taken {\emph{explicitly}} into account through Eqs.~(\ref{eq:res-stress}, \ref{eq:lambda-1}), in order to reliably evaluate the stress field %in the wrinkled sheet 
and thereby the indentation force.    \\         
%
%A close inspection of the stress profiles obtained in Subsecs.~\ref{subsec:both-wrinkling},\ref{subsec:geo-limit}, allows us to understand the difference between the parameter regimes {\emph{(ii)}} $\beta \ll 1$ and {\emph{(iii)}} $1 < \beta \ll \tdelta^2$. In the former, the isotropic (tensile) stress in the pre-indented state, $\sigma_0$, is already much larger in magnitude than the residual hoop stress $|\sqq^{(res)}| \approx |2B/\lambda^2|$, hence the wrinkle-induced contribution to stress can be safely neglected when computing the tension field stress in response to indentation. In contrast, in the latter case, the fact that $\beta \ll \tdelta^2$ implies that $\sqq^{(res)}$ can be ignored only in the vicinity of the hole's edge, where the radial stress (and correspondingly a hoop stress in an unwrinkled state) is governed by indentation, $\sim Y (\delta/R)^2$; however, sufficiently far from the hole's edge, the radial component of the tension field stress is $\sim \sigma_0$, and since $\beta >1$ it is smaller than the residual hoop compression. Hence, in the parameter regime  {\emph{(iii)}} $1 < \beta \ll \tdelta^2$, the stress in the indented sheet is affected directly by the residual scale $2\sqrt{B\Ksub}$ and so is the mechanical response, $F(\delta)$. 

One may find the stress and indentation force in regime {\emph{(iii)}} 
%of intermediate sheet-substrate attachment 
by applying a generalized version of tension field theory \cite{Davidovitch19}. Rather than neglecting the contribution of the residual hoop compression to the radial stress altogether, %one may include 
%the residual value of the compressive hoop stress in the supported wrinkled zone 
Eq.~(\ref{eq:res-stress})  
%$\sqq^{(res)} \approx -2\sqrt{B\Ksub}$, 
is taken as 
a non-homogenous source 
in the radial force balance equation~(\ref{eq:FvK-dim-2}), yielding for $r>R$:  
%yielding: 
%
%it is possible to show that incorporating Eq.~(\ref{eq:res-stress}) into the radial force balance yields in the wrinkled zone ($\Li<r<L_O$): 
\begin{equation}
\psi(r)  = \psi_0 - 2\sqrt{\Ksub B}{r} \ \ \ \ \Rightarrow  \ \ \ \Psi(\rho) = \Psi_0 - \beta \rho \ , 
\label{eq:Airy-new}
\end{equation}  
where $\Psi_0$ is a constant determined through matching conditions with the unwrinkled zones at $r<\Li$ and $r>L_O$, similarly to the analysis in Sec.~\ref{sec:clamping_sliding}. The Airy potential (\ref{eq:Airy-new}) describes a {\emph{bending-induced}} radial tension \cite{hure12,Davidovitch19,Tobasco20}, 
which can be conveniently expressed as: 
\begin{equation}
\srr(r) %= \frac{\psi}{r} 
=  (\srr(L_O) + 2\sqrt{B\Ksub})\frac{L_O}{r} - 2\sqrt{\Ksub B} \ . 
%\psi(r)  = \psi_0 - 2\sqrt{\Ksub B}{r} \ \ \ \ \Rightarrow  \ \ \ \Psi(\rho) = \Psi_0 - \beta \rho \ , 
\label{eq:srr-new}
\end{equation}  
%The above expression (with unknown parameters $L_O$ and $\srr(L_O)$, is a general solution to the radial force balance equation, $\partial_r (r\srr) - \sqq = 0$, where the residual hoop stress in the wrinkled zone, $\sqq$, is not neglected, but rather assumes the constant value~(\ref{eq:res-stress}), thus acting as a non-homogenous source for a {\emph{bending-induced}} radial tensile stress \cite{hure12,Davidovitch19,Tobasco20}. 
%As a result, the radial stress is the wrinkled zone may be approximated as, $\srr \approx (\sigma_0 + 2\sqrt{B\Ksub})L/r$ (where $L$ is the extent of the wrinkled zone and we omitted an $r$-independent constant). This expression reveals the two sources that underlie the     

Recalling that we focus here on regime {\emph{(iii)}}, $1\ll \beta \ll \tdelta^2$, and expecting that $\srr(L_O) \!\sim\! \sigma_0$ (since the region $r\!>\!L_O$ is under nearly isotropic tension),  Eq.~(\ref{eq:srr-new}) can be simplified
in $R<r<L_O$ to $\srr(r) \approx 2\sqrt{B\Ksub} L_O/r$. 
%
%by the expression 
%
%the above expression reveals the crucial role of the parameter $\beta = 2\sqrt{B\Ksub}/\sigma_0$ in regime {\emph{(iii)}})
%, $1\ll \beta \ll \tdelta^2$, 
%upon simplifying   
%
%. If $\beta \ll 1$ (parameter regime {\emph{(ii)}}), the bending-induced contributions to $\srr(r)$ in the wrinkled zone are negligible (and so is the contribution to the indentation force). In contrast, if $1\ll \beta \ll \tdelta^2$ (parameter regime {\emph{(iii)}}), 
%Eq.~(\ref{eq:srr-new}) %may be simplified in the region $\Ri<r<L_O$ to $\srr(r) \approx 2\sqrt{B\Ksub} L_O/r$ in $\Ri<r<L_O$. 
Contrasting this simplified expression with Eq.~(\ref{eq:Lame-TFT-2-a}) or Eq.~(\ref{eq:TFT-22}), we notice that 
%the characterization of 
the stress field and thereby the indentation force in regime {\emph{(iii)}} may be determined in an analogous manner to the analysis of regime {\emph{(ii)}} in Subsecs.~\ref{subsec:both-wrinkling},\ref{subsec:geo-limit}, upon replacing in the definition of the dimensionless variable $\Psi$, Eq.~(\ref{eq:dimensionless-var}): % (and  $\zeta$): 
\begin{equation}
\sigma_0 \to \beta\sigma_0 = 2\sqrt{B\Ksub} \ . 
\label{eq:replace-1}
\end{equation} 
Hence, at this level of approximation, %which we expect to be correct 
expected to be valid up to corrections of $O(\beta^{-1}) \ll 1$, the mechanics in regime {\emph{(iii)}}  $\tdelta^2 \!\gg \!\beta\!\gg\! 1$, is described by the mechanics of regime {\emph{(ii)}}  ($\beta \!\ll\! 1 \  \& \ \tdelta \!\gg\! 1$, with the replacement~(\ref{eq:replace-1}). This observation underlies the last row of Table II.

\section{Hindered sliding: clamping the sheet's edge \label{sec:Expansion}} 
In the previous sections, we assumed that the far-edge, $r=\Rf$, is subjected to a fixed tensile load, $\sigma_0$. Here
%In Sec.\ref{sec:FT_geometric} 
we consider another basic boundary condition, which may be of interest to an experimenter, whereby the far edge is clamped. Mathematically, this amounts to replacing the BC $\srr(\Rf) = \sigma_0$, with: 
\begin{equation}
\rmur(\Rf) =(1-\nu)\cdot \frac{\sigma_0}{Y}\cdot \Rf \ , 
\label{eq:new-BC-Rf}
\end{equation} 
where $\sigma_0$ is now understood as an isotropic pre-tension in the sheet {\emph{prior}} to clamping its far edge, $r=\Rf$ (and prior to indenting its center) \cite{Vella18}. Clamping the sheet at its far edge hinders its sliding inwards, %($\rmur(r) <0$), 
which is necessary to release the radial strain induced by indentation. Thus, for a given $\tdelta \gg 1$, the in-plane stress in this version of the problem 
%in the far-edge clamped sheet 
is larger in comparison to a sheet under fixed tensile load, $\srr(\Rf) = \sigma_0$, and so is the indentation force. 
This effect is elucidated by contrasting the corresponding versions of the  
%contrasting the corresponding versions of the 
Lam\'e problem. % (where we continue to assume ${\cal R} \gg 1$ for simplicity). % of an annulus, $R<r<\Rf$, under a tensile load $\srr(R)$ at its inner edge. 
In the first version, which was the basis for our analysis in the preceding sections, the far edge is under a given radial tension, $\srr(\Rf) = \sigma_0$, but otherwise is free to slide on the substrate ($\rmur(\Rf)< 0$), %For this version of the problem, 
the stress field of the planar, unwrinkled state, is given by Eq.~(\ref{eq:Lame-stress}), and the tension field solution of the wrinkled state is given by Eqs.~(\ref{eq:Lame-TFT-2-a}-\ref{eq:Lame-TFT-2-c}). In the second version of the Lam\'e problem, the BC at the far edge is given by Eq.~(\ref{eq:new-BC-Rf}), the stress field of the planar (unwrinkled) state is: % given by:  
%\lipsum[1]
\begin{widetext}
\begin{eqnarray}
%\resizebox{0.5\textwidth}{!}{$
\srr(r) &=& \frac{{\cal R}^2}{{\cal R}^2 (1-\nu) 
 + (1+\nu)} \cdot \left\{
[\sigma_0 (1-\nu) + \srr(R)(1+\nu)] %\right. \nonumber  \\\left. 
+ [(1-\nu)[\srr(R) - \sigma_0](\frac{R}{r})^2 \right\} 
\label{eq:Lame-srr-clamped} \\ 
&\approx& \frac{1}{1-\nu} 
%\frac{{\cal R}^2}{{\cal R}^2 (1-\nu)  + (1+\nu)} 
 \cdot \left\{
[\sigma_0 (1-\nu) + \srr(R)(1+\nu)] %\right. \nonumber  \\\left. 
+ [(1-\nu)[\srr(R) - \sigma_0](\frac{R}{r})^2 \right\} 
\nonumber \\
%\end{equation}
%\lipsum[2-4]
%\begin{equation}
%\resizebox{0.5\textwidth}{!}{$
\sqq (r) &=& \frac{{\cal R}^2}{{\cal R}^2 (1-\nu) \!+\! (1+\nu)} \cdot \left\{
[\sigma_0 (1-\nu) \!+\! \srr(R)(1+\nu)] %\right. \nonumber  \\\left. 
\!-\! [(1-\nu)[\srr(R) \!-\! \sigma_0](\frac{R}{r})^2 \right\}  \ . 
\label{eq:Lame-sqq-clamped} \\
&\approx& \frac{1}{1-\nu}
%\frac{{\cal R}^2}{{\cal R}^2 (1-\nu) \!+\! (1+\nu)} 
\cdot \left\{
[\sigma_0 (1-\nu) \!+\! \srr(R)(1+\nu)] %\right. \nonumber  \\\left. 
\!-\! [(1-\nu)[\srr(R) \!-\! \sigma_0](\frac{R}{r})^2 \right\}  \ ,  \nonumber
\end{eqnarray}
\end{widetext}
where the second lines in the above equations are valid for ${\cal R} \gg 1$. 
As a result, the TFT solution is characterized by a compression-free stress in the wrinkled zone: %is given by:
\begin{equation}
%&\mbox{}& {\rm wrinkled \ zone} \ 
R<r<L_O : \left\{  
\begin{array}{c}
   \srr (r) = \srr(R)\frac{R}{r}    \\
    \sqq (r) = 0   
\end{array} 
%\srr (r) = \srr(R)\frac{R}{r}  \  \ \ ; \ \ \   \sqq (r) = 0  
\right.
%\ \ \ ;  \ \ \ 
\label{eq:Lame-TFT-2-clamped}
\end{equation}
with
\begin{equation} 
\resizebox{0.4\textwidth}{!}{$
L_O \!=\!R \left( 
\sqrt{\frac{1-\nu}{1+\nu}} \!\cdot\! \sqrt{1\!+\! (\frac{{\cal R}}{\Psi(1)})^2\frac{1-\nu}{1+\nu}}\!\cdot\! {\cal R}  -  
\frac{{\cal R}^2}{\Psi(1)}\frac{1-\nu}{1+\nu} \right)  $} 
%\frac{\srr(R)}{2\sigma_0}R=\frac{\Psi(1)}{2}R 
\label{eq:Lame-TFT-2-clamped-Lo} 
\end{equation}
\vspace{-1.0cm}
$$
\approx \frac{1}{2}R \Psi(1)    \ \ \ ({\rm for} \ \ {\cal R} \gg 1)  \ , $$
where $\Psi(1) = \srr(R)/\sigma_0$, and the stress components in the unwrinkled zone, $L_O<r<\Rf$, are given by 
Eqs.~(\ref{eq:Lame-srr-clamped},\ref{eq:Lame-sqq-clamped}) upon replacing: 
$R \!\to\! L_O, \srr(R) \!\to\! \srr(L_O) \!= \!\srr(R)\!\cdot\! R/L_O$, and ${\cal R} \!\to\! \Rf/L_O \!=\! {\cal R}\!\cdot\! R/L_O$. 
%The radial displacement is given by: \begin{equation}\rmur(r) = -r \frac{\srr(R)}{Y} \cdot \left(\nu \ + \ \log\frac{L_O}{r} \right) \end{equation}
%
%
%\nonumber \\{\rm unwrinkled \ zone}: \ \srr (r) &=& \frac{{\cal R}^2}{{\cal R}^2 (1-\nu) + (1+\nu)} \cdot \left([\sigma_0 (1-\nu) + \srr(R)(1+\nu)] + [(1-\nu)[\srr(R) - \sigma_0](\frac{R}{r})^2 \right)
%\sigma_0 + \left(\srr(R) - \sigma_0\right)\frac{R^2}{r^2}  
%\nonumber \\\sqq (r) &=& \frac{{\cal R}^2}{{\cal R}^2 (1-\nu) + (1+\nu)} \cdot \left([\sigma_0 (1-\nu) + \srr(R)(1+\nu)] - [(1-\nu)[\srr(R) - \sigma_0](\frac{R}{r})^2 \right)
%
%&:& \ \  \ \srr (r) = \sigma_0 + \left(\srr(L_O) - \sigma_0\right)\frac{L_O^2}{r^2} \  \  ; \ \    \sqq (r) = \sigma_0 - \left(\srr(L_O) - \sigma_0\right)\frac{L_O^2}{r^2} \ ,
%\label{eq:Lame-TFT-2-clamped}
%\end{eqnarray} 
%(see App.?? for derivation of above equations).

The primary effect of the BC~(\ref{eq:new-BC-Rf}) is elucidated by considering a fixed ${\cal R}\!\gg\! 1$, and using the above expressions to evaluate $\srr(\Rf)$ for $\Psi(1) \!= \!\srr(R)/\sigma_0 \!\to\! \infty$.
For both planar state and wrinkled state, we find that the far-edge stress $\srr(\Rf)$ it proportional to the stress at the hole's edge $\srr(R)$. More specifically, we find that for the planar state $\srr(\Rf)/\srr(R) \propto {\cal R}^{-2}$, 
%\approx \tfrac{2}{1-\nu} {\cal R}^{-2} \srr(R)$, 
whereas for the wrinkled state $\srr(\Rf)/\srr(R) \propto {\cal R}^{-1} $. 
%\approx \tfrac{1}{2}(\sqrt{\tfrac{1+\nu}{1-\nu}} +\sqrt{\tfrac{1-\nu}{1+\nu}}) {\cal R}^{-1} \srr(R)$.
% (where we simplified the expressions by taking ${\cal R} \gg 1$).
This means that in order to keep the far edge from sliding inwards under the influence of the large radial stress $\srr(R)$ that pulls at the inner edge, the clamp must exert a comparable radial load on the far edge, hence $\srr(\Rf) \sim \srr(R) \gg \sigma_0$. 
This observation is rather intuitive, indicating 
%The fact that the radial stress throughout the sheet scales with $\srr(R)$ (either in the unwrinkled planar state or in the wrinkled state) indicates 
that the elastic energy needed to deform a sheet clamped at its far edge 
is much larger than the energy required to deform a sheet whose far edge is free to slide. As a consequence, the indentation force $F(\delta)$
%(at a given $\tdelta$) to be 
is larger in comparison to the response we found in the preceding sections for a sheet subjected to a fixed boundary load.  

We find the indentation force $F(\delta)$ by following the tracks of our analysis in Subsecs.~\ref{subsec:II-wrinkling},\ref{subsec:both-wrinkling}, 
assuming the sheet is wrinkled in an azimuthally-confined zone, $\Li<r<L_O$ (with $\Li <R$ and $L_O >R$) and unwrinkled in $r<\Li$ and $L_O<r<\Rf$. 
Considering the first two zones, we notice that the displacement and stress are given by expressions identical to Eqs.~(\ref{eq:wrinkled-I}) and the BCs~(\ref{eq:BC-sliding-wrinkling-I}) for the nonlinear FvK equations (\ref{eq:FvK-nondim-2},\ref{eq:FvK-nondim-1}) in the unwrinkled core, albeit with a different triplet of constants $\Psi(1),\ta, \tL$, that must be determined by matching the radial displacement at the hole's edge $\rmur(R)$ with the wrinkled state at the exterior of the hole. Thus, exactly as we found %in our analysis 
in Subsec.~\ref{subsec:both-wrinkling}, two equations among the three %equations 
that specify the constants $\Psi(1),\ta, \tL$, %two equations 
are identical to their counterparts in Eq.~(\ref{eq:twomore}), and the third equation reflects a continuity of the radial displacement at $r=R$. Employing Eqs.~(\ref{eq:Lame-srr-clamped}-\ref{eq:Lame-TFT-2-clamped-Lo}) and Eq.~(\ref{eq:ur-core}) we obtain an equation that replaces Eq.~(\ref{eq:thirdmatching-1}):
\begin{widetext}
\begin{equation}
\Psi(1)\cdot \log\left(-\frac{{\cal R}^2}{\Psi(1)\tL}\frac{1-\nu}{1+\nu} \ + \frac{{\cal R}}{\tL}\sqrt{\frac{1-\nu}{1+\nu}}\sqrt{1+\frac{1-\nu}{1+\nu}(\frac{{\cal R}}{\Psi(1)})^2} \right) 
%\frac{\Psi(1)}{2\tL} 
%
- \frac{1}{2}\ta^2\cdot(1-\tL) = 0 \ , 
\label{eq:thirdmatching-2}
\end{equation}
\vspace{-0.5 cm}
%which can be approximated at ${\cal R} \gg 1$ by: 
$$ \Longrightarrow \Psi(1)\cdot \log\left( \frac{{\cal R} \Psi(1)}{2 \tL} \right) - \frac{1}{2}\ta^2\cdot(1-\tL) \approx 0  \ \ \ 
({\rm for} \ {\cal R} \gg 1)
\ .  $$
\end{widetext}

The values of the unknowns $\Psi(1), \tL, \ta$, for any given $\tdelta > 3.3$ are obtained from the numerical solution of the three algebraic equations that are derived from Eqs.~(\ref{eq:twomore},\ref{eq:thirdmatching-2}) and the exact solutions of the FvK equations (\ref{eq:FvK-nondim-2},\ref{eq:FvK-nondim-1}), under the BCs~(\ref{eq:BC-sliding-wrinkling-I}); see details in App.~\ref{sub-app:wrinkling}. 
%
%The results are shown by the green curves in Figs.~\ref{fig:response-1}-\ref{fig:stress-profile}. 
As anticipated by the above discussion, we notice that if the far-edge is clamped, the asymptotic response at large indentation depth, $\tdelta \to \infty$, is $F \sim (Y/R^2)\cdot \delta^3$, hence the system does not reach the extreme wrinkle-assisted softening obtained upon exerting a fixed load at the far edge. Nevertheless, the asymptotic value of the constant, $F/\delta^3$, scales as $1/\log({\cal R})$ (second row of Table II), so that as ${\cal R}$ is increased, the wrinkle-induced suppression of the indentation force becomes more and more effective.  

%We note that the presence of wrinkles on the substrate underlies a sub-cubic asymptotic response, namely $F/\delta^3 \sim \log^{-1}(\tdelta) \to 0$ as $\tdelta \to \infty$, reflecting a logarithmic suppression of the radial stress at the hole's edge with respect to the bare indentation-induced stress: $\srr(R) \sim \log^{-1}(\tdelta)\cdot Y\cdot(\delta/R)^2$. The expansion of wrinkles on the supported zone of the sheet affects strongly also the displacement field, where the slope at the hole's edge now approaches asymptotically the ``natural'' cone angle: $a \to \tfrac{\delta}{R}\cdot[1- O(\log^{-1}(\tdelta))]$, and the size of the unwrinkled core vanishes, $\tL \sim \log^{-1}(\tdelta)$.   

\section{%Relieving compression in a supported sheet: \\ 
Wrinkling {\emph{vs.}} delamination} \label{sec:wrink-delam}
In our model we assume that relieving compression in the supported part of the sheet does not require the formation of delaminated zones, in which the sheet-substrate distance $d$ exceeds the width of the VdW potential well (Fig.~\ref{fig:schem-slide}d), but merely tiny deviations of $d$ from the thermodynamic equilibrium value $d_{\rm min}$. The crucial distinction between these deformation types stems from the 
respective %different 
energetic costs (per area) of sheet-substrate attachment: %associated with each of them: 
\begin{gather}
{\rm delamination}:  \usub \!\approx\! V({d_{\rm min}})  % {\rm (independent \ on} \ d)} 
\label{eq:VdWmodel-1} \\ 
{\rm Zhang\!-\!Witten}: % \ deflection}:  
\usub(d) \!\approx\! \tfrac{1}{2}V''({d_{\rm min}})\!\cdot\! (d\!-\!d_{\rm min})^2  \label{eq:VdWmodel-2}
\end{gather}
With the Zhang-Witten stiffness, $\Ksub = V''({d_{\rm min}})$, %Eq.~(\ref{eq:KsubWitten}), 
a rigid substrate that supports a thin sheet is merely an example of a ``Winkler foundation'' \cite{TimoshenkoBook}, hence the response of the sheet to compression is analogous to other examples of this basic model, such as a sheet floating on a liquid bath (where $\Ksub = \rho_{liq} g$ with $\rho_{liq}$ being the liquid's mass density). For Winkler-like problems, planar deformations are unstable to wrinkling -- periodic undulations characterized by a {\emph{single wavelength}} $\lambda \sim (B/\Ksub)^{1/4}$ (see Eq.~\ref{eq:local-lam-law-1}) -- which emerges through a supercritical (second order) instability of the planar state, not involving any energy barrier.

In contrast, the finite, $d$-independent energy $V(d_{\rm min})$ associated with delamination, Eq.~(\ref{eq:VdWmodel-1}), which one may view as a surface energy penalty, entails a strictly different instability of the planar state. This instability is sub-critical (first order), and therefore requires the crossing of an energy barrier, which in turn gives rise hysteresis loops. Furthermore, the basic deformation mode \cite{Wagner12,Davidovitch20} is a {\emph{single delaminated zone}}, which may accommodate any excess length by increasing the sheet-substrate distance $d$ without further energy cost, as is indicated by Eq.~(\ref{eq:VdWmodel-1}), rather than by forming multiple delamination zones. Even though periodic delamination patterns have been observed under certain circumstances (such as uniaxial compression of a sheet attached to compliant substrate \cite{Vella09}), those patterns are characterized by two length scales, whereby the width of each delaminated zone is much smaller than the distance between them (where the sheet remains fully laminated). Hence, even if the  indentation-induced hoop compression leads to delamination instability, 
the number of blisters at a given distance $r$ should be $\ll 2\pi r/\lambda$, where $\lambda$ is the average width of an individual blister. This suggests that a recent attempt to describe such a delamination pattern by a wrinkling-like sinusoidal profile, characterized by single wavelength $\lambda$ \cite{Dai2020}, is nonphysical.    

In order to determine which of the two deformation types, described by Eqs.~(\ref{eq:VdWmodel-1}) and (\ref{eq:VdWmodel-2}), is likely to relieve hoop compression in a given indentation experiment, 
%While the above paragraphs suggest that the Zhang-Witten response, described by Eqs.~(\ref{eq:KsubWitten},\ref{eq:VdWmodel-2}), is a plausible mechanism for relaxing indentation-induced compression of a sheet supported by a rigid substrate, we do expect that delamination may be unavoidble at a certain parameter regime. In order to elucidate this, 
we note two necessary conditions for a wrinkle pattern to be physically realizable. 

{\emph{(i)}} The wrinkle wavelength $\lambda$ (Eq.~\ref{eq:local-lam-law-1}, with $\Keff=\Ksub$, Eq.~\ref{eq:KsubWitten}) must exceed the length $\ell_{bend} = \sqrt{B/Y}$, otherwise the bending energy would be too large, rendering wrinkles energetically unfavorable. In terms of the parameters of our model this condition reads: 
\begin{equation}
\ell_{bend} \ll \ell_{VdW}  \ , 
%V''(d_{min}) \ll Y/\ell_{bend}^2 
\label{eq:cond-wrink-1}
\end{equation}
where we defined the length scale: 
\begin{equation}
\ell_{VdW} \equiv \sqrt{Y/V''(d_{min})} \ , 
    \label{eq:ell-vdw}
\end{equation}

{\emph{(ii)}} The wrinkle amplitude $d$ must not exceed a length $d_{max}$ above which the sheet ``escapes'' from the attractive zone of the VdW potential (see schematic Fig.~\ref{fig:schem-slide}d), and the energetic cost transitions from Eq.~(\ref{eq:VdWmodel-2}) to Eq.~(\ref{eq:VdWmodel-1}). Noting that the ratio between the wrinkle amplitude and wavelength is ``slaved'' to the excess hoop length, 
$(|d-d_{min}|/\lambda)^2 \sim -\rmur/r$ (such that the wrinkly undulations ``waste'' just the right arclength necessary to suppress hoop compression \cite{Davidovitch11}), and using the estimate $\rmur \sim - \delta^2/R$ (Eq.~\ref{eq:radial-scale}), we obtain the second condition: 
\begin{equation}
\frac{\delta}{R} \ll \frac{|d_{max}-d_{min}|}{\sqrt{\ell_{bend} \cdot \ell_{VdW}}} \ . 
    \label{eq:cond-wrink-2}
\end{equation}
The two conditions~(\ref{eq:cond-wrink-1},\ref{eq:cond-wrink-2}) define a parameter regime in which we expect the wrinkle patterns assumed in our model to be a feasible, energetically-favorable mechanism for relaxing the hoop compression induced by indentation and sliding. If condition~(\ref{eq:cond-wrink-1}) is violated, an axisymmetric (unwrinkled) deformation in the supported portion of the sheet (SubSec.~\ref{subsec:II-sliding}) is stable against wrinkling, and delamination may occur, through a sub-critical instability, at some large  indentation depth directly from the planar state. If condition~(\ref{eq:cond-wrink-1}) is satisfied, the supported portion of the sheet becomes unstable to wrinkling at $\tdeltast(\beta)$ (Eq.~\ref{eq:threshold-stst}), and delamination is expected to occur when the indentation depth $\delta$ reaches ${ R (d_{max}-d_{min})}/{\sqrt{\ell_{bend} \cdot \ell_{VdW}}}$. 

A crude estimate of the various lengths in the conditions~(\ref{eq:cond-wrink-1},\ref{eq:cond-wrink-2}) may be obtained by assuming 
$V''(d_{min}) \sim V(d_{min})/d_{min}^2$, and $ 0.1 {\rm nm} <d_{min} < d_{max} < 1 {\rm nm}$. For Graphene (on SI or BN), we use the values $Y \sim 300 N/m \ , \ \ell_{bend} \sim 0.1 {\rm nm}$, and $V(d_{min}) \sim 0.1 N/m$. With these values, 
%and recalling that the maximal indentation depths are $\ll R$, 
we find that both conditions~(\ref{eq:cond-wrink-1}) and (\ref{eq:cond-wrink-2}) are satisfied for $\delta/R < 0.1$, % smaller than $0.1$, 
suggesting the relevance of a wrinkle-assisted compression-relieving mechanism for experiments, at least at indentation depths $\delta \lesssim 100$ nm.

\section{Discussion \label{sec:discussion}} 

\subsection{The non-perturbative macroscale effect of bending rigidity}
%The primary lesson that may be drawn from analyzing our theoretical model is the subtle yet crucial way by which bending rigidity affects the mechanics of a highly bendable sheet, even though the deformation is governed by tensile strain and the characteristic scale of deformation is $\gg \ell_{bend}=\sqrt{B/Y}$. 
Employing standard TFT (Secs.~\ref{sec:clamping_sliding},\ref{sec:Expansion}) or its recently generalized version (Sec.~\ref{sec:residual_compression}) we showed that, as long as there is %deformation gives rise to 
compressive stress somewhere within the indented sheet, the ability to relax it by energetically-inexpensive wrinkles acts to suppress considerably the elastic energy. Our results, summarized in Tables I and II, show that the wrinkle-assisted reduction of elastic energy and the consequent suppression of the indentation force $F(\delta)$ is a {\emph{non-perturbative}} phenomenon, which is not sensitive to the specific value of the bending modulus, but rather stems from its mere smallness ({\emph{i.e. $\epsilon \ll 1$}}). That is, for specific BCs ({\emph{e.g.}} sliding at $r=R$ and a constant tensile load at $r=\Rf$), we find that the error incurred by ignoring the effect of wrinkles on the indentation force is $O(Y/R^2)\delta^3$, as one can see by comparing the second row of Table I (which ignores the effects of wrinkles, describing a mechanically-unstable state 
for $\tdelta \!>\!\tdelta_c$) with the third row of Table I or the first two rows of Table II.   

While we focused our study on the pointwise indentation problem, the above lesson is general and applies to any situation in which a confining geometry or loading conditions give rise to compressive stress within a thin, highly bendable sheet. One example, which has attracted some interest lately, is the strain induced in a 2D solid sheet, supported on a smooth substrate, by      
%This was studied in a related example, which addressed the (the shape of 
high-pressure ``bubbles'' confined between the sheet and the substrate \cite{Khestanova16}. Such bubbles cause radial stretching of the sheet around the bubble axis, and -- similarly to the indentation problem (with sliding BCs) -- a hoop compression emerges in the sheet at the vicinity of the bubble's edge. While a wrinkle-assisted suppression of hoop compression may not have a pronounced effect on the bubble's shape or the pressure within it \cite{Khestanova16}, the strain components in the sheet are strongly affected by the presence of wrinkles. This effect, however, has been overlooked in a recent paper \cite{Dai18}, where the authors computed the strain tensor by assuming a mechanically-unstable (unwrinkled, axisymmetric) deformation of the sheet.

\subsection{Beyond ideal mechanics -- substrate roughness and thermal fluctuations} 
Our model assumes a smooth, homogeneous substrate, such that the only energetic cost of sliding stems from the consequent hoop compression. From a pure mechanical perspective, a roughness of the substrate may give rise to localized or extended zones in which the sheet is pinned to the substrate, hindering its sliding inwards. A simple, effective-medium-theory approach to incorporate surface roughness into our model may be to replace the {\emph{control parameters}} $\geff$ and ${\cal R}$ in the last two rows of Table II with {\emph{effective parameters}} that account for the excess radial tensile and clamping (away from the hole's edge), associated with the hindrance of sliding. A more thorough study of the effects of surface roughness, as well as thermal fluctuations, on the indentation force, should account for the anomalous elasticity\cite{NP87,AL88,LR92,Radzihovsky18} that has been predicted for 2D solid membranes such as Graphene at room temperature \cite{Blees14,Kosmrlj17,KatsnelsonBook}.

%\subsection{Summary}

\subsection{Summary}
%$\bullet$ Summary -- clamping versus sliding. 

%$\bullet$ Use slope at $r=R$ to determine nature of BC (sliding or not). 

%Despite its ideal nature, the 
The main purpose of the 
%The motivation underlying the 
ideal model we introduced in this paper is to elucidate the crucial assumptions one has to make in order to extract the stretching modulus of a suspended sheet from indentation experiments. In this context, the central outcome of our analysis is that %since it shows that 
sliding and wrinkling of the sheet affect significantly the commonly-assumed cubic dependence of the indentation force, $F/\delta^3 \propto (Y/R^2)$; the assumption of clamping at the edges of the suspended sheet gives a lower bound to the value of the Young modulus. If the membrane can slide over the non suspended zone, the force required to achieve a given deformation can be significantly lower than in the case of clamping.
%ignoring this effect may lead one to underestimating the stretching modulus. 
%dependence of the indentation force $F (\delta )$ on the stretching modulus $Y$ may not be described by a simple cubic law. 
%This %central observation 
This message is illustrated most conspicuously % vividly 
in the geometry-dominated nature of the pseudo-linear response, Eq.~(\ref{eq:pseudo-linear-1}), where $F/ \delta$ may depend on a pre-tension $\sigma_0$ or a bending-induced tension $\geff = 2\sqrt{B\Ksub}$, as well as on the radii $R$ (of the hole) and $\Rf$ (of the whole sheet), but not on the stretching modulus $Y$! Such a stretching-independent %, pseudo-linear 
response may be avoided if the attachment to the substrate is sufficiently strong, or if the sheet is clamped at the far edge ($r = \Rf \gg R$). But also in such cases 
%(third row of Table I and second row of Table II, respectively), 
sliding and wrinkling have a significant effect on the indentation force, which must be considered %accounted for 
in order to properly extract the stretching modulus $Y$ from the measured response.  %%%%%%%%%%%%%%%%%%%%%%%%%%%%%%%

Our theoretical model is quite elementary and does not include effects which may be important for experimental set-ups of 2D membranes, such as pinning, spatial disorder, and thermal fluctuations. We suspect that further theoretical progress is required, possibly along the directions outlined above, in order to render our model applicable for a quantitative description of actual experiments. Nevertheless, some basic predictions may be sufficient to test the relevance (or lack thereof) of sliding and wrinkling. Specifically, measuring the slope ($\approx \theta$) of the suspended sheet in the vicinity of the hole's edge may provide a robust, indirect probe for this purpose. A slope that is close to $63\%$ should indicate that the sheet is practically clamped at the hole's edge. A larger slope should indicate a substantial sliding and wrinkling of the sheet in the suspended part and possibly also on the substrate. %in the supported part.              

Beyond its relevance to metrology %the measurement of stretching modulus 
and to studying sliding and wrinkling phenomena, our model highlights the complexity that is often ignored by one's perception of 2D solid membranes as being ``nearly inextensible, highly bendable'' objects, whose resistance to bending can be ignored in analyzing macro-scale, tension-dominated deformations. Instead, our study     
%features which are attributed to graphene as well as other thin sheets -- 
illuminates the subtle role played by both stretching and bending rigidity in the response to such external stimuli.

%%%%%%%%%%%%%%%%%%%%%%%%%%%%%%%%
\begin{acknowledgements}
We thank A. Geim, M. Katsnelson, K. Novoselov, D. Vella, and participants of the program ``Geometry and Elasticity of 2D Soft Matter'' at the Kavli Institute for Theoretical Physics Santa Barbara 2016, where we began working on this manuscript, 
%the work on this manuscript began, 
for many useful discussions. We thank D. Vella for a thorough, critical review of the manuscript.  
We acknowledge support by the National Science Foundation under grants NSF-DMR-CAREER-1151780 and NSF-DMR-1822439 (BD), and by the European Commission, under the Graphene Flagship, Core 3, grant no. 881603, and by the grants NMAT2D (Comunidad de Madrid, Spain), SprQuMat and SEV-2016-0686, (Ministerio de Ciencia e Innovacion, Spain) (FG).  
\end{acknowledgements}

\begin{appendix}
\section{Boundary conditions at the hole's edge and the negligibly of radial curvature} \label{app:krr-BC}
%In Sec.~\ref{sec:clamping_sliding} we assumed that when sliding occurs the radial stress is continuous at the hole's edge, Eq.~(\ref{eq:matchingsrr}). 
In order to elucidate the neglect of the radial bending force $B\partial^4z/\partial r^4$ in Eq.~(\ref{eq:FvK-dim-1}), often referred to as a ``membrane approximation'', as well as the BCs at the hole's edge, let us recall that for a sheet with finite (albeit small) bending modulus $B$ the tangent $\hat{t}$ to the sheet's plane must be a continuous function of the radial distance $r$. A discontinuity of $\hat{t}$ implies a divergence of the radial curvature, $\kappa_{rr} \approx |\partial\hat{t}/\partial r|$, and hence an infinite bending energy, regardless of how small $B$ is. In fact, the vicinity of the hole's edge, where the tangent $\hat{t}$ varies sharply, is the only zone where the radial curvature has to be considered, since it is required to regularize this divergence.
%the only place where the radial bending force    
Specifically, the characteristic length over which occurs the necessary change from $\hat{t}(r\to R^-) \!=\! \cos\theta \hat{r} \!-\! \sin\theta \hat{z}$ to $\hat{t}(r\to R^+) \!=\! \hat{r}$, is the ``local bendo-capillary'' length \cite{Vella18}: 
\begin{equation} \ellbc^* \approx \sqrt{B/\srr(R)}  \ . \label{eq:bendo-cap}
\end{equation}
On one hand, we have that $\ellbc^*\ll R$, since the sheet is highly bendable ({\emph{i.e.}} $\epsilon\ll 1$, see Eq.~(\ref{eq:FvK-nondim-2}) and the following paragraphs); on the other hand we assume $\ellbc^*$ is much larger than the atomic scale (over which the corner in the substrate is ``smoothed out''), see schematic Fig.~\ref{fig:schem-slide}b-c.

In our analysis of the FvK equations, either of the unwrinkled state in Subsec.~\ref{subsec:II-sliding} or the wrinkled state in Subsec.~\ref{subsec:II-wrinkling} and the rest of the paper, we exploited the fact that $\ellbc^* \ll R$, and considered the narrow annulus, 
$R-\ellbc^* < r< R$, as a ``boundary layer'', whose energetic cost may be ignored. More precisely, this excess energy can be estimated as 
$\sim B(\theta/\ellbc^*)^2 R \sim \sqrt{B Y} \delta^3/R^2$, and an inspection of Tables I-II reveals that it is smaller by a factor $\sqrt{\epsilon}$, Eq.(~\ref{eq:DG-0}), than %effect on 
the elastic energy evaluated in Secs.~\ref{sec:clamping_sliding}-\ref{sec:Expansion}.  
%and the indentation force derived from that energy can be safely ignored. 
Hence, neglecting the explicit energetic cost of that boundary 
layer amounts to evaluating the leading order of the elastic energy (and the indentation force derived from it) in an expansion whose small parameter is $\sqrt{\epsilon}$.    
Mathematically, since the radial bending force, $B\partial^4z/\partial^4r$, is significant only in this narrow zone, our analysis has been greatly simplified by omitting this term from the $1^{st}$ FvK Eq.~(\ref{eq:FvK-dim-1}), rendering it -- along with Eq.~(\ref{eq:FvK-dim-2}) -- a coupled set of $2^{nd}$ order ODEs for $\psi(r)$ and $z(r)$, and allowing for a discontinuity of $z'(r)$ at $r=R$.  

The boundary layer approach %({\emph{i.e.}} $\ellbc^* \ll R$) 
implies that the radial and vertical components of the displacement may be considered continuous at $r=R$ yielding the BCs~(\ref{eq:matchingur})  and~(\ref{eq:BC-nondim-sliding-axi}iii), while the derivative of the latter is allowed to be discontinuous ($[z'(r)]_{R^-}^{R+} \approx \theta$). At the same time, the mere existence of the boundary layer underlies the continuity of the radial stress component (even though one may naively view it as a violating a force balance in the horizontal direction at $r=R$), as is illustrated in the schematic Fig.~\ref{fig:schem-slide}. We note that these continuity BCs remain valid even if a small portion of the sheet slides vertically in order to gain some surface energy by contacting the hole's walls (contrast panels b and c in Fig.~\ref{fig:schem-slide}), as long the sheet does not get pinned to the substrate. A detailed discussion of this effect will be discussed elsewhere.   

%, which express continuity of the radial displacement and the vertical displacements (but not its derivative !) at the hole's edge. However, in invoking continuity of the radial stress,  Eq.~(\ref{eq:matchingsrr}), we made the further assumption that the sheet does not contact the hole's walls (Fig.~\ref{fig:hole-edge}a), such that the substrate does not exert any load tangential to the sheet. Another possibility is that a small portion of the sheet sticks to the walls, such that the substrate does exert a tangential load in the vicinity of the hole's edge, pulling the sheet downward (Fig.~\ref{fig:hole-edge}b). In such a case, the BC~(\ref{eq:matchingsrr}) must be replaced by: 

%The physical assumption underlying this BC is that the substrate exerts forces only in a normal direction to the sheet. More precisely,      

%addressed the suitable BCs at the hole's edge in the presence of sliding   

%$\bullet$ The boundary conditions at the hole's edge, $r=R$. If the substrate's shape is smoothed over a scale $r_{tip} \gtrsim \sqrt{B/\srr(R)}$, then the force exerted by the substrate is not restricted to the $\hat{z}$ direction, but may contribute also an external  tangential component ??? in such a case, the assumption of continuity of radial stress may have to be revised ...

%$\bullet$ Give schematic figure that discuss BC in case the sheet slides into the hole.  

%\section{Analytic solution of FvK equation for 
\section{General analysis of the unwrinkled core \label{app:axisymmetric}}
Here we describe the steps underlying an analytic solution for an axisymmetric (unwrinkled) solution the nonlinear FvK equations (\ref{eq:FvK-nondim-2},\ref{eq:FvK-nondim-1}). This solution, with distinct types of BCs, is used to characterize a purely tensile ``core'' around the indenter, which exists under all various conditions (clamping/sliding at the hole's edge, and various parameter regimes, Eqs.~(\ref{eq:regime-i},\ref{eq:regime-ii})). Our exposition follows closely Ref. \cite{Vella17} and the Supplementary Information of Ref. \cite{Vella15} 
\\

We start by integrating the $1^{st}$ FvK equation (\ref{eq:FvK-nondim-1}), and obtain: 
\begin{equation}
\Psi \frac{d \zeta}{d \rho} = {\cal F}
\label{eq:integrate-1}
\end{equation}
Next, we introduce the variable transformation \cite{Bhatia68}:
\begin{equation} 
\Phi = \rho \Psi \ \ , \ \ \eta = \rho^2  \ , 
\label{eq:transform-1}
\end{equation}
such that: $\Psi = \tfrac{\Phi}{\sqrt{\eta}}$, and $\tfrac{d \Psi}{d \rho} = \left( 2 \tfrac{d \Phi}{d \eta} - \frac{\Phi}{\eta} \right)$.
With this transformation, the $2^{nd}$ FvK equation (\ref{eq:FvK-nondim-2}) becomes:  
\begin{equation}
\Phi'' = - \frac{{\cal F}^2}{8 \Phi^2} \ , 
\label{eq:FvK-nondim-tran-2}
\end{equation} 
which can be integrated once, obtaining: 
\begin{align}
\Phi' &= \frac{\cal F}{2} \frac{\sqrt{1 + A \Phi}}{\sqrt{\Phi}}
\label{eq:FvK-nondim-tran-integrate-2}
\end{align}
where $A$ is a constant of integration. Evaluating Eq.~(\ref{eq:FvK-nondim-tran-integrate-2}) at $\eta=1$, we obtain a first equation that involves the unknowns $\Phi'(1), \Phi(1), A$, and ${\cal F}$: 
\begin{align}
\Phi'(1) &= \frac{\cal F}{2} \frac{\sqrt{1 + A \Phi(1)}}{\sqrt{\Phi(1)}}
\label{eq:FvK-nondim-tran-integrate-2-edge}
\end{align} 
%, whose value must be determined separately for each of the physical conditions discussed in our paper (clamping/sliding at the hole's edge, absence of radial wrinkles, or their presence outside of a core of radius $L<R$). 

Integrating now Eq.~(\ref{eq:FvK-nondim-tran-integrate-2}), we obtain an explicit expression between the variable $\eta$ and the function $\Phi(\eta)$: 
\begin{align}
\frac{\sqrt{\Phi(1 + A \Phi)}}{A}   - \sqrt{\frac{1+A\Phi}{A^3}} \sinh^{-1}[\sqrt{A\Phi}] &=  \frac{\cal F}{2} \eta  \ , 
\label{eq:FvK-nondim-tran-integrate-22}
\end{align}
(where we used the BC $\Phi(0) = 0$, which is valid for all cases addressed here). Evaluating the above equation at the hole's edge ($\eta = 1$), we obtain a second equation that involves the unknowns $\Phi(1),A$, and ${\cal F}$: 
 \begin{align}
\frac{\sqrt{\Phi(1)(1 + A \Phi(1))}}{A}   - \sqrt{\frac{1+A\Phi(1)}{A^3}} \sinh^{-1}[\sqrt{A\Phi(1)}] &=  \frac{\cal F}{2}  \  . 
\label{eq:FvK-nondim-tran-integrate-22-edge}
\end{align} 

Turning to the integrated form of the $1^{st}$ FvK Eq.~(\ref{eq:integrate-1}), we re-parametrize the function $\zeta(\rho) \to \zeta[\Phi(\eta)]$. With the aid of Eq.~(\ref{eq:FvK-nondim-tran-integrate-2}), and integration (over $\Phi$), we obtain an explicit form for the shape: 
\begin{equation}
\zeta(\Phi) - \zeta(0) = \frac{2}{\sqrt{A}} \sinh^{-1}[\sqrt{A\Phi}]  \ . 
\label{eq:FvK-nondim-tran-integrate-1}
\end{equation} 
\\

Equations~(\ref{eq:FvK-nondim-tran-integrate-2-edge},\ref{eq:FvK-nondim-tran-integrate-22-edge}) constitute  two equations for the four unknowns: $\Phi'(1), \Phi(1), A$, and ${\cal F}$. These two equations are common to all cases we study in this paper. The other two equations must come from the BCs that reflect the various physical conditions discussed in our paper (clamping/sliding at the hole's edge, absence/presence of wrinkles). 
%of radial wrinkles, or their presence outside of a core of radius $\Li<R$).   

Once the four constants ($\Phi'(1), \Phi(1), A ,{\cal F}$) are determined, Eqs.~(\ref{eq:FvK-nondim-tran-integrate-22},\ref{eq:FvK-nondim-tran-integrate-1}) provide explicit expressions for the functions $\Phi(\eta), \zeta(\Phi)$, which can be directly transformed (through Eqs.\ref{eq:transform-1},\ref{eq:define-Airy}) to the shape, $\zeta(\rho)$, and the stress components: $\sigma_{rr}(r),\sigma_{\theta\theta}(\rho)$.

\section{Clamping at the hole's edge} %(Subsec.\ref{subsec:II-clamping}) 
\label{sub-app:clamping}  
For the clamped case, Subsec.~\ref{subsec:II-clamping}, the BCs (\ref{eq:BC-nondim-clamping}) become: 
\begin{align} 
\eta =0 \ &:  \, \, \, \,  (i ) \, \zeta = -\tdelta \, \, \, \, \, \, \, \, \,(ii ) \Phi = 0 \nonumber \\ 
\eta =1 \ &:  \, \, \, \,  (iii) \, \zeta = 0 \, \, \, \, \, \, \, \, \, (iv) \, 2\Phi'  = (1-\nu) + (1+\nu) \Phi  \ . 
\label{eq:BC-nondim-trans-clamping} 
\end{align}  
Among these BCs, {\emph{(ii)}} was used already to obtain Eq.~(\ref{eq:FvK-nondim-tran-integrate-22}). Since the FvK equations (\ref{eq:FvK-nondim-2},\ref{eq:FvK-nondim-1}) are invariant under: $\zeta \to \zeta + c$, only the difference $\zeta(1) - \zeta(0)$ can affect the physics, and hence the three remaining BCs in (\ref{eq:BC-nondim-trans-clamping}) give rise to two equations that involve the unknowns ($\Phi'(1), \Phi(1), A,{\cal F}$). The first equation is simply BC {\emph{(iv)}}: 
\begin{equation} 
2\Phi'(1)  = (1-\nu) + (1+\nu) \Phi(1)  \ ,  
\label{eq:BC-iv-clamping}
\end{equation}    
and the second equation is obtained by evaluating Eq.~(\ref{eq:FvK-nondim-tran-integrate-1}) at $\Phi(1)$, and  
substituting for the difference: $\zeta(\Phi(1))-\zeta(\Phi(0)) = \tdelta$:
\begin{equation} 
\tdelta = \frac{2}{\sqrt{A}} \sinh^{-1}[\sqrt{A\Phi(1)}]  \ . 
\label{eq:FvK-nondim-tran-integrate-1-edge}
\end{equation}

Solving the four algebraic equations (\ref{eq:FvK-nondim-tran-integrate-2-edge},\ref{eq:FvK-nondim-tran-integrate-22-edge},\ref{eq:BC-iv-clamping},\ref{eq:FvK-nondim-tran-integrate-1-edge}), is straightforward ({\emph{e.g.}} using Mathematica's ``FindRoot"), and allows us to obtain the constants 
$\Phi'(1), \Phi(1), A,{\cal F}$, as a function of the single dimensionless parameter $\tdelta$. The response function ${\cal F} (\tdelta)$, the deformed shape, and the stress profile (which are evaluated with the aid of
Eqs.~(\ref{eq:FvK-nondim-tran-integrate-22},\ref{eq:FvK-nondim-tran-integrate-1},\ref{eq:transform-1},\ref{eq:define-Airy}), are shown in the gray curves in Figs.~\ref{fig:response-1}-\ref{fig:shape}.

\section{Sliding (no wrinkling)} 
%(Subsec.~\ref{subsec:II-sliding}) 
\label{sub-app:sliding}

The BCs that corresponds to an axisymmetric (unwrinkled) state, for which the sheet can slide on the substrate were derived in Subsec.~\ref{subsec:II-sliding}. The difference between clamped-edge and sliding boils down to replacing the BC {\emph{(iv)}} in  Eq.~(\ref{eq:BC-nondim-clamping}) with the corresponding BC in Eq.~(\ref{eq:BC-nondim-sliding-axi}). Hence, the algebraic equations for the four unknowns $\Phi'(1), \Phi(1), A,{\cal F}$ are Eqs.~(\ref{eq:FvK-nondim-tran-integrate-2-edge},\ref{eq:FvK-nondim-tran-integrate-22-edge},\ref{eq:FvK-nondim-tran-integrate-1-edge}), and: 
 \begin{equation} 
\Phi'(1)  = 1  \ .  
\label{eq:BC-iv-sliding-axi}
\end{equation}   
The response function ${\cal F} (\tdelta)$, the deformed shape, and the stress profile 
%(which are evaluated with the aid ofEqs.~(\ref{eq:FvK-nondim-tran-integrate-22},\ref{eq:FvK-nondim-tran-integrate-1},\ref{eq:transform-1},\ref{eq:define-Airy}), 
that correspond to this solution,  
are shown in the blue curves in Figs.~\ref{fig:response-1}-\ref{fig:shape}.
 
%%%%%%%%%%%%%%%%%%%%%%%%%%%%%%%%%%%%%%%%%%%%%%
\section{Sliding and Wrinkling % in suspended sheet 
\label{sub-app:wrinkling}}  
%%%%%%%%%%%%%%%%%%%%%%%%%%%%%%%%%%%%%%%%%%%%%%
If the sheet can slide at the hole's edge, hoop compression evolves around the hole's edge and the compressed zone expands upon increasing indentation depth,  in a manner that depends on the sheet-substrate attachment (through the parameter $\beta$, Subsecs.~\ref{subsec:II-wrinkling},\ref{subsec:both-wrinkling}), the sheet's size (through the parameter ${\cal R}$, Subsec. \ref{subsec:geo-limit}), and the boundary conditions at the far edge (Sec.~\ref{sec:Expansion}). Central to  all of these cases is the presence of a purely tensile, unwrinkled core, $0<r<\Li$, around the indenter, where the deformation is described by solving   
%The solution of this problem is conveniently addressed by considering first the solution of 
the axisymmetric FvK equations (\ref{eq:FvK-nondim-2},\ref{eq:FvK-nondim-1}), subject to Eq.~(\ref{eq:wrinkled-I}) and the BCs~(\ref{eq:BC-sliding-wrinkling-I}), that yield two equations (\ref{eq:twomore}) for the three unknowns $\Psi(1),\ta, \tL$. The various cases in Subsecs.~\ref{subsec:II-wrinkling},\ref{subsec:both-wrinkling},\ref{subsec:geo-limit}, and Sec.~\ref{sec:Expansion} differ only in the final equation that connects $\Psi(1),\ta, \tL$, which stems from the continuity of radial displacement at the hole's edge (Eqs.~\ref{eq:thirdmatching}, \ref{eq:thirdmatching-1},\ref{eq:thirdmatching-1a},\ref{eq:thirdmatching-2}, respectively). In the following, we obtain the first two algebraic equations for $\Psi(1),\ta, \tL$, that are common to all of these cases.      

%Considering first the last two parts, one notices that the displacement and stress are given by expressions identical to their counterparts in Subsec.~{\ref{subsec:II-wrinkling}}, Eqs.~(\ref{eq:wrinkled-I}) and the BCs~(\ref{eq:BC-sliding-wrinkling-I}) for the nonlinear FvK equations (\ref{eq:FvK-nondim-2},\ref{eq:FvK-nondim-1}) in the unwrinkled core, albeit with a different triplet of constants $\Psi(1),\ta, \tL$, that have to be determined by matching the radial displacement at the hole's edge $\rmur(R)$ with the wrinkled sheet at the exterior of the hole. Thus, among the three equations that specify the constants $\Psi(1),\ta, \tL$, two equations (that stem from continuity of the displacement and stress at $r=\Li$) are identical to their counterparts in Eq.~(\ref{eq:twomore}).  

%In order to find the remaining equation that relates the constants $\Psi(1),\ta, \tL$, 

 %, at the core zone, $0<r<\Li$. 
Following Ref.~\cite{Vella15} (Sec.~3 of {\emph{Supplementary information}}), it is convenient to replace the dimensionless variables (\ref{eq:dimensionless-var}) with: 
\begin{gather} 
\brho = \frac{r}{\Li} = \frac{\rho}{\tL}  \ \ ; \ \ \bPsi = \frac{\psi}{\srr(R)R} = \frac{\Psi}{\Psi(1)} \ \ ; \nonumber \\
\bzeta = \frac{z}{\sqrt{RL}} \sqrt{\frac{Y}{\srr(R)}}  = 
\frac{\zeta}{\sqrt{\tL \Psi(1)}}
% \sqrt{\frac{Y}{\srr(R)}} 
%=\frac{z}{R}\tL^{-1/2} \sqrt{\frac{Y}{\srr(R)}} \ , 
\label{eq:dimensionless-var-new}
\end{gather}
and the dimensionless force %parameter 
%$\bar
${\cal F}$ (\ref{eq:dimensionless-param}) with: 
\begin{equation}
\bar{\cal F} %= \frac{1}{2 \pi R} \frac{F}{\sigma_0 \sqrt{\frac{\sigma_0}{Y}}} 
=  \frac{1}{2 \pi R} 
\sqrt{\frac{\Li}{R}} \sqrt{\frac{Y}{\srr(R)^3}} \ F   \ = \
{\cal F}\sqrt{\frac{\tL}{\Psi(1)^3}}  \ , 
%\frac{\tL}{\sqrt{\Psi(1)^3 \tL}} = 
%\frac{1}{2 \pi R} \sqrt{\frac{Y}{\srr(R)^3}} \tL^{1/2} \ F 
% , \, \, \, \tilde{\delta} =  \frac{1}{R} \sqrt{\frac{Y}{\sigma_0}} \ \delta \ . 
 \label{eq:dimensionless-param-new}
\end{equation}
such that the BCs in (\ref{eq:BC-sliding-wrinkling-I},\ref{eq:twomore}) that involve explicitly the function $\bPsi(\brho)$ are:
\begin{equation}
\bPsi(\brho=1) =1 \ \ ; \ \ \bPsi(\brho=0)=0 \ \ ; \ \  \bPsi'(\brho=1)=0 \ .  
\label{eq:BC-new-new}
\end{equation}  
Using a similar manipulation to the one employed earlier, 
%in Subsec.~\ref{app:axisymmetric}, 
we make the additional transformation:   
\begin{gather}
\Phi  = \brho \bPsi \ \ ; \ \  \eta = \brho^2 \ ,  
\end{gather}
with which the BCs~(\ref{eq:BC-new-new}) become $\Phi(\eta = 0)=0 \ , \ \Phi(\eta = 1)=1 \ , \ \Phi'(\eta = 1)=\tfrac{1}{2}$, and the implicit expression for $\Phi(\eta)$, Eq.~(\ref{eq:FvK-nondim-tran-integrate-22}), is fully satisfied by the  
numerical constants $A, \bar{\cal F}$, through the algebraic equations (\ref{eq:FvK-nondim-tran-integrate-2-edge},\ref{eq:FvK-nondim-tran-integrate-22-edge}). Solution of these equations yield the numerical values: 
\begin{equation}
A \approx -0.697 \ \ ; \ \ \bar{\cal F} \approx 1.815 \ , \label{eq:nn11}
\end{equation}    
which were found already in \cite{Vella15}. Equation~(\ref{eq:FvK-nondim-tran-integrate-1}), with $\zeta\to \bzeta$, together with the BCs for $\zeta$ in Eq.~(\ref{eq:BC-sliding-wrinkling-I}) yield: % the equation: 
\begin{gather}
\left(\ta\cdot (\tL-1) + \tdelta\right)\cdot \left(\tL \cdot \Psi(1)\right)^{-1/2} \nonumber \\
=\frac{2}{\sqrt{A}} \sinh^{-1}(\sqrt{A}) \approx 2.367 \ , 
\label{eq:new-new-11a}
\end{gather}
and the BC for the slope (\ref{eq:twomore}) becomes:
\begin{equation}
\ta\cdot \left(\tL/\Psi(1)\right)^{1/2} =  (1+A)^{-1/2} \approx 1.815  \ .
\label{eq:new-new-11b}
\end{equation} 
For any value of the control parameter $\tdelta \gtrsim 3.3$, Eqs.~(\ref{eq:new-new-11a},\ref{eq:new-new-11b}), together with Eq.~(\ref{eq:thirdmatching}) for Subsec.~\ref{subsec:II-wrinkling}, or Eq.~(\ref{eq:thirdmatching-1}) for Subsec.~\ref{subsec:both-wrinkling}, or Eq.~(\ref{eq:thirdmatching-1a}) for Subsec.~\ref{subsec:geo-limit}, or Eq.~(\ref{eq:thirdmatching-2}) for Sec.~\ref{sec:Expansion}, form a set of 3 nonlinear algebraic equations for the three unknowns, $\Psi(1),\tL,\ta$. The solutions of these equations 
%(which we obtained with the aid of Mathematica, by using the ``FindRoot" package) 
fully characterize the shape and stress of the deformed sheet in each case, and the corresponding indentation force is obtained with the aid of Eqs.~(\ref{eq:dimensionless-param-new},\ref{eq:nn11}). 
%These solution are shown in the brown, purple and .. curves in Figs.~. .

%%%%%%%%%%%%%%%%%%%%%%%%%%%%%%%%%%%%%%%%%%%%%%
%\subsection{Sliding and Wrinkling in suspended and supported portions \label{sub-app:wrinkling-beyond}}  
%%%%%%%%%%%%%%%%%%%%%%%%%%%%%%%%%%%%%%%%%%%%%%
%Finally, 

\end{appendix}
\bibliography{bib_suspended_2}
\newpage
%%%%%%%%%%%%%%%%%%%%%%%%%%%%
%%Figures 

\end{document}